\documentclass[aps,prx,twocolumn,secnumarabic,amsmath,amssymb,superscriptaddress,longbibliography]{revtex4-2}
\usepackage[makeroom]{cancel}
\usepackage{titlesec}
\usepackage{bm}
\usepackage{comment}
\usepackage{verbatim}

\usepackage{bbm}

\usepackage{physics}

\usepackage{graphicx}
\usepackage{pict2e}
\usepackage{subfigure}
\usepackage{tabularx}

\usepackage{diagbox}

\makeatletter
\newcommand{\thickhline}{%
    \noalign {\ifnum 0=`}\fi \hrule height 1.25pt
    \futurelet \reserved@a \@xhline
}
\newcolumntype{[}{@{\vrule width 1pt\hspace{6pt}}} 
\newcolumntype{]}{@{\hspace{6pt}\vrule width 1pt}} 
\newcolumntype{\thv}{@{\hskip\tabcolsep\vrule width 1pt\hskip\tabcolsep}}
\makeatother

\usepackage{dcolumn}% Align table columns on decimal point

\usepackage[mathscr]{euscript}

\usepackage{booktabs}

\usepackage{epsf}
\usepackage{epsfig}
\usepackage{color}
\usepackage{graphicx}
\usepackage{rotating}

\usepackage{color}
\usepackage[colorlinks,bookmarks=false,citecolor=blue,linkcolor=red,urlcolor=blue]{hyperref}

\definecolor{darkred}{rgb}{0.5,0.0,0.0}
\def\cdr{\color{darkred}}
\definecolor{darkblue}{rgb}{0,0.02,0.45}

\definecolor{darkgreen}{rgb}{0.02,0.45,0.0}

\definecolor{violet}{rgb}{0.8,0.2,0.6}

\definecolor{mygray}{gray}{0.3}

\def\da{\downarrow}

\def\be{\begin{equation}}
\def\ee{\end{equation}}
\def\bea{\begin{eqnarray}}
\def\eea{\end{eqnarray}}

\def\mc{\mathcal}

\usepackage{times}

\usepackage{orcidlink}

\begin{document}
\date{\today}

\title{Spin-$S\,$ Kitaev-Heisenberg model on the honeycomb lattice:  A 
high-order treatment via the many-body coupled cluster method}

\author{M.~Georgiou\,\orcidlink{0000-0003-0135-7389}}
\email{M.Georgiou@lboro.ac.uk}
\affiliation{Department of Physics, Loughborough University, Leicestershire LE11 3TU, United Kingdom}

\author{I.~Rousochatzakis\,\orcidlink{0000-0002-5517-8389}}
\email{I.Rousochatzakis@lboro.ac.uk}
\affiliation{Department of Physics, Loughborough University, Leicestershire LE11 3TU, United Kingdom}

\author{D.~J.~J.~Farnell\,\orcidlink{0000-0003-0662-1927}}
\email{FarnellD@cardiff.ac.uk}
\affiliation{School of Dentistry, Cardiff University, Cardiff CF14 4XY, Wales, United Kingdom}

\author{J.~Richter\,\orcidlink{0000-0002-5630-3786}}
\email{Johannes.Richter@Physik.Uni-Magdeburg.DE}
\affiliation{Institut f{\"u}r Physik, Otto-von-Guericke-Universit{\"a}t Magdeburg, P.O. Box 4120, 39016 Magdeburg, Germany}

\author{R.~F.~Bishop\,\orcidlink{0000-0001-5565-0658}}
\email{raymond.bishop@manchester.ac.uk}
\affiliation{Department of Physics and Astronomy, The University of Manchester, Manchester M13 9PL, United Kingdom}
\affiliation{Department of Physics, Loughborough University, Leicestershire LE11 3TU, United Kingdom}

\begin{abstract}
We study the spin-$S$ Kitaev-Heisenberg model on the honeycomb lattice for $S\!=\!1/2$, $1$ and $3/2$, by using the coupled cluster method (CCM) of microscopic quantum many-body theory. This system is one of the earliest extensions of the Kitaev model and is believed to contain two extended spin liquid phases for any value of the spin quantum number $S$.
We show that the CCM delivers accurate estimates for the phase boundaries of these spin liquid phases, as well as other transition points in the phase diagram. 
Moreover, we find evidence of two unexpected narrow phases for $S\!=\!1/2$, one sandwiched between the zigzag and ferromagnetic phases and the other between the N\'eel and the stripy phases.
The results establish the CCM as a versatile numerical technique that can capture the strong quantum-mechanical fluctuations that are inherently present in generalized Kitaev models with competing bond-dependent anisotropies.
\end{abstract}

\maketitle

%\pagebreak

\section{Introduction}\label{sec:intro}
Mott insulators with strong spin-orbit coupling (SOC) have attracted a great deal of interest in recent years as a promising playground for exploring unconventional states of quantum magnetism, most notably quantum spin liquids (QSLs)~\cite{Kitaev2006,Jackeli2009PRL,Chaloupka2010PRL,Krempa2014ARCMP,Rau2016ARCMP,Winter2016PRB,Winter2017r,Knolle2017ARCMP,Takagi2019NRP,Janssen2019,Motome2019JPSJ,Takayama2021JPSJ,Trebst2022,Tsirlin2022,RousochatzakisRoPP2024}. In these systems, the synergy of the strong SOC with crystal field effects and strong electron-electron interactions gives rise to effective, low-energy pseudospin degrees of freedom that are characterised by entangled spin-orbital wavefunctions. The inherent anisotropy of these degrees of freedom is manifested in the form of highly anisotropic, bond-dependent exchange interactions~\cite{Khaliullin2005,Jackeli2009PRL,Chaloupka2010PRL}. 

\begin{figure}[!b]
\includegraphics[width=0.55\linewidth]{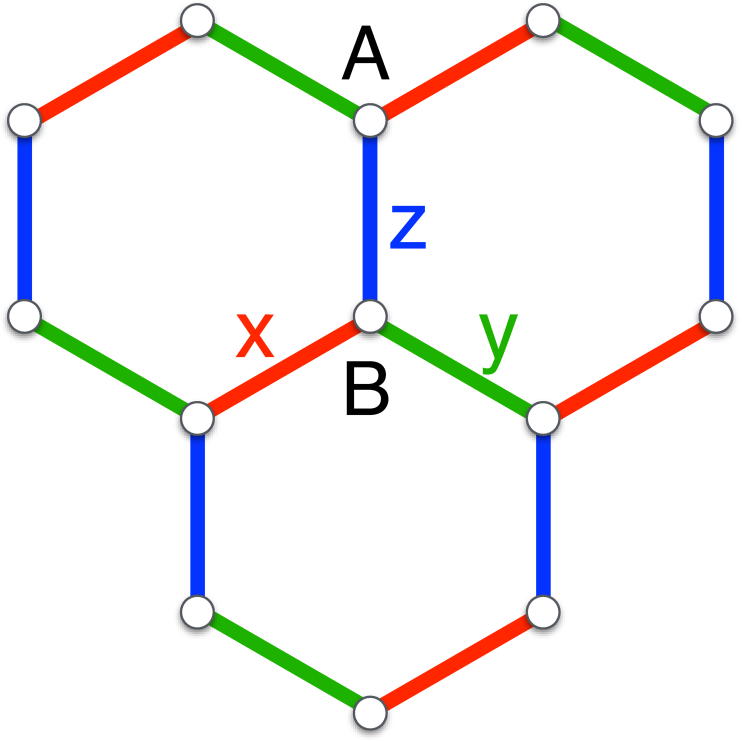}
\caption{The 2D honeycomb lattice, with `$x$', `$y$' and `$z$' labelling the three types of NN bonds, and A and B denoting the two sublattices.}\label{fig:Lattice}
\end{figure}

Theoretical and experimental activities have focused on a family of two- (2D) and three-dimensional (3D) tri-coordinated 4$d$ and 5$d$ materials that appear to be proximate to the celebrated Kitaev model, one of the few exactly solvable models with gapped and gapless QSL ground states~\cite{Kitaev2006,Chen2008,Mandal2009PRB,Obrien2016}. 
The nearest-neighbor (NN) bonds on these tri-coordinated lattices split into three types, generally labelled by `$x$', `$y$' and `$z$',  see Fig.~\ref{fig:Lattice}.
In the Kitaev model, the pseudospin degrees of freedom residing on these bonds interact with each other via bond-dependent, Ising-like couplings, of the form $S_i^{\alpha_{ij}}S_j^{\alpha_{ij}}$, where $\alpha_{ij}=x$, $y$ or $z$ for bonds of type `$x$', `$y$' or `$z$', respectively. More explicitly, the Kitaev model takes the form~\cite{Kitaev2006} 
\be
\mc{H}_{K} = K \Big\{\sum_{\langle ij\rangle_x} S_i^x S_j^x+ \sum_{\langle ij\rangle_y}S_i^y S_j^y
+\sum_{\langle ij\rangle_z}S_i^z S_j^z
\Big\}\,,
\ee
where $K$ denotes the strength of the Kitaev coupling, and $\langle ij\rangle_\alpha$ indicates a NN bond of type $\alpha$, with $\alpha = x,y,z$. 
A realistic description of known Kitaev materials necessitates adding further types of bond-dependent interactions, such as the $\Gamma$ and $\Gamma'$ couplings (related to the symmetric part of the exchange anisotropy), as well as the isotropic Heisenberg exchange $J$~\cite{Jackeli2009PRL,Chaloupka2010PRL,Krempa2014ARCMP,Rau2016ARCMP,Winter2016PRB,Winter2017r,Knolle2017ARCMP,Takagi2019NRP,Janssen2019,Motome2019JPSJ,Takayama2021JPSJ,Trebst2022,Tsirlin2022,RousochatzakisRoPP2024}.

The physics of these bond-dependent models has been explored with various state-of-the-art numerical techniques, including exact diagonalization (ED) and cluster mean field theory (CMFT) calculations~\cite{Chaloupka2010PRL,Chaloupka2013PRL,RousochatzakisPRX2015,Gotfryd2017,Catuneanu2018npj,Hickey2020PRR,Rousochatzakis2019PRB}, density-matrix renormalization group (DMRG)~\cite{Jiang2011PRB,Shinjo2015PRB,Gohlke2017PRL,Gohlke2018PRBa,Gohlke2018PRBb,Gordon2019NC,Dong2020PRB,Gohlke2020PRR}, tensor-network methods~\cite{Osorio2014,Czarnik2019,Lee2020NC}, functional renormalization group (fRG)~\cite{Reuther2011PRB,RousochatzakisPRX2015,Buessen2021PRB,Fukui2022}, slave-particle mean-field  theories~\cite{Burnell2011PRB,Schaffer2012PRB,Johannes2018PRB}, Schwinger boson mean field theory~\cite{Ralko2024}, variational Monte Carlo (VMC)~\cite{Normand2019} and machine learning~\cite{Rau2021PRR,Liu2021PRR}.
Quite generally, these investigations are challenging for various reasons, most notably: (i) the inherent complexity and low symmetry of the materials (which mirror the complex interplay of electron correlations, crystal field, SOC, and structural characteristics), (ii) the strong frustration associated with the bond-dependent interactions, (iii) the rich phase diagrams and general fragility of QSL phases, and (iv) the necessity to treat both quantum and thermal fluctuations on an equal footing~\cite{Rau2016ARCMP,Winter2016PRB,Winter2017r,Takagi2019NRP,Trebst2022,Tsirlin2022,RousochatzakisRoPP2024}.
Importantly, these challenges extend beyond the more interesting spin liquid regions (where the bond-dependent terms compete the most), even in regions where long-range magnetic order settles in at low energy scales. In these regions, standard semiclassical approaches may be inadequate to correctly capture the effect of quantum fluctuations, due to strong magnon-magnon interactions and decay processes~\cite{Winter2017,Smit2020,Maksimov2020PRR}, which become further amplified in the vicinity of the QSL regions.

In this work, we establish that such strong quantum fluctuations, which are present inherently in generalized Kitaev-like models, can be captured accurately and in a systematic way by using the coupled cluster method (CCM)~\cite{Kummel_Luhr-Zab_1978,Bishop-Luhrmann_1978,Emrich_1981a,Emrich_1981b,Bishop-Luhrmann_1982,Arponen_1983,Bishop-Kummel_1987,Arp-Bish-Paj_1987a,Arp-Bish-Paj_1987b,Bishop_1991,Arponen-Bishop_1991,Arponen-Bishop_1993a,Bishop_1998}, which is a versatile method of quantum many-body theory.
Here we consider the Kitaev-Heisenberg (KH) model on the honeycomb lattice, which is one of the earliest extensions of the Kitaev model~\cite{Jackeli2009PRL}. This model has been crucial in establishing that Kitaev QSLs can survive in an extended parameter region. % (and not just at the special Kitaev points). 
We also extend our investigation to higher spin quantum numbers, namely, $S=1$ and $3/2$, in order not only to show the versatility of the method, but also because such higher-spin ($S>1/2$) Kitaev models have attracted much attention in recent years~\cite{Baskaran2008PRB,Chandra2010,Rousochatzakis2018NC,Koga2018JPSJ,Oitmaa2018PRB,Minakawa2019,Zhu2020PRR,Hickey2020PRR,LeePRB2021,Khait2021PRR,Jin2022NC,ChenPRB2022,Bradley2022,GordonPRR2022,Fukui2022,MaPRL2023,Cen2023,liu2023symmetries}. This recent interest has also been prompted by possible material realizations~\cite{Yamada2018,Lee20220PRL,Xu2020PRL,Stavropoulos2019PRL,Stavropoulos2021PRR,Samarakoon2021}.

\begin{figure}[!t]
\includegraphics[width=\linewidth]{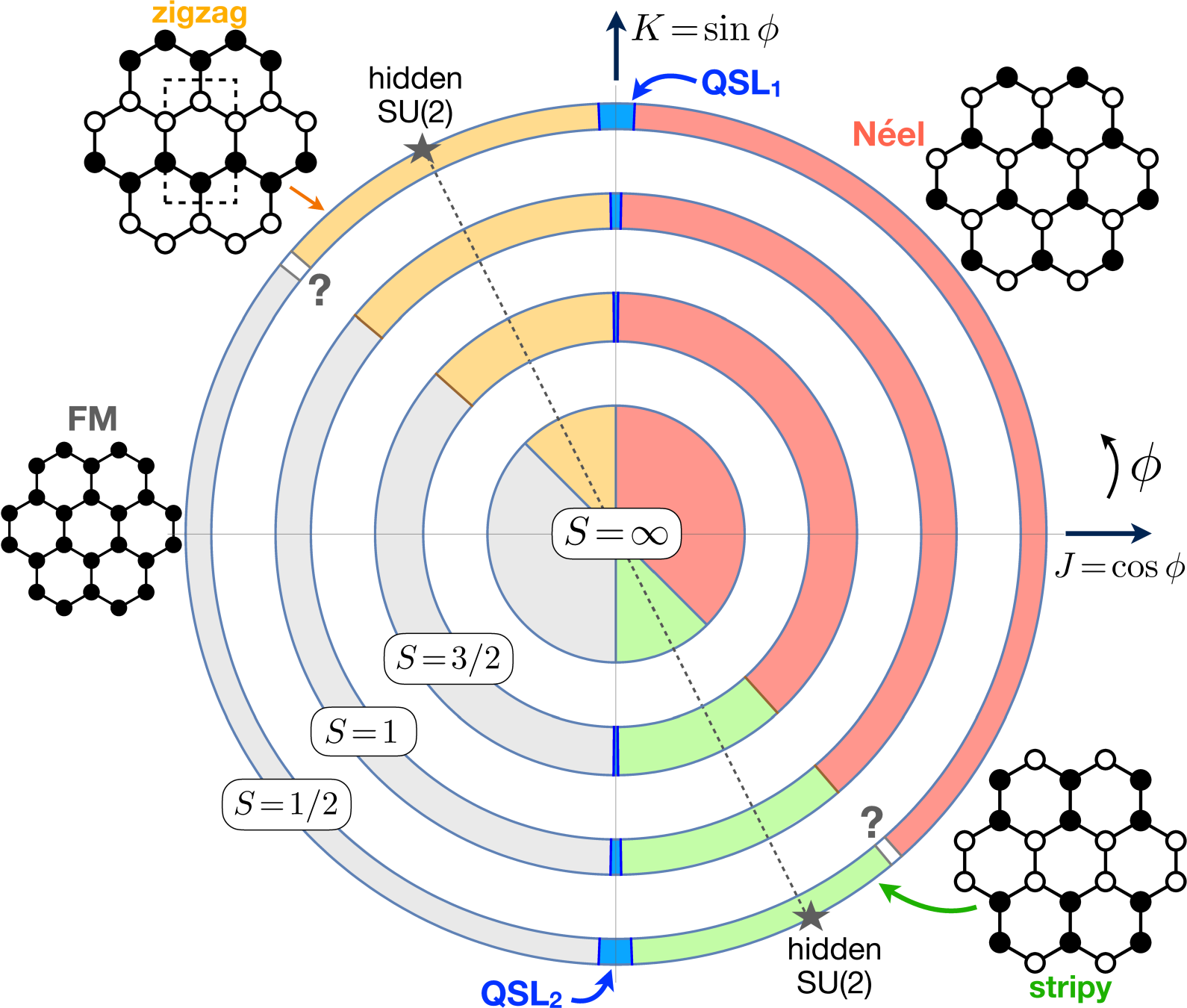}
\caption{Classical ($S=\infty$,~\cite{Chaloupka2013PRL}) and quantum phase diagrams of the KH model (with $J\!=\!\cos\phi$  and $K\!=\!\sin\phi$) for spin $S=3/2$, $1$ and $1/2$, as obtained here from the CCM. Star symbols indicate the two hidden SU(2) points. The (blue) shaded QSL$_1$ and QSL$_2$ regions denote the two quantum spin liquid phases, while the question marks against the unshaded regions denote the enigmatic intermediate phases discussed in Sec.~\ref{sec:intphases}.
The classical states are shown in the insets, where the open and filled black symbols denote spins pointing along the global $+z_s$ and  $-z_s$ spin-space directions, respectively.}
\label{fig:PD}
\end{figure}

We use the well-established SUB$m$--$m$ truncation scheme, where the index $m$ is the order of truncation  (and with $m$ up to $10$ for $S\!=\!1/2$ and up to $8$ for $S\!=\!1$ and $3/2$), to obtain numerical predictions for the ground-state energy and local order parameters for the four magnetically ordered phases of the model for $S\!=\!1/2$, $1$ and $3/2$. We also establish the existence of extended QSL phases from the appearance of CCM termination points to SUB$m$--$m$ equations, which occur before the crossing of the CCM energies of the surrounding magnetic phases. These termination points yield an accurate determination of the phase boundaries of the two Kitaev QSL phases, and their evolution with $S$, see Fig.~\ref{fig:PD}. Quite surprisingly, we find CCM termination points in two more regions for $S\!=\!1/2$, one of which is sandwiched between the zigzag and the ferromagnetic phases, and the other between the N\'eel and the stripy phases. These results provide evidence for two narrow intermediate phases that are completely unexpected from the classical limit or from previous calculations.  
Finally, we use the four-sublattice duality transformation of the model to benchmark the positions of dual phase boundaries with independent CCM calculations around the respective dual phases.

The remainder of the article is organized as follows. In Sec.~\ref{sec:model} we introduce the spin Hamiltonian of the KH model, discuss the duality transformation $\mc{T}_4$ of the model and provide a brief summary of some of the main known facts about this model. 
In Sec.~\ref{sec:CCM}, we discuss general aspects of the CCM formalism (in Sec.~\ref{sec:CCMformal}), the basic steps in applying it to the four magnetically ordered states of the model (in Sec.~\ref{sec:SettingUpCCM}), the approximations schemes that we use (in Sec,~\ref{sec:ApproxSchemes}), and the extrapolation schemes used for the energies and the order parameters (in Sec.~\ref{sec:ExtrSchemes}). In Sec.~\ref{sec:results}, we present and analyze our CCM results for the ground-state energies and order parameters of the four ordered states of the model,  the numerical evidence for the existence of the extended spin liquid phases from an analysis of the CCM termination points, and the final phase diagram for $S\!=\!1/2$, $1$ and $3/2$. 
Finally, we provide our conclusions and a broader overview of our study in Sec.~\ref{sec:Conclusions}.

\section{The spin-$S$ Kitaev-Heisenberg model}\label{sec:model}

\subsection{The KH model and the duality transformation $\mc{T}_4$}
{\it Spin Hamiltonian.} 
The Kitaev-Heisenberg (KH) model on the two-dimensional (2D) honeycomb lattice is described by the spin Hamiltonian
\be
\mc{H}=\sum_{\langle ij\rangle} \left(
J~ {\bf S}_i\cdot{\bf S}_j+K S_i^{\alpha_{ij}}S_j^{\alpha_{ij}}
\right)\,,
\label{eq:KH-Hamiltonian}
\ee
where the sum on $\langle ij \rangle$ is over all bonds of NN sites $i$ and $j$ of the honeycomb lattice, counting each bond once and once only; ${\bf S}_i \equiv (S_i^x,S_i^y,S_i^z)$ and ${\bf S}_j\equiv(S_j^x,S_j^y,S_j^z)$ denote the corresponding spin-$S$ operators residing on these sites, which obey the standard SU(2) commutation relations,
\be\label{eq:SU2-commutators}
[S_i^z,S_j^{\pm}]=\pm S_i^{\pm}\delta_{ij}\,; \quad
[S_i^+,S_j^-]=2S_i^z\delta_{ij}\,,
\ee
with $S_k^{\pm}\equiv S_k^x \pm iS_k^y\,$; $\alpha_{ij}=x$, $y$ or $z$ if the bond $(ij)$ is of type $x$, $y$ or $z$, respectively, as mentioned above; and the constants $J$ and $K$ are the Heisenberg and Kitaev coupling parameters, respectively. Henceforth, we use the standard parametrization (see, e.g., Ref.~\cite{Gotfryd2017}), given by 
\be\label{eq:Param}
J\equiv\cos\phi,\quad K\equiv\sin\phi\,,
\ee
with $\phi\in[0,2\pi)$, where we have used an energy scale in which $\sqrt{J^2+K^2}\equiv 1$.

{\it Duality mapping.}  
The KH model features a 4-sublattice duality transformation (denoted by $\mc{T}_4$ in Refs.~\cite{Chaloupka2010PRL,Chaloupka2013PRL,Chaloupka2015PRB}), in which the spins ${\bf S}_i$ are replaced by corresponding spins $\widetilde{{\bf S}}_i$ in each of which the signs of two appropriate Cartesian components (depending on the sublattice index) are changed.  This unitary transformation leaves the SU(2) commutation relations unchanged, but it results in a new Hamiltonian in terms of the $\widetilde{{\bf S}}_i$ spins of the same form as that in Eq.~(\ref{eq:KH-Hamiltonian}) in terms of the original ${\bf S}_i$ spins.  In particular, this duality transformation
maps the general parameter point $(J,K)$ to $(\widetilde{J},\widetilde{K})\!\equiv\!(-J,K\!+\!2J)$.   Equivalently, in terms of the angular variables, the angle $\phi$ is mapped to 
$\widetilde{\phi}$, where $\tan\widetilde{\phi}=-\tan\phi -2$. This mapping allows one to deduce the phase diagram in half of the parameter space from that in the other half. Moreover, the mapping reveals that the two
points where $\widetilde{K}\!=\!0$ (or, equivalently, $K\!=\!-2J$), namely, 
$\phi\!=\!-63.43^\circ \!\equiv\! 296.57^\circ$ and $\phi\! = \! 116.57^\circ$, are hidden SU(2) points, i.e., the 
Hamiltonian in the rotated frame takes the Heisenberg form. Altogether then, 
the KH model features four SU(2) points, $\phi=(0, 116.57^\circ, 180^\circ, 296.57^\circ$), which govern much of the phase diagram, as we discuss below.

\subsection{Brief summary of known results}
Before we present the results from our CCM study, let us summarize some of the main results that have been reported about the model.

{\it Classical phase diagram.} 
At the classical level, the isolated Kitaev points $\phi=\pm\frac{\pi}{2}$, which are self-dual under the duality transformation $\mc{T}_4$, have an infinite number of classical ground states~\cite{Baskaran2008PRB,Chandra2010,Rousochatzakis2018NC}. Henceforth, we denote these points as the (antiferromagnetic) AF ($\phi = \frac{\pi}{2}$) and (ferromagnetic) FM ($\phi = -\frac{\pi}{2}$) Kitaev points, respectively.

Away from these points, the KH model features four extended collinear magnetic phases (see Fig.~\ref{fig:PD}, inner disc), each surrounding or bounding one of the four SU(2) points~\cite{Chaloupka2013PRL}.  These are the N\'eel  antiferromagnetic (AFM) phase [$\phi\in(-\frac{\pi}{4},\frac{\pi}{2})$], the so-called zigzag phase [$\phi\in(\frac{\pi}{2},\frac{3\pi}{4})$], the ferromagnetic (FM) phase [$\phi\in(\frac{3\pi}{4},\frac{3\pi}{2})$], and the so-called stripy phase [$\phi\in(\frac{3\pi}{2},\frac{7\pi}{4})$]. The zigzag and stripy phases are the dual phases (under  $\mc{T}_4$) of the N\'eel and FM phases, respectively. Thus, for example, a knowledge of the
right-hand side ($J>0$) of the phase diagram of Fig.~\ref{fig:PD} implies a knowledge of the left-hand side ($J<0$), and vice versa.

{\it Quantum regime.}
The most notable impact of quantum fluctuations is that they introduce two QSL phases (denoted by QSL$_1$ and QSL$_2$ in the following, see also Fig.~\ref{fig:PD}) which remain stable in extended regions around the two Kitaev points $\phi=\pm\pi/2$.  
In particular, according to finite-size ED (respectively, CMFT) calculations~\cite{Gotfryd2017}, the QSL$_1$ phase is stable over the regime $88.92^\circ \!< \! \phi < 91.08^\circ$ (respectively, $89.28^\circ \!<\! \phi < 90.9^\circ$), whereas the QSL$_2$ phase is stable for $260.64^\circ \! <\! \phi < 277.02^\circ$ (respectively, $266.04^\circ\! < \! \phi < 273.42^\circ$).

The nature of the two Kitaev QSL phases is known exactly only for $S\!=\!1/2$~\cite{Kitaev2006,Chen2008,Mandal2009PRB,Obrien2016}. Their persistence for $S\!>\!1/2$ is anticipated on  general grounds~\cite{Rousochatzakis2018NC} from the presence of local symmetries~\cite{Baskaran2008PRB} and Elitzur's theorem~\cite{Elitzur1975}. The precise nature of these QSL phases for $S\!>\!1/2$ has been the focus of recent studies~\cite{Rousochatzakis2018NC,Dong2020PRB,Lee2020PRR,Jin2022NC,Natori2023}, but much less is known about their stability under perturbations, such as the Heisenberg coupling $J$ (see, e.g., Ref.~\cite{Dong2020PRB} for $S\!=\!1$ and \cite{Fukui2022} for $S\!=\!1$ and $3/2$).

The impact of quantum fluctuations on the four ordered phases is multifold: 

(i) While the direction of the moments in these phases is not fixed at the classical level (which is an example of accidental ground-state degeneracy), a number of works~\cite{Chaloupka2010PRL,Chaloupka2013PRL,Sizyuk2016,Peter2017,Sela2014PRB,Chaloupka2016,Janssen2016,Gotfryd2017,Chern2017}) has shown that fluctuations select one of the Cartesian axes ($\pm\hat{{\bf x}}$, $\pm\hat{{\bf y}}$ or $\pm\hat{{\bf z}}$), restoring the discrete threefold rotational symmetry of the model around the $[111]$ axis (which is perpendicular to the plane of the spins). 

(ii) Quantum fluctuations also affect the spin gap. At the quadratic level, the linear spin-wave spectrum has a quasi-Goldstone mode at the ordering wavevectors, even away from the four SU(2) points, despite the absence of continuous symmetry, due to the accidental degeneracy mentioned above. Eventually a nonzero spin gap is recovered at the anharmonic level~\cite{Chaloupka2010PRL,Chaloupka2013PRL}.

(iii) The boundaries between the four extended magnetic phases are also affected by quantum fluctuations.  For $S=1/2$, for example, the boundary between the FM and the zigzag phases shifts from $\phi\!=\!135^\circ$ to $146.52^\circ$ (respectively, $148.5^\circ$), and the boundary between the stripy and the N\'eel phase shifts from $\phi\!=\!315^\circ$ to $306.72^\circ$ (respectively, $305.82^\circ$), according to finite-size ED (respectively, CMFT) calculations~\cite{Gotfryd2017}.

\section{The coupled cluster method (CCM)}\label{sec:CCM}

\subsection{Formal aspects of the CCM}\label{sec:CCMformal}

The CCM~\cite{Kummel_Luhr-Zab_1978,Bishop-Luhrmann_1978,Emrich_1981a,Emrich_1981b,Bishop-Luhrmann_1982,Arponen_1983,Bishop-Kummel_1987,Arp-Bish-Paj_1987a,Arp-Bish-Paj_1987b,Bishop_1991,Arponen-Bishop_1991,Arponen-Bishop_1993a,Bishop_1998} is one of the most versatile and most accurate of all modern {\it ab initio} techniques of quantum many-body theory at attainable levels of computational implementation. It has been applied to a wider array of both finite and extended strongly-interacting systems when compared to other competing approximate methods. Indeed, the CCM has been applied either on a spatial continuum or on a regular discrete lattice, and in fields ranging from the electron gas~\cite{Bishop-Luhrmann_1978,Bishop-Luhrmann_1982} and quantum chemistry~\cite{Bartlett-Musial_2007} to nuclear physics~\cite{Hagen_2014}, sub-nuclear physics~\cite{Kuemmel_1983,Hasberg-Kuemmel_1986} and quantum field theory~\cite{Kaulfuss_1985,Altenokum-Kuemmel_1985,Baker-Bishop-Davidson_1996,Ligterink-Walet-Bishop_1998,Ligterink-Walet-Bishop_2000}. Of particular relevance here, it has also been applied very successfully in recent years to a wide variety of highly frustrated spin-lattice models in quantum magnetism (see, e.g., Refs.~\cite{Zeng_et-al_1998,Bishop2000,Kruger2000,Farnell-et-al_2001,Farnell-Bishop_2004,Darradi2005,Bishop2008,Bishop-et-al_2008,Darradi-et-al_2008,Bishop2011,Farnell2011,Reuther-et-al_2011,Gotze2011,Gotze-et-al_2012,Bishop-Li-Campbell_2014,Farnell-et-al_2014,Li-Bishop_2016,Bishop-et-al_2019,Farnell-et-al_2019,Li-Bishop_2022} and references cited therein).

In order to implement the CCM in its ``single-reference'' version, one first needs to choose a suitable normalized reference (or model) state  $|\Phi_0\rangle$ in the full Hilbert space $\mathscr{H}$, the main condition on which is that it has a nonzero overlap, $\langle\Phi_0|\Psi_0\rangle \neq 0$, with the target state $|\Psi_0\rangle \!\in \! \mathscr{H}$ of the quantum $N$-body problem (typically with $N \to \infty$) under study.  We also require that the model state $|\Phi_0\rangle$ should be a cyclic vector (or, equivalently, a generalized vacuum state), with respect to which all of the states in $\mathscr{H}$ can be expressed in terms of suitably defined, mutually commuting, many-body (multiconfigurational) creation operators $C_I^+$ acting on $|\Phi_0\rangle$. Thus, the algebra of all operators in $\mathscr{H}$ and its adjoint space $\mathscr{H}^*$ is spanned by the two Abelian subalgebras of multiconfigurational creation operators $\{C_I^+\}$ and their Hermitian-adjoint counterparts, namely the multiconfigurational destruction operators $\{C_I^- \!\equiv \!(C_I^+)^\dagger\}$. Both sets of operators are defined with respect to the given model state $|\Phi_0\rangle$.  The index $I\,$ in this compact notation is a set-index, comprising a set of single-particle labels (in some suitable single-particle basis), which completely characterizes a given many-body configuration in this basis. Usually one characterizes the single-particle labels contained in the set-index $I\,$ by choosing only those that are needed to describe single-particle states that differ from those occupied in the model state $|\Phi_0\rangle$. Accordingly, it is convenient to introduce the notation,
\be\label{eq:C_0}
C_0^+ \equiv \mathbbm{1} = C_0^-\,,
\ee
where $\mathbbm{1}$ is the unit operator in $\mathscr{H}$.

To summarize, the set $\{|\Phi_0\rangle;C_I^+\}$ is thus required to satisfy the conditions,
\be\label{eq:C_I-commutators}
[C_I^+ , C_J^+] = 0 = [C_I^- , C_J^-]\,,
\ee
\be\label{eq:C_I-destroy-vacuum-relations}
\langle\Phi|C_I^+ = 0 = C_I^-|\Phi\rangle\,, \quad \forall \, I \neq 0\,.
\ee
The two subalgebras and the state $|\Phi_0\rangle$ are also required  to be cyclic in the following sense,
 \begin{subequations}
 \begin{align}
 |\Psi\rangle &= \sum_I \psi_I C_I^+ |\Phi_0\rangle
 \,; \quad \forall \;|\Psi\rangle \in \mathscr{H}\,,
 \label{eq:cyclic-ket}\\
 \langle\widetilde{\Psi}|&= \sum_I \widetilde{\psi}_I \langle \Phi_0|C_I^-
 \,; \quad \forall \; \langle\widetilde{\Psi}| \in \mathscr{H}^* \,,
 \label{eq:cyclic-bra}
 \end{align}
 \end{subequations}
in terms, respectively, of some sets of $c$-number expansion coefficients $\{\psi_I\}$ and $\{\widetilde{\psi}_I\}$. The configuration-label space $\mc{I} \! \equiv \! \{I\}$ must thus be complete (for a given model state $|\Phi_0 \rangle$) with respect to the possible many-body configurations. It is also convenient, but not necessary, to choose the states $\{C_I^+|\Phi_0\rangle\}$ that now span $\mathscr{H}$ to be an orthonormalized set,
\be
\langle\Phi_0|C_I^-C_J^+|\Phi_0\rangle = \delta_{I,J}\,,
\label{eq:C_I-orthonormal}
\ee
where $\delta_{I,J}$ is a suitably generalized Kronecker symbol that implies equality between the sets of single-particle labels $I$ and $J$ (i.e., equality under at least one permutation).  If we assume that Eq.~(\ref{eq:C_I-orthonormal}) holds, we hence have the following completeness relation in $\mathscr{H}$,
\be
\sum_I C_I^+|\Phi_0\rangle\langle\Phi_0|C_I^- = \mathbbm{1} = 
|\Phi_0\rangle\langle\Phi_0| + \sum_{I \neq 0} 
C_I^+|\Phi_0\rangle\langle\Phi_0|C_I^- \,.
\label{eq:C_I-completeness}
\ee

We now consider the exact many-body ground-state ket and bra states, $|\Psi_0\rangle$ and $\langle\widetilde{\Psi}_0|\,\, (\equiv \!\langle \Psi_0|/\langle \Psi_0|\Psi_0\rangle)$, which satisfy the respective ground-state Schr\"odinger equations,
\be\label{GS-Schroedinger-eqs}
\mc{H}|\Psi_0\rangle = E_0|\Psi_0\rangle\,, \quad
\langle\widetilde{\Psi}_0|\mc{H} = E_0\langle\widetilde{\Psi}_0|\,,
\ee
where we require that $|\Psi_0\rangle$ satisfies the intermediate normalization condition, $\langle\Phi_0|\Psi_0\rangle \!=\! 1 = \langle\Phi_0|\Phi_0\rangle$, and we also have that $\langle\widetilde{\Psi}_0|\Psi_0\rangle = 1$, by construction. These ground-state wave functions are now parametrized {\it independently\,} (i.e., not requiring manifest Hermiticity between the ground-state bra and ket states) within the CCM with respect to the model state $|\Phi_0\rangle$ as follows,
 \begin{subequations}
 \begin{align}
 |\Psi_0\rangle = \mathrm{e}^S|\Phi_0\rangle & \,;
 \quad S = \sum_{I \neq 0} \mc{S}_I C_I^+ \,,
 \label{eq:CCM_GS-ket}\\
 \langle\widetilde{\Psi}_0| = \langle\Phi_0|\widetilde{S}\mathrm{e}^{-S} &\,;
 \quad \widetilde{S} = \mathbbm{1} + \sum_{I \neq 0} \widetilde{\mc{S}}_I C_I^-
 \,,
 \label{eq:CCM_GS-bra}
 \end{align}
 \end{subequations}
in which the distinctive exponentiated forms of the creation correlation operator $S$ are one of the hallmarks of the method.

If no further approximations are made, the destruction correlation operator $\widetilde{S}$ will be related to its creation counterpart $S\,$ by Hermiticity,
\be
\langle\Phi_0|\widetilde{S} = \frac{\langle\Phi_0|\mathrm{e}^{S^\dagger}
\mathrm{e}^S}{\langle\Phi_0|\mathrm{e}^{S^\dagger}\mathrm{e}^S|\Phi_0\rangle} \,.
\label{eq:CCM-GS-correlators_hermiticity}
\ee
However, in practical implementations of the CCM when truncations are made in the sums over the set-index $I$ in Eqs.~(\ref{eq:CCM_GS-ket}) and (\ref{eq:CCM_GS-bra}), as described more fully below, Eq.~(\ref{eq:CCM-GS-correlators_hermiticity}) will likely be violated. Nevertheless, any loss of explicit Hermiticity is far outweighed in practice by the fact that the very important Hellmann-Feynman theorem~\cite{Hellmann_1935,Feynman_1939} is now manifestly obeyed at {\it all\,} such levels of truncation~\cite{Bishop_1998}. Indeed, it can be proven~\cite{Bishop_1998} that if the CCM parametrization of Eq.~(\ref{eq:CCM_GS-ket}) is chosen for the ground ket state, then the ground bra state parametrization of Eq.~(\ref{eq:CCM_GS-bra}) is actually derivable from the Hellmann-Feynman theorem, when the ground-state energy is calculated from the simple relation,
\be
E_0 = E_0(\mc{S}_I) = \langle\Phi_0|\mathrm{e}^{-S}\mc{H}\mathrm{e}^S|\Phi_0\rangle \,,
\label{eq:CCM_GS-energy-simple}
\ee
which follows trivially from Eqs.~(\ref{GS-Schroedinger-eqs}) and (\ref{eq:CCM_GS-ket}).

Clearly, the set of real $c$-number CCM correlation coefficients $\{\mc{S}_I,\widetilde{\mc{S}}_I\}$ completely determines any property of the many-body ground state under consideration.  They are themselves calculated by inserting the parametrizations of Eqs.~(\ref{eq:CCM_GS-ket}) and (\ref{eq:CCM_GS-bra}) into the respective Schr\"odinger equations (\ref{GS-Schroedinger-eqs}), and then projecting in turn onto the complete sets of states $\{\langle\Phi_0|C_I^-\}$ and $\{C_I^+|\Phi_0\rangle\}$.  A completely equivalent procedure to derive the CCM ground-state correlation coefficients is to extremize the ground-state energy expectation value functional, $\overline{\mc{H}}=\overline{\mc{H}}(\mc{S}_I,\widetilde{\mc{S}}_I)$,
defined as follows,
\be
\overline{\mc{H}} \equiv \langle\widetilde{\Psi}_0|\mc{H}|\Psi_0\rangle =
\langle\Phi_0|\widetilde{S}\mathrm{e}^{-S}\mc{H}\mathrm{e}^S|\Phi_0\rangle \,,
\label{eq:CCM_GS-energy-functional}
\ee
with respect to every member of the parameter set 
$\{\widetilde{\mc{S}}_I,\mc{S}_I\}$ in turn. Both methods readily yield the sets of equations,
 \begin{subequations}
 \begin{align}
\langle\Phi_0|C_I^-\mathrm{e}^{-S}\mc{H}\mathrm{e}^S|\Phi_0\rangle = 0 \,, 
\quad
\forall\, I \neq 0\,;
 \label{eq:eqs-for-CCM_GS-ket-coeffts}\\
 \langle\Phi_0|\widetilde{S}\mathrm{e}^{-S}
 [\mc{H},C_I^+]\mathrm{e}^S|\Phi_0\rangle =0 \,, \quad
\forall\, I \neq 0 \,.
 \label{eq:eqs-for-CCM_GS-bra-coeffts}
 \end{align}
 \end{subequations}
 
By making use of Eqs.~(\ref{eq:CCM_GS-bra}) and (\ref{eq:eqs-for-CCM_GS-bra-coeffts}), we readily observe that when $\overline{\mc{H}} \!=\!\overline{\mc{H}}(\mc{S}_I,\widetilde{\mc{S}}_I)$ from Eq.~(\ref{eq:CCM_GS-energy-functional}) is evaluated at the stationary point, the resulting expression for the ground-state energy is just that given by Eq.~(\ref{eq:CCM_GS-energy-simple}).  Similarly, by making use of the fact that, by construction, the operators $C_I^+$ commute with the CCM correlation operator $S$, one can easily show that Eq.~(\ref{eq:eqs-for-CCM_GS-bra-coeffts}) takes the equivalent form,
\be
 \langle\Phi_0|\widetilde{S}(\mathrm{e}^{-S}
 \mc{H}\mathrm{e}^S - E_0)C_I^+|\Phi_0\rangle =0 \,, \quad
 \forall\, I \neq 0 \,.
\label{eq:eqs-for-CCM_GS-bra-coeffts_alternate-form}
\ee
Equation (\ref{eq:eqs-for-CCM_GS-ket-coeffts}) is a coupled set of highly nonlinear, multinomial equations for the ground ket-state CCM correlation coefficients $\{\mc{S}_I\}$, which contains as many equations as there are coefficients to be solved for. By contrast, Eq.~(\ref{eq:eqs-for-CCM_GS-bra-coeffts}), or equivalently Eq.~(\ref{eq:eqs-for-CCM_GS-bra-coeffts_alternate-form}), is just a linear set of generalized eigenvalue equations for the ground bra-state CCM correlation coefficients $\{\widetilde{\mc{S}}_I\}$ once the known coefficients $\{\mc{S}_I\}$ are used as input, again with as many equations as there are coefficients to be solved for. Whereas the ground-state energy $E_0$ may be expressed from Eq.~(\ref{eq:CCM_GS-energy-simple}) in terms only of the set of CCM creation coefficients $\{\mc{S}_I\}$, the ground-state expectation value, $\overline{\mc{A}}\equiv \langle \widetilde{\Psi}_0|\mc{A}|\Psi_0\rangle$, of any operator $\mc{A}$ other than the Hamiltonian $\mc{H}$ requires a knowledge
of both the CCM creation and destruction coefficients for its evaluation,
\be
\overline{\mc{A}} = \overline{\mc{A}}(\mc{S}_I,\widetilde{\mc{S}}_I)
= \langle\Phi_0|\widetilde{S}\mathrm{e}^{-S}\mc{A}\mathrm{e}^S|\Phi_0\rangle\,.
\label{eq:CCM-GS-expectaion-value-for-arbitrary-operator}
\ee

A noteworthy feature of the CCM is that its characteristic exponentiated operators $\mathrm{e}^{\pm S}$ only  ever enter the formalism in the form of a similarity transform, $\mathrm{e}^{-S}\mc{A}\mathrm{e}^S$, for some  operator $\mc{A}$, as in Eq.~(\ref{eq:CCM-GS-expectaion-value-for-arbitrary-operator}). More importantly, this is also true, for the special case when $\mc{A} \! \to \! \mc{H}$, in Eqs.~(\ref{eq:eqs-for-CCM_GS-ket-coeffts}) and (\ref{eq:eqs-for-CCM_GS-bra-coeffts}) [or \ref{eq:eqs-for-CCM_GS-bra-coeffts_alternate-form}], which are just the equations that need to be solved for the complete CCM description of the ground state.  In all such calculations we make use of the nested commutator expansion,
\be
{\mathrm e}^{-S}\mc{A}{\mathrm e}^{S}=\sum_{n=0}^{\infty}\frac{1}{n!}[\mc{A},S]_{n}\,,  
\label{eq:similarity_transform_expansion}
\ee
in terms of the $n$-fold nested commutators $[\mc{A},S]_{n}$, which are themselves defined recursively as
\be
[\mc{A},S]_{n} \equiv [[\mc{A},S]_{n-1},S]\,; 
\quad [\mc{A},S]_{0}=\mc{A}\,.
\label{eq:nested-commutator-def}
\ee

At first sight one might expect that approximations would need to be made to truncate the infinite sum in Eq.~(\ref{eq:similarity_transform_expansion}) in order to make computations in practice.  However, it is an important feature of the specific choices of the CCM parametrizations in Eqs.~(\ref{eq:CCM_GS-ket}) and (\ref{eq:CCM_GS-bra}) that the otherwise infinite sum in Eq.~(\ref{eq:similarity_transform_expansion}) actually terminates exactly at some finite order in practice for all operators $\mc{A}$ that contain only finite-order multinomials in the single-particle operators, as in the present case.  The reason for this is twofold.  Firstly, all components in the expansion of Eq.~({\ref{eq:CCM_GS-ket}}) for the CCM operator $S$ commute among themselves. Secondly, in general, the algebra of the single-particle operators is closed under commutation. In particular, for the present case of spin Hamiltonians, we shall see that for the choices of the set $\{C_I^+;| \Phi \rangle \}$ that we make here, namely, where the operators $C_I^+$ are formed as products of spin-raising operators, $S_j^+$, on various sites $j$, the sum in Eq.~(\ref{eq:similarity_transform_expansion}) simply terminates at the term with $n=2$, since all higher-order nested commutators with $n>2$ vanish identically, due to the SU(2) commutation relations of Eq.~(\ref{eq:SU2-commutators}).

For the same reason as already mentioned above, namely, that all components for the CCM operator $S$ in the expansion of Eq.~({\ref{eq:CCM_GS-ket}}) commute among themselves, it is clear that all terms in the linked commutator expansion for $\overline{\mc{H}}$ in Eq.~(\ref{eq:CCM_GS-energy-functional}) are linked.  Thus, no unlinked terms (i.e., any terms that are not themselves linked to the Hamiltonian) can ever appear in the CCM formalism. For that reason, the CCM automatically obeys the Goldstone theorem~\cite{Goldstone_1957} at {\it all\,} levels of truncation in the expansions of Eqs.~(\ref{eq:CCM_GS-ket}) and (\ref{eq:CCM_GS-bra}) for the CCM correlation operators $S$ and $\widetilde{S}$. In turn, this guarantees the size-extensivity of the CCM in all practical calculations. 

In light of the above discussion, it is clear that the CCM has the distinct advantage that one can work from the outset in the thermodynamic limit ($N \to \infty$), thereby obviating the need for any finite-size scaling of the results performed on various lattices with a different number $N$ of sites, and the consequent errors associated with extrapolating to the infinite lattice. Furthermore, as we have shown, the {\it only\,} approximation that is ever needed for practical implementations of the CCM, once a model state $|\Phi_0\rangle$ has been selected, is to choose which set of configurations $\{I\}$ to retain in the expansions of Eqs.~(\ref{eq:CCM_GS-ket}) and (\ref{eq:CCM_GS-bra}), and then how to extrapolate the subsequently obtained values for physical ground-state parameters to the exact limit where all configurations are retained.

We now first discuss the selection of suitable CCM model states for the Kitaev-Heisenberg model under study in Sec.~\ref{sec:SettingUpCCM}.  Subsequently, in Secs.~\ref{sec:ApproxSchemes} and \ref{sec:ExtrSchemes}, we discuss, respectively, both the approximation schemes that we adopt and the associated extrapolation schemes for the ground-state energy and order parameter.

\subsection{Setting up the CCM for the four magnetically ordered reference states}\label{sec:SettingUpCCM}
The most elementary class of model states for spin-lattice problems comprises independent-spin product states, for which the spin projection of the spin on each site (along some quantization axis, which can be separately defined for each site) is specified independently. Clearly, the four collinear states (FM, N\'eel, zigzag, and stripy) shown in insets of Fig.~\ref{fig:PD}, where open (respectively, filled black) symbols denote spins pointing along {\it global} $+z_s$ (respectively, $-z_s$) spin-coordinate axes, belong to this class. Completely generally, so do {\it all\,} such (quasiclassical) states with perfect magnetic long-range order.

For computational purposes, we carry out a passive rotation of each spin (i.e., by choosing local spin quantization axes independently on every lattice site) such that all spins now point along the negative $z_s$-direction. All sites become equivalent to one another for any such chosen form of the quasiclassical model state $|\Phi_0\rangle$. They all take the universal fully polarized form, $|\Phi_0\rangle = |\! \da \da \da \cdots \da \rangle$, in their own {\it local} spin-coordinate frames. We carry out this rotation for each of the model states separately, which means that the resulting Hamiltonians are different for each model state with respect to these local spin axes. However, such unitary transformations also preserve the underlying SU(2) commutation relations of Eq.~(\ref{eq:SU2-commutators}).   

With any such choice of model state $|\Phi_0\rangle$, it is now  straightforward
to make it a fiducial vector with respect to a suitable set of mutually commuting multiconfigurational (many-body) creation operators, $\{C_I^+\}$, as required. Thus, they are now constructed as simple products of single spin-raising operators, $C_I^+ \to S^+_{k_1} S^+_{k_2} \cdots S^+_{k_n};\, n=1,2,\cdots ,2SN$, where $N$ ($\to \infty$) is the number of lattice sites. The set index $I$ thus becomes a set of lattice-site indices, $I \to \{k_1,k_2,\cdots k_n; \, n=1,2,\cdots,2SN \}$, in which any individual site may appear no more than $2S$ times.

The magnetic order parameter, $M$, (i.e., the sublattice magnetization) with respect to the local set of rotated spin axes takes the same form for all of the model states considered here, namely,
\be
M = -\frac{1}{N} \sum_{k=1}^N \langle\Phi_0|\widetilde{S}\mathrm{e}^{-S}S_k^z\mathrm{e}^S|\Phi_0\rangle\,.
\label{eq;magnetic-order-parameter}
\ee
As before, the term $\mathrm{e}^{-S}S_k^z\mathrm{e}^S$ is evaluated {\it exactly}, using a nested commutator expansion of the form of Eq.~(\ref{eq:similarity_transform_expansion}), which now terminates after the term with $n=1$.

We base our CCM calculations on all four collinear magnetic 
states of the model. 
In principle, this is not needed as the duality transformation $\mc{T}_4$ allows us to restrict ourselves to, e.g., the N\'eel and FM phases only. We have nevertheless carried out independent CCM calculations around the zigzag and stripy phases as well, for benchmark purposes. 

For the reference classical states we pick the ones where spins point along the $\pm z_s$ global spin-space directions, which are among the six states that are selected by quantum fluctuations and which are related to each other via threefold rotation symmetry in the combined spin-orbit space. The reference states are shown in the insets of Fig.~\ref{fig:PD}, where the open and filled black symbols denote spins pointing along the global $+z_s$ and  $-z_s$ spin-space directions, respectively. 
Thus, we choose that all spins point along the global $-z_s$ direction for the FM state. For the N\'eel state, we take `up' spins to lie on the $A$-sublattice and `down' spins to be on the $B$-sublattice. We set the `$z$' bonds to feature antiparallel spins for the zigzag state, whereas `$x$' and `$y$' bonds have parallel spins. Finally, `$z$' bonds feature parallel spins for the stripy state, whereas `$x$' and `$y$' bonds have antiparallel spins.

For the N\'eel, zigzag and stripy phases, we now perform the sublattice rotations so that we map the corresponding classical reference states to the fully polarized state along the $-z_s$ direction in the respective locally chosen spin-coordinate axes, as indicated above.
This means that for any spin on site $k$ pointing along the global spin space $+z_s$ direction in the quasiclassical reference state we perform a passive rotation about the $y_s$ axis, under which the components of the spin $\mathbf{S}_k$ on the site transforms as follows,
\be
(S_k^x, S_k^y, S_k^z) \mapsto (-S_k^x, S_k^y, -S_k^z)\,.
\label{eq:Rpiy}
\ee
This rotation changes the form of the Hamiltonian, such that $\mc{H} \mapsto \widetilde{\mc{H}}$. 

For the N\'eel state, for example, carrying out the above rotation on the $A$-sublattice sites changes the Hamiltonian to the following form,
\bea
\widetilde{\mc{H}}_{\text{N\'eel}} \!&\!=\!&\! 
J \sum_{\langle ij \rangle} \{ -S_i^x S_j^x + S_i^y S_j^y - S_i^z S_j^z \}    \nonumber \\    \!&\!+\!&\! K \Big\{\!-\!\sum_{\langle ij \rangle_x} S_i^x S_j^x \!+\! \sum_{\langle ij \rangle_y} S_i^y S_j^y \!-\! \sum_{\langle ij \rangle_z} S_i^z S_j^z \Big\}.~~~    
\label{eq:HNeeltilde}
\eea
The rotation of local spin axes affects only those terms on the `$z$' bonds in the Hamiltonian for the zigzag state, such that the Hamiltonian is now given by
\bea
\!\!\widetilde{\mc{H}}_{\text{zigzag}} \!&\!=\!&\!\! 
\sum_{\langle ij \rangle_x} \{ (K\!+\!J) S_i^x S_j^x \!+\! J S_i^y S_j^y \!+\!J S_i^z S_j^z \} \nonumber \\
\!&+&\!\!\!\sum_{\langle ij \rangle_y} \{ J S_i^x S_j^x \!+\! (K\!+\!J) S_i^y S_j^y +J S_i^z S_j^z \} \nonumber \\
\!&+&\!\!\!\sum_{\langle ij \rangle_z} \{ -J S_i^x S_j^x \!+\! J S_i^y S_j^y \!-\!(K+J) S_i^z S_j^z\}.~\label{eq:Hzztilde}
\eea
Similarly, the rotation only affects the terms on the `$x$' and `$y$' bonds for the stripy phase, such that the Hamiltonian takes the form,
\bea
\widetilde{\mc{H}}_{\text{stripy}} \!&\!=\!&\!\!\sum_{\langle ij \rangle_x} \{-(K\!+\!J) S_i^x S_j^x \!+\! J S_i^y S_j^y \!-\!J S_i^z S_j^z \} \nonumber \\
\!&+&\!\!\sum_{\langle ij \rangle_y} \{-J S_i^x S_j^x \!+\!(K\!+\!J) S_i^y S_j^y \!-\!J S_i^z S_j^z\} \nonumber \\
&+& \sum_{\langle ij \rangle_z} \{J S_i^x S_j^x \!+\! J S_i^y S_j^y \!+\!(K\!+\!J) S_i^z S_j^z \}.~\label{eq:Hstrtilde}
\eea
Finally, all spins point along the $-z_s$ spin-space direction already for the FM model state
%, $\widetilde{\mc{H}}_{\text{FM}}=\mc{H}$, 
and so there is no need for any rotations of local spin axes. 

Importantly,  $\widetilde{\mc{H}}_{\text{N\'eel}}$, $\widetilde{\mc{H}}_{\text{zigzag}}$ and $\widetilde{\mc{H}}_{\text{stripy}}$ retain the full translational symmetry of $\mc{H}$, as well as the real-space inversion through the midpoint of a NN bond. The presence of these symmetries significantly reduces the number of independent CCM amplitudes $\mc{S}_I$ (and, correspondingly, $\widetilde{\mc{S}}_I$) contained in the CCM correlation operators, $S$ (and $\widetilde{S}$), as we explain more fully in Sec. \ref{sec:ApproxSchemes}.
Furthermore, we note that, as $\mc{H}$ contains only products of two spin operators, the same is true for $\widetilde{\mc{H}}$. As seen in similar cases before (such as the XXZ model on the square lattice~\cite{Zeng_et-al_1998}), this allows us to restrict clusters in ${S}$ and $\widetilde{S}$ to contain only even numbers of spin operators. This considerably reduces the computational cost of carrying out high-order CCM calculations.

\subsection{The LSUB$\bm{m}$ and SUB$\bm{n}$--$\bm{m}$ approximation schemes}
\label{sec:ApproxSchemes}
A widely used CCM approximation scheme for spin-lattice models is the so-called SUB$n$--$m$ scheme. It has been applied with equal success to both unfrustrated and highly frustrated quantum magnets.  It retains, for specified values of the pair of (positive-definite) integral truncation indices $n$ and $m$, only those multispin configurations $I$, discussed above in Sec.~\ref{sec:SettingUpCCM}, which involve $n$ or fewer spin flips that span a range of no more than $m$ contiguous lattice sites.  For these purposes, a single spin flip is defined to require the operation of a spin-raising operator $S_k^+$ acting once, and a set of lattice sites is said to be contiguous if every site in the set is NN (in the specified lattice  geometry) to at least one other site in the set.  It is obvious that the SUB$n$--$m$ approximation so defined becomes exact as both truncation indices $n$ and $m$ approach infinity.  One can define different sub-schemes according to any specified relations between the two truncation indices, or as to how each index separately approaches infinity. 

For example, if we put $n\!=\!2Sm$, we arrive at the localized (or lattice-animal-based) subsystem sub-scheme, abbreviated as the LSUB$m$ approximation scheme~\cite{Zeng_et-al_1998,Farnell-Bishop_2004}, which has been very widely used, especially for spin-1/2 systems.  Clearly, for $S\!=\!1/2$ (only), SUB$m$--$m$$\;\equiv\;$LSUB$m$, whereas for $S > 1/2$ we have SUB$m$--$m$$\;\subset\;$LSUB$m$.  Clearly, the LSUB$m$ approximation is equivalently defined to retain all clusters of spins described by multispin-flip configurations $\{I\}$ in the sums in Eqs.~(\ref{eq:CCM_GS-ket}) and (\ref{eq:CCM_GS-bra}) for the CCM correlation operators, $S$ and $\widetilde{S}$, respectively, that span no more than $m$ contiguous lattice sites.

In general, the space- and point-group symmetries of the lattice and the pertinent CCM model state $| \Phi_0 \rangle$ being employed, are used (together with any applicable conservation laws) to reduce the number of {\it independent\,} configurations retained at any given order of approximation. We define $N_f(m)$ to be the minimal number of distinct nonzero such multispin-flip configurations that are thereby retained at a given $m$th level of LSUB$m$ or SUB$m$--$m$ approximation. Since $N_f(m)$ increases much more rapidly with the truncation index $m$ for the LSUB$m$ approximation than for its SUB$m$--$m$ counterpart, the former scheme has typically only been employed for spin-half systems, whereas the latter has mostly been employed for spin-$S$ systems with $S \ge 1$.

Here we perform high-level CCM calculations for the infinite spin-$S\,$ Kitaev-Heisenberg model on the honeycomb lattice, based on each of the four quasiclassical model states discussed above, and the SUB$m$--$m$ scheme with $m=2,4,6,8,10$ for $S=1/2$, and $m=2,4,6,8$ for $S=1$, $3/2$. In order to achieve the highest levels of approximation, we use a purpose-built computer-algebra package~\footnote{We use the program package CCCM of D.~J.~J.~Farnell and J. Schulenburg, see \url{https://www-e.ovgu.de/jschulen/ccm/}.} to derive and solve the CCM equations (\ref{eq:eqs-for-CCM_GS-ket-coeffts}) and (\ref{eq:eqs-for-CCM_GS-bra-coeffts}),  
using direct iteration or the Newton-Raphson method.

\subsection{CCM extrapolation schemes}\label{sec:ExtrSchemes}
The SUB$m$--$m$ scheme becomes exact only in the $m\to\infty$ limit, so one needs to perform suitable extrapolations. For the ground-state energy, $E_0$ we use the well-established extrapolation scheme~\cite{Gotze2011,Kruger2000,Farnell-et-al_2001,Bishop2008,Farnell2011,Farnell-et-al_2014,Bishop-Li-Campbell_2014,Li-Bishop_2016,Bishop-et-al_2019,Li-Bishop_2022},
\be 
\text{e1:}~~
E_0(m)/N = E_0(m =\infty)/N +a_1/m^2 +a_2/m^4 \,,
\label{eq:e1-extrapolation-scheme}
\ee
where $E_0(m)$ is the value obtained at the $m$th level of approximation, $N$ is the total number of sites and $a_{1,2}$ are constants. For the magnetic order parameter $M$, choosing an appropriate extrapolation for the $m$th-order approximants, $M(m)$, is more subtle. Thus, for unfrustrated or only very mildly frustrated systems, where the magnetic order is stable even in the presence of (hence, relatively weak) quantum fluctuations, an extrapolation scheme for $M(m)$ with leading power $1/m$ (rather than $1/m^2$, as for the ground-state energy),
\be
\text{e2}:~~
M(m) = M_{\text{e2}}(m =\infty)+\frac{b_1}{m}+\frac{b_2}{m^2}\,,
\label{eq:e2-extrapolation-scheme}
\ee
(where $b_{1,2}$ are constants) leads to excellent results~\cite{Farnell-et-al_2001,Darradi2005,Farnell2011,Farnell-et-al_2014,Li-Bishop_2016}.  On the other hand, for magnetic states with strong quantum fluctuations, where the magnetic order is only weakly established, a scheme with leading power $1/m^{1/2}$,
\be
\text{e3}:~~
M(m) = M_{\text{e3}}(m =\infty) +\frac{c_1}{m^{1/2}} + \frac{c_2}{m^{3/2}}\,,
\label{eq:e3-extrapolation-scheme}
\ee
(where $c_{1,2}$ are constants) is always more favorable~\cite{Gotze2011,Bishop2008,Bishop-et-al_2008,Darradi-et-al_2008,Reuther-et-al_2011,Gotze-et-al_2012,Bishop-Li-Campbell_2014,Farnell-et-al_2014,Li-Bishop_2016,Bishop-et-al_2019,Li-Bishop_2022}. In particular, the latter scheme $\mathrm{e}3$ is always appropriate for systems with an order-disorder transition, or for those which are either close to a quantum critical point or for which $M$ is close to zero.

It is important to be aware from the outset of any ``staggering'' effects that might be present in the LSUB$m$ or SUB$m$--$m$ sequences of approximants.  In this context, it is well known in $m$th-order perturbation theory calculations, for example, that there often exists such an even/odd [or $2n/(2n-1)$, where $n \in \mathbbm{Z}^+$ is a positive integer] staggering effect in the corresponding sequences of approximants for various physical properties. In such cases, where {\it exact\,} extrapolation schemes are often known, both the (even) $m=2n$ and (odd) $m=(2n-1)$ subsequences obey a scheme of the same form (i.e., with the same exponents in the series) but with possibly differing values of the respective coefficients in all terms other than that corresponding to the limiting ($n \to \infty$) value itself. Clearly, in such cases, which now include CCM calculations using either the LSUB$m$ or SUB$m$--$m$ approximation scheme on a wide variety of spin-lattice models, one should not mix even-order and odd-order approximants in a single extrapolation scheme, without incorporating the staggering effect explicitly.  The latter is usually difficult to perform robustly, and it is usually better to extrapolate the even and odd subsequences separately.  As already noted, since our Hamiltonian is bilinear in the single-site spin operators, it is natural, as has been done here, to confine ourselves to even-order CCM LSUB$m$ or SUB$m$--$m$ approximations with $m=2n$.

\begin{figure}
\includegraphics[width=\linewidth]{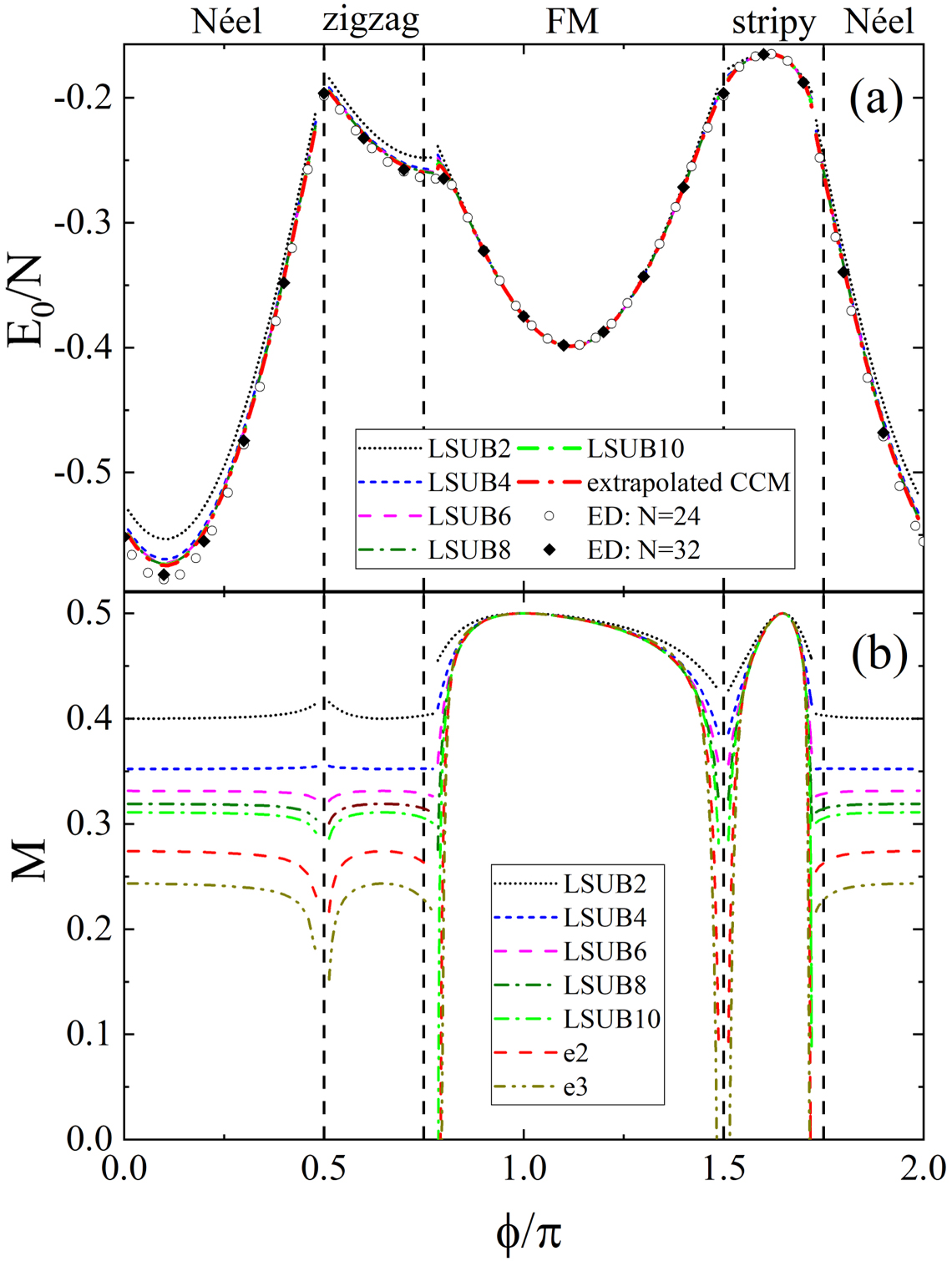}
\caption{{\bf CCM convergence.} Results for (a) the ground-state energy per site $E_0/N$ and (b) the local order parameter $M$  in the four magnetic phases of the spin-half KH model, as obtained from the LSUB$m$ scheme (with $m=2$, $4$, $6$, $8$ and $10$). We also show the corresponding extrapolated CCM values (using data from the LSUB4, LSUB6, LSUB8, and LSUB10 approximations), based on (a) the extrapolation scheme $\mathrm{e}1$ for the ground-state energy and (b) the $\mathrm{e}2$ and $\mathrm{e}3$ schemes for the order parameter. For comparison, we include exact ED results from the $N=24$- and $32$-site clusters. Vertical dashed lines designate the classical phase boundaries.}
\label{fig:ccm_convergence}
\end{figure}

\begin{figure}[!t]
\includegraphics[width=\linewidth]{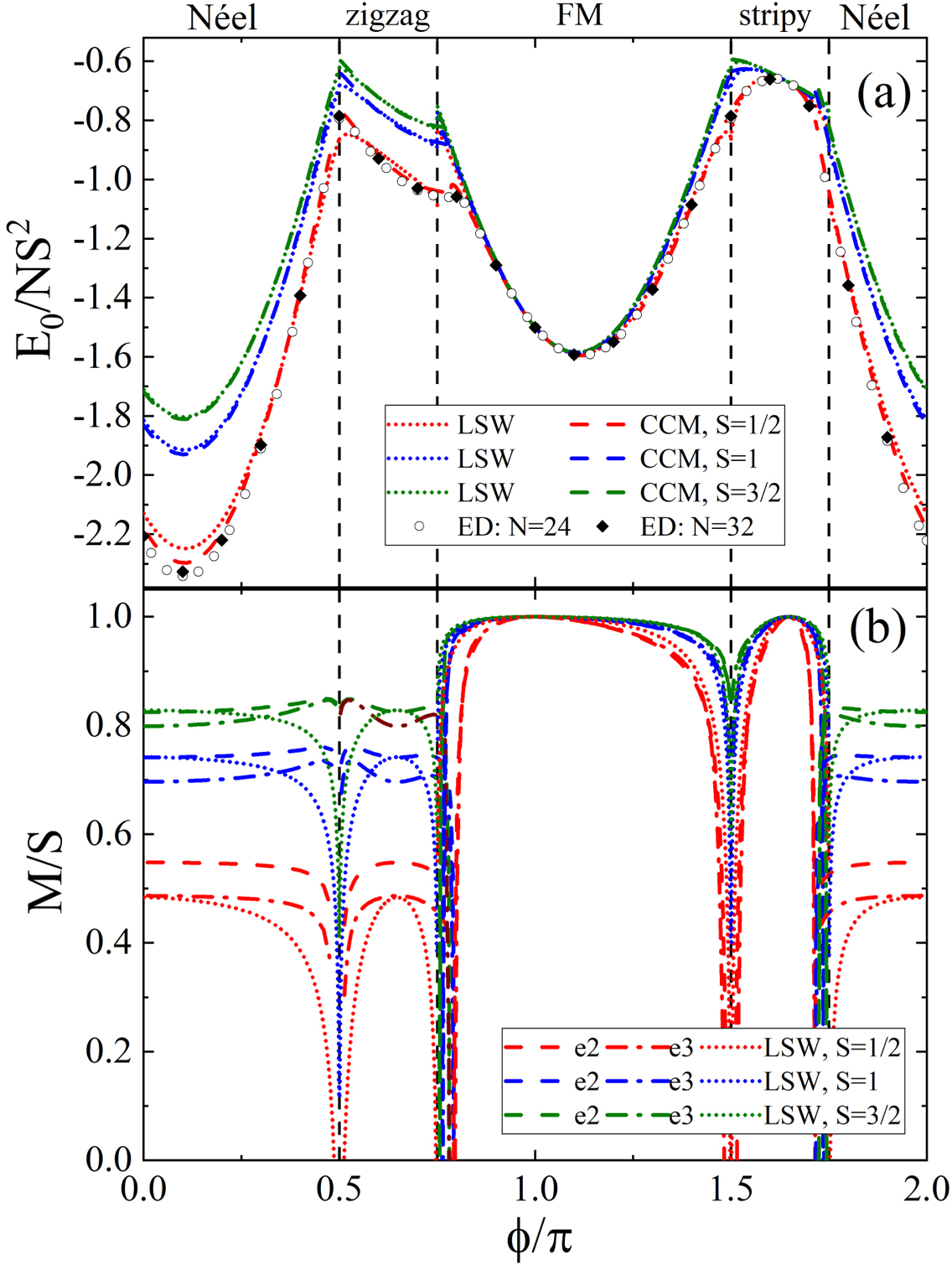}
\caption{{\bf Variation with $\vb*{S}$.} Extrapolated CCM results for (a) the rescaled ground-state energy $E_0/(NS^2)$, and (b) the rescaled order parameter $M/S$, for the KH model for spin $S=1/2$, $1$ and $3/2$, as obtained from extrapolated CCM results (using SUB$m$--$m$ data with $m=4,6,8,10$ for $S=1/2$, and $m=4,6,8$ for $S=1, 3/2$), based on (a) the $\mathrm{e}1$ extrapolation scheme for the energy and (b) the $\mathrm{e}2$ and $\mathrm{e}3$ schemes for the order parameter). For comparison, we include $S=1/2$ ED results from the $24$- and $32$-site clusters, and results from linear spin-wave (LSW) theory for each value of $S$. The vertical dashed lines designate the classical phase boundaries.}
\label{fig:Sdependence}
\end{figure}

The above $2n/(2n-1)$ staggering in LSUB$m$ or SUB$m$--$m$ sequences of CCM approximants for physical quantities is common to essentially all spin-lattice models on all lattices. However, one should note that an  additional $4n/(4n-2)$ staggering has been observed on honeycomb-lattice models (and see, e.g., Refs.~\cite{Li-Bishop_2016,Li-Bishop_2022} and references cited therein) in even ($m=2n$) CCM subsequences themselves.  It has been postulated that such additional staggering arises from the non-Bravais nature of the honeycomb lattice, which is itself composed of two interlocking triangular Bravais lattices. Thus, each triangular lattice separately exhibits the $2n/(2n-1)$ staggering, and the effect is then magnified twofold into a $4n/(4n-2)$ staggering in the even ($m=2n$) subsequences and, presumably, a corresponding $(4n-1)/(4n-3)$ staggering in the odd ($m=2n-1$) subsequences.

In the present case, we have not observed any marked $4n/(4n-2)$ staggering of the above sort in any of our calculations, and hence we feel confident in extrapolating our LSUB$m$ and SUB$m$--$m$ results using all of the even ($m=2n$) approximants. However, when using any of the above extrapolation schemes, we always exclude the lowest-order approximants with $m=2$, since they are too far removed from the corresponding asymptotic ($m \to \infty$) limits. Thus, for the spin-$1/2$ case, all of our extrapolated results are based on LSUB$m$ data sets with $m=\{4,6,8,10\}$, whereas for the spin-1 and spin-$3/2$ cases they are based on SUB$m$--$m$ data sets with $m=\{4,6,8\}$.

\section{Results}\label{sec:results}

\subsection{CCM convergence} 
We first examine the convergence of the CCM results, by looking at the ground-state energy and the order parameter in the four magnetic phases of the spin-half KH model, as displayed in Fig.~\ref{fig:ccm_convergence}. 
Results for the ground-state energy are shown in Fig.~\ref{fig:ccm_convergence}\,(a). LSUB$m$ results are found to converge very rapidly with increasing level of approximation $m$, and the differences in energies between LSUB8 and LSUB10  levels of approximation are broadly of order $10^{-4}$ for all values of $K$ and $J$ for this system. 

The extrapolated ground-state energy values, based on the e1 scheme, compare well with ED results from two finite-size clusters with 24 and 32 sites and periodic boundary conditions. In some regions, the ED energies are noticeably lower than those of the CCM, which can be attributed to finite-size effects. 

The CCM results for the order parameter in the four magnetic phases of the spin-half model are shown in Fig.~\ref{fig:ccm_convergence}\,(b). LSUB$m$ results are again found to converge very rapidly with increasing level of approximation $m$, and differences in the order parameter between LSUB8 and LSUB10 levels of approximation are broadly less than about $0.05$ for all values of the parameter $\phi$. Extrapolated results are shown using both the e2 and e3 extrapolation schemes.

\subsection{Variation with $S$}
In addition to the $S\!=\!1/2$ calculations, we have also carried out CCM calculations for $S\!=\!1$ and $3/2$. Figure~\ref{fig:Sdependence} shows the extrapolated CCM results for (a) the rescaled ground-state energy $E_0/(N S^2)$, and (b) the rescaled order parameters $M/S$, for all values of $S$, including the $S\!=\!1/2$ data of Fig.~{\ref{fig:ccm_convergence}}. Although not shown explicitly here, convergence of the SUB$m$--$m$ data is rapid for all values of $S$ and for all values of $\phi$, for both the ground-state energy and the order parameters. 
For comparison, we also show the $S\!=\!1/2$ ED data from the 24- and 32-site clusters shown previously, as well as results from linear spin-wave (LSW) theory calculations (which agree with previously published data~\cite{Gotfryd2017,Consoli2020}). The correspondence with the CCM data is again good. 

The rescaled order parameters $M/S$ clearly increase with $S$ for all parameters $\phi$, which is consistent with the expectation that quantum fluctuations weaken with increasing $S$.
Furthermore, we observe that, for $S\!=\!1$ and $3/2$, the extrapolated CCM data based on the $e2$ scheme agree well with the LSW data far away from the spin liquid regions. This is consistent with the general expectations mentioned in Sec.~\ref{sec:ExtrSchemes} that the $e2$ scheme is the appropriate one in regions with weak quantum fluctuations.

\begin{figure}[!t]
\includegraphics[width=\linewidth]{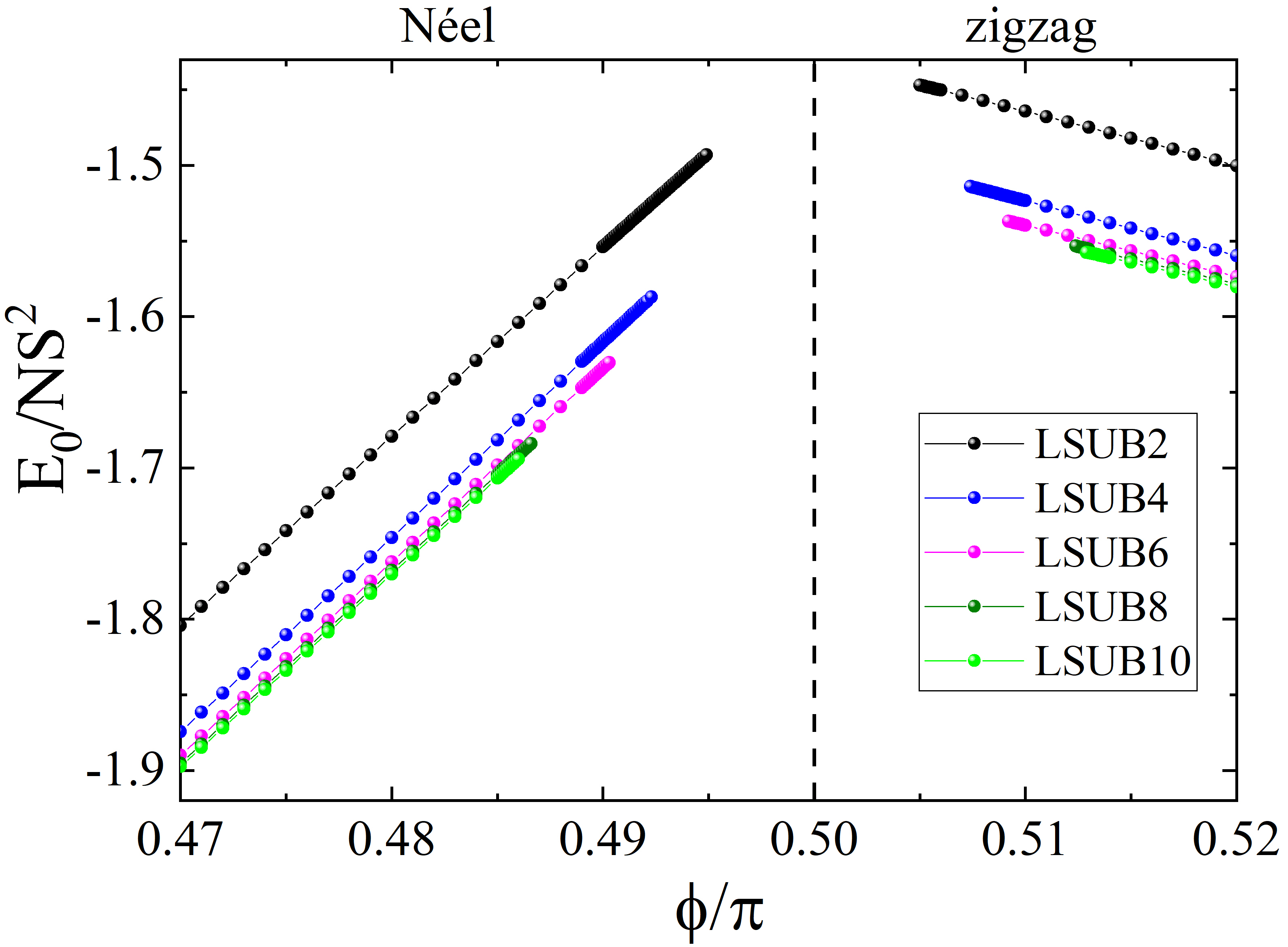}
\caption{{\bf Convergence of termination points with $\vb*{m}$.} Evolution of the rescaled ground-state energy $E_0/(NS^2)$ of the spin-$1/2$ KH model with $\phi$, in the vicinity of the AF Kitaev point $\phi=\pi/2$ (vertical dashed line), for different  different levels of LSUB$m$ approximation ($m=2, 4, 6, 8, 10$).}
\label{fig:ConvTermPoints}
\end{figure}

\subsection{Termination points}
As we show below, we find a number of termination points in the vicinity of the classical transition points $\phi=\pm\pi/2$, $3\pi/4$ and $7\pi/4$ for $S\!=\!1/2$, and in the vicinity of $\phi=\pm\pi/2$ for $S\!=\!1, 3/2$. At these points, the CCM equations cease to converge before results for the order parameters of the two adjacent semiclassical phases can either go to zero or intersect one another. 
The presence of such termination points is common in spin models and their physical origin is well understood  (see e.g., Ref.~\cite{Farnell-Bishop_2004}). They are manifestations of quantum phase transitions, i.e., they signal the presence of intermediate phases, sandwiched between the two neighboring model states around which we perform the CCM calculations.

More specifically, termination points have been seen in many numerical implementations of the CCM for spin-lattice problems, using either the LSUB$m$ or SUB$m$--$m$ approximation schemes. Interestingly, and as is seen here in particular, the $m$th-order calculations (with a fixed finite value of $m$), based on a specific quasiclassical ordered state, extend beyond the exact ($m \to \infty$) transition point for the phase in question, out to some corresponding termination points, beyond which no real solution exists for the respective coupled sets of nonlinear LSUB$m$ or SUB$m$--$m$ equations for the ket-state  CCM coefficients, as given by Eq.~(\ref{eq:eqs-for-CCM_GS-ket-coeffts}). 

We note that the positions of the termination points depend on the index $m$. This is demonstrated in  Fig.~\ref{fig:ConvTermPoints} for the termination points occurring in the neighborhood of the AF Kitaev   point ($\phi\!=\!\pi/2$). In particular, these $m$th-order termination points can be seen to shift away systematically from the classical boundary with increasing $m$. This reflects the fact that higher level of truncations incorporate more and more quantum fluctuations, which in turn tend to destabilize the classical orders further.

Additionally, the $m$th-order termination points can be seen to converge uniformly to their $m \to \infty$ limit with increasing values of $m$. Such behavior has been observed in many previous CCM calculations for spin-lattice models (see, e.g., Refs.~\cite{Farnell-Bishop_2004,Li-Bishop_2016,Li-Bishop_2022}).

\begin{figure}[!t]
\includegraphics[width=\linewidth]{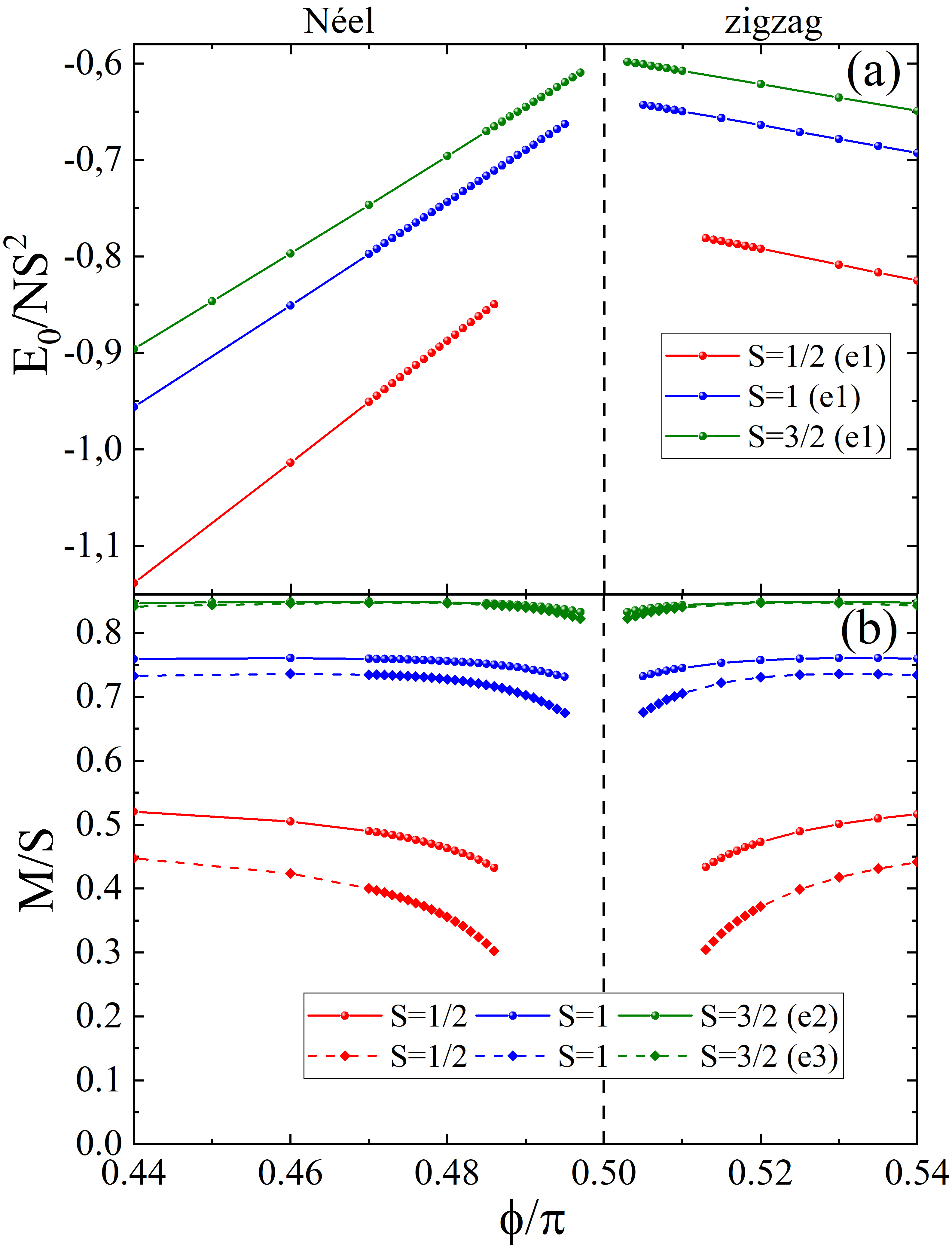}
\caption{{\bf Boundaries of the AF Kitaev QSL$_1$ phase from CCM termination points.} Extrapolated results for (a) the rescaled ground-state energy $E_0/(NS^2)$, and (b) the rescaled order parameter $M/S$, of the spin-$S$ KH model, with $S=1/2$ (red), $1$ (blue) and $3/2$ (olive), obtained from CCM expansions around the N\'eel and zigzag reference states, in the vicinity of the AF Kitaev point $\phi=\pi /2$ (vertical dashed line).}\label{fig:AFKitaev}
\end{figure}

\begin{figure}[!t]
\includegraphics[width=0.98\linewidth]{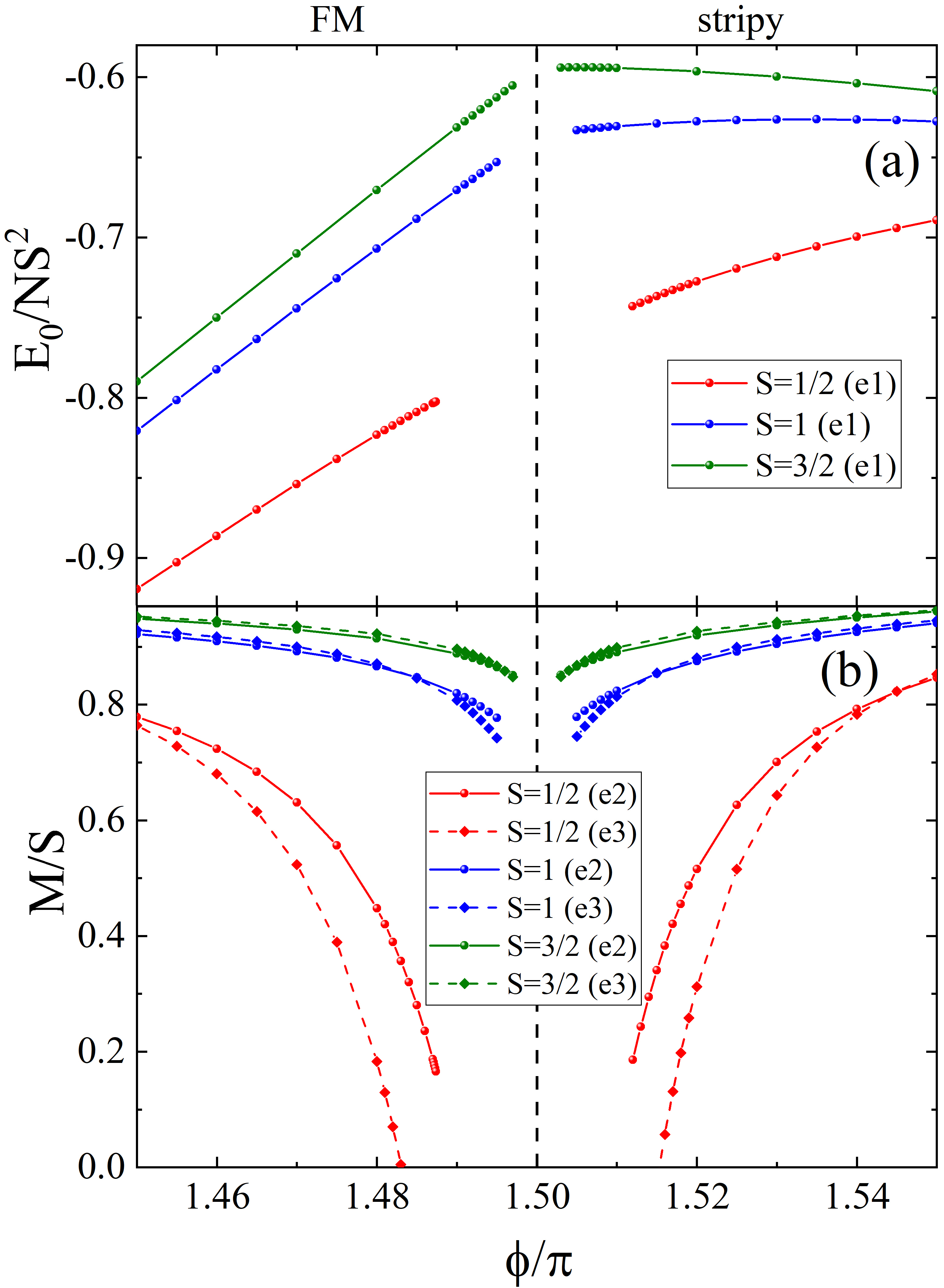}
\caption{{\bf Boundaries of the FM Kitaev QSL$_2$ phase from CCM termination points.} Extrapolated results for (a) the rescaled ground-state energy $E_0/(NS^2)$, and (b) the rescaled order parameter $M/S$, of the spin-$S$ KH model, with $S=1/2$ (red), $1$ (blue) and $3/2$ (olive), obtained from CCM expansions around the FM and stripy reference states, in the vicinity of the FM Kitaev point $\phi =3\pi /2$ (vertical dashed line).}
\label{fig:FMKitaev}
\end{figure}

\begin{figure}[!t]
\includegraphics[width=\linewidth]{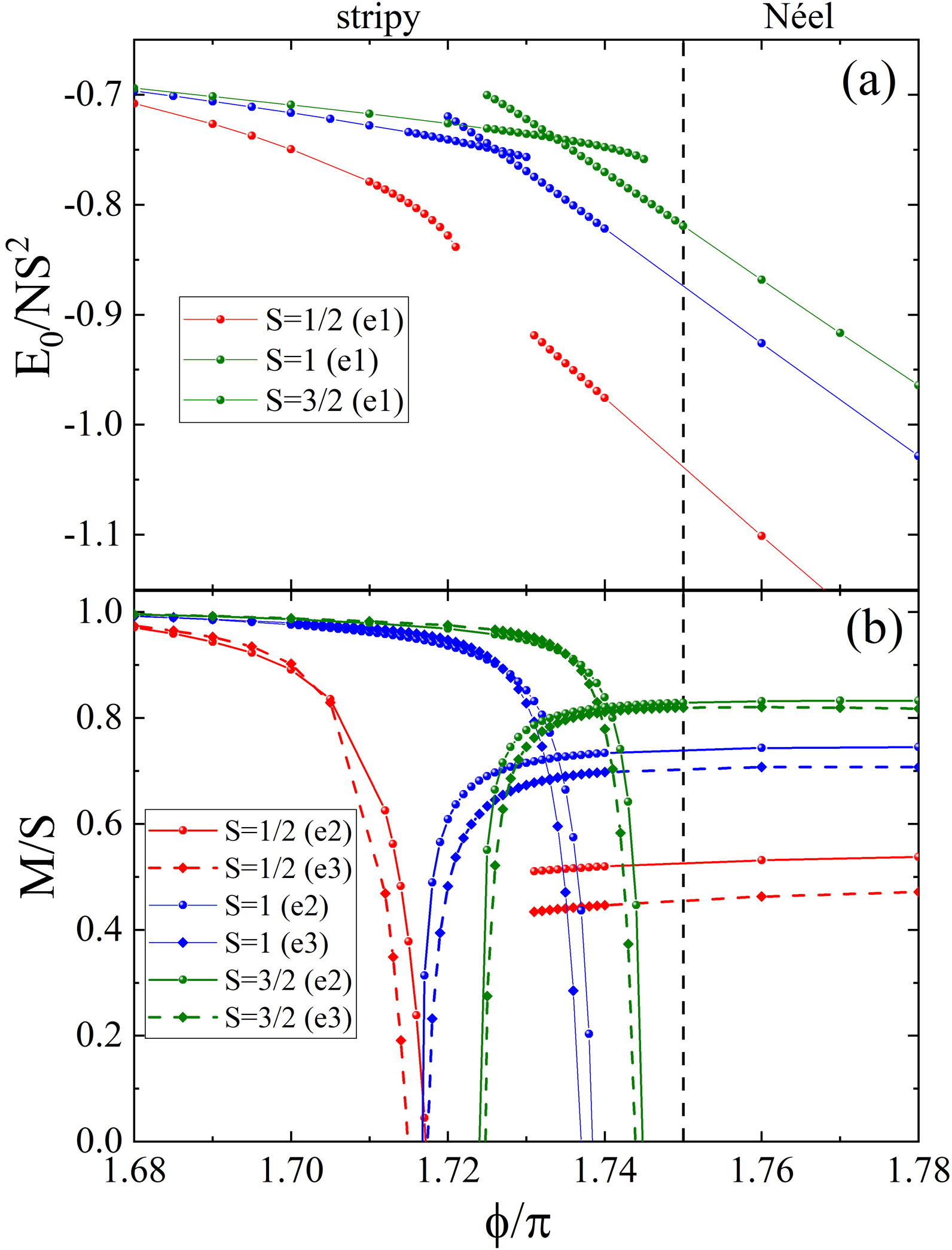}
\caption{{\bf Transition region between the stripy and the N\'eel phases.} Extrapolated results for (a) the rescaled ground-state energy $E_0/(NS^2)$, and (b) the rescaled order parameter $M/S$, of the spin-$S$ KH model, with $S=1/2$ (red), $1$ (blue) and $3/2$ (olive), obtained from CCM expansions around the stripy and N\'eel reference states, in the vicinity of the classical transition point $\phi=7\pi/4$ (vertical dashed line).}
\label{fig:stripy-Neel}
\end{figure}

\begin{figure}[t]
\includegraphics[width=\linewidth]{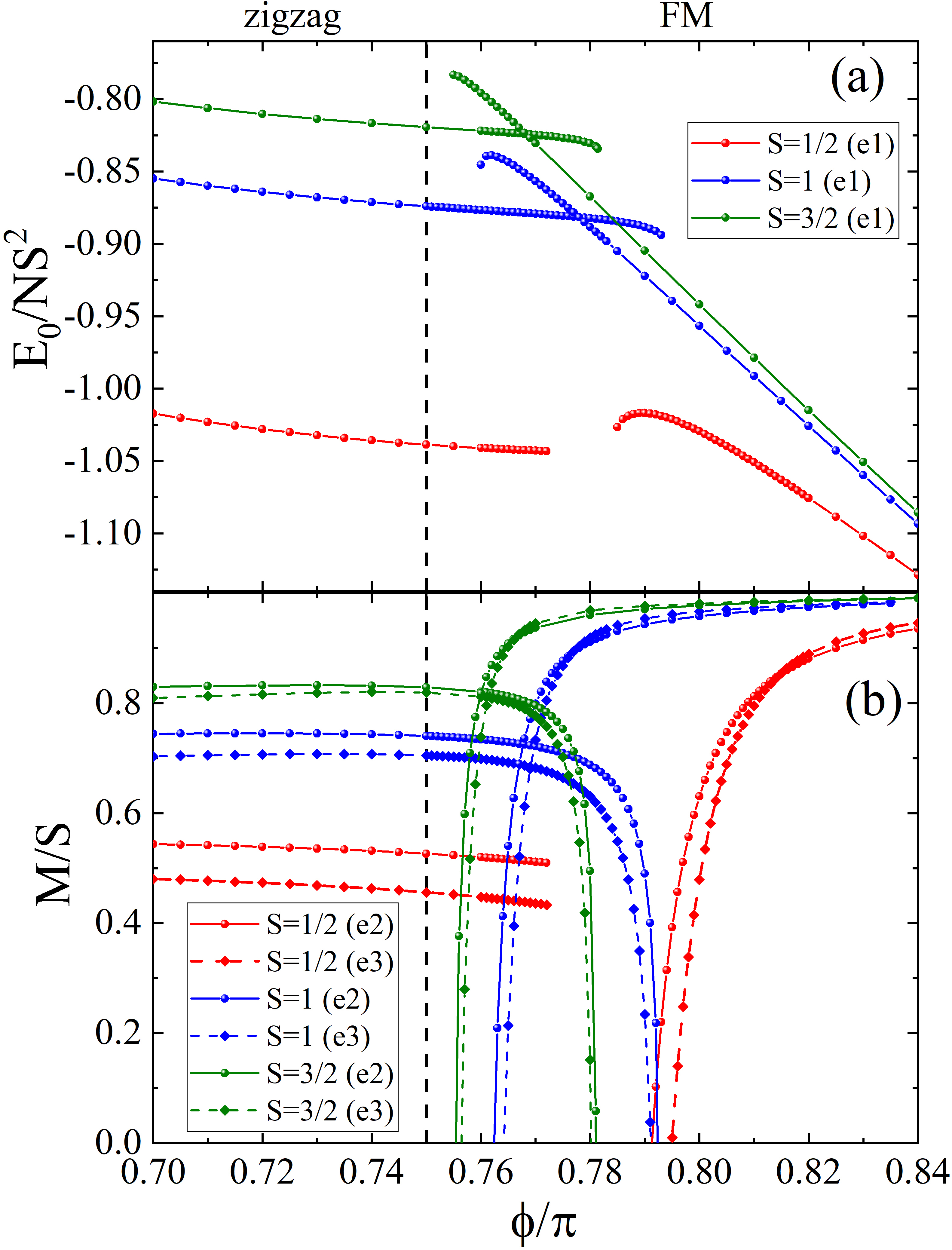}
\caption{{\bf Transition region between the FM and zigzag phases.} Extrapolated results for (a) the rescaled ground-state energy $E_0/(NS^2)$, and (b) the rescaled order parameter $M/S$, of the spin-$S$ KH model, with $S=1/2$ (red), $1$ (blue) and $3/2$ (olive), obtained from CCM expansions around the FM and zigzag phases, in the vicinity of the classical transition point $\phi=3\pi/4$ (vertical dashed line).} \label{fig:FM-zz}
\end{figure}

\subsection{Existence of the two Kitaev QSL phases and their stability region from CCM}
Next, we focus on the behavior of the extrapolated CCM data for (a) the ground-state energy and (b) the scaled order parameter $M/S$, in the vicinity of the two Kitaev points in Figs.~\ref{fig:AFKitaev} and \ref{fig:FMKitaev}. 
The most salient feature is the existence of termination points, occurring before the ground-state energies of the two neighboring ordered phases cross each other. This occurs in both the region between the N\'eel and the zigzag phase (Fig.~\ref{fig:AFKitaev}) and the region between the ferromagnetic and the stripy phases (Fig.~\ref{fig:FMKitaev}).  
As mentioned above, such termination points are  manifestations of quantum phase transitions. 
In the present case they signal the existence of the two Kitaev spin liquid phases, which are expected to survive in an extended region around the pure Kitaev points $\phi\!=\!\pm\pi/2$~\cite{Kitaev2006,Chen2008,Mandal2009PRB,Obrien2016,Baskaran2008PRB,Rousochatzakis2018NC,Dong2020PRB,Lee2020PRR,Jin2022NC,Natori2023}.

While we do not address the nature of the two Kitaev QSL phases for $S\!=\!1$ and $3/2$, the CCM can give accurate predictions for the stability range of these phases for all values of $S$, including $S\!=\!1/2$. 
The numerical values of the parameter $\phi$ associated with the onsets of the two QSL phases, as extracted from the e1 extrapolation scheme of the CCM energies at the corresponding termination points, are provided in Table~\ref{tab:PD}. 
As expected, the extent of each of the two Kitaev QSL phases shrinks quickly with increasing $S$, which is consistent with fRG calculations~\cite{Fukui2022}. 

A surprising result for the $S\!=\!1/2$ case is that the CCM prediction for the extent of the QSL$_2$ phase is appreciably narrower than the one found by ED on the symmetric 24-site cluster~\cite{Gotfryd2017}, or the one found by fRG~\cite{Fukui2022}
(see Table~\ref{tab:PD}). It appears therefore that these methods overestimate the extent of the QSL$_2$ phase. The CCM results are, however, much more consistent with CMFT calculations on the same (24-site) cluster by the same group~\cite{Gotfryd2017} (see Table~\ref{tab:PD}). 

Let us now comment on the nature of phase transitions between the two Kitaev spin liquid phases and their neighboring semiclassical orders. 
As discussed above, the extrapolation rule e2 for the order parameters only applies to cases of weak or mild frustration, whereas rule e3 applies to cases of high frustration, and especially in regions where order breaks down. As such, in order to determine with accuracy any phase terminations, the extrapolation rule e3 for the local order parameters should take complete precedence over rule e2 at or near phase transitions. 
With this in mind, the data shown in Figs.~\ref{fig:AFKitaev} and \ref{fig:FMKitaev} deliver the following insights:

(i) The transitions between the Kitaev QSL$_1$ state and the surrounding N\'eel and zigzag states are all first-order for $S\!=\!3/2$, $1$ and $1/2$ (and perhaps weakly first-order for $S\!=\!1/2$).

(ii) The transitions between the Kitaev QSL$_2$ state and the surrounding FM and stripy states are first-order for $S\!=\!3/2$ and $1$ but seem to be continuous for $S\!=\!1/2$. This latter observation is consistent with numerical studies based on DMRG~\cite{Shinjo2015PRB} and ED~\cite{Gotfryd2017} methods, which suggest that the transition between the QSL$_2$ state and the stripy state is continuous~\cite{Chaloupka2013PRL} or weakly first-order.

\subsection{Stripy-N\'eel and zigzag-FM transition regions: \\ Evidence for intermediate phases}\label{sec:intphases}
In Figs.~\ref{fig:stripy-Neel} and \ref{fig:FM-zz} we zoom in on the behavior of the extrapolated CCM data for the ground-state energy and the order parameter in the transition region between the stripy and the N\'eel phases and the transition region between the zigzag and the FM phases, respectively. These two regions map to each other under the duality transformation $\mc{T}_4$.

For $S\!=\!1$ and $3/2$, we find that the CCM energies [Figs.~\ref{fig:stripy-Neel}\,(a) and \ref{fig:FM-zz}\,(a)] of the respective ground states cross each other, and a similar crossing is found in the CCM order parameter data [Fig.~\ref{fig:stripy-Neel}\,(b) and ~\ref{fig:FM-zz}\,(b)]. These crossings signal first-order phase transitions, consistent with general expectations based on the classical limit of the model and the different symmetries of the states.  

\begin{figure}[!t]
\includegraphics[width=\linewidth]{
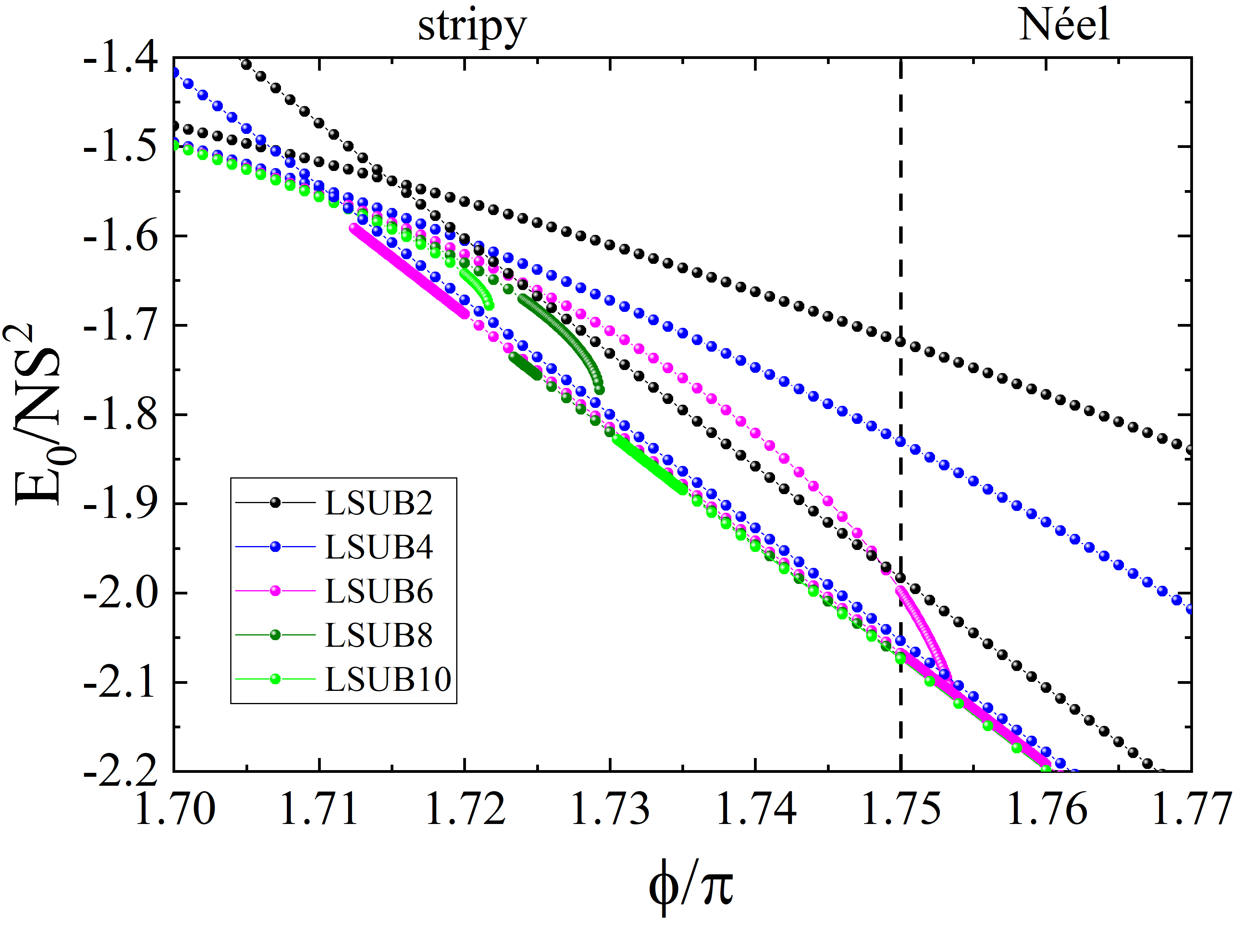}
\caption{{\bf Transition region between the stripy and the N\'eel phases, bare CCM data,} $\vb*{S=1/2}$. Evolution of the rescaled ground-state energy $E_0/(NS^2)$ of the spin-$1/2$ KH model
with $\phi$ in the vicinity of the classical stripy-N\'eel transition point ($\phi\!=\!7\pi/4$, vertical dashed line) for different levels of LSUB$m$ approximation ($m=2, 4, 6, 8, 10$).}
\label{fig:Fig10}
\end{figure}

The situation for $S\!=\!1/2$ is qualitatively different.  Here, the CCM data for the ground-state energies do not cross (except for the lowest LSUB$m$ levels with $m=2,4$), but show termination points instead. 
In particular, unlike the behavior seen in Fig.~\ref{fig:ConvTermPoints}, here one needs to go up to LSUB$m$ levels with $m\!\geq\!10$ to find two clearly separated termination points, see Fig.~\ref{fig:Fig10}. At lower truncation levels, $m\leq 8$, we find either no termination points at all, at least in the range shown in Fig.~\ref{fig:Fig10} ($m\!=\!2$ and $4$, where we see crossing points), or two termination points that have passed each other, i.e., if we denote by $\phi_t^{\text{phase}}$ the termination point of a specific phase, then $\phi_t^{\text{stripy}} > \phi_t^{\text{N\'eel}}$ ($m\!=\!6$ and $8$).  It is only for the LSUB$m$ data shown with $m\!=\!10$ that finally $\phi_t^{\text{stripy}}<\phi_t^{\text{N\'eel}}$, as expected on physical grounds for an intermediate phase.  Clearly, such behavior is then expected in similar LSUB$m$ data $\forall\, m\geq 10$.

Turning to the extrapolated order parameter data [Figs.~\ref{fig:stripy-Neel}\,(b) and \ref{fig:FM-zz}\,(b)], we find a weak variation with $\phi$ on the N\'eel side (respectively, the zigzag side) with our data simply terminating at the termination point, $\phi_t^{\text{N\'eel}}$, of the LSUB$10$ curve, and a rapid disappearance of the order parameter on the stripy (respectively, the FM) side. 
So, at the level of the LSUB$m$ scheme, with $m\leq 10$, the results suggest the presence of a narrow intermediate phase, sandwiched between the N\'eel and the stripy phase (Fig.~\ref{fig:stripy-Neel}), and (precisely as demanded by duality) a similar intermediate phase sandwiched between the zigzag and the FM phase (Fig.~\ref{fig:FM-zz}). The results also show that the transition to the stripy phase (and similarly the transition to the FM phase) is continuous, whereas the transition to the N\'eel phase (and similarly the transition to the zigzag phase) is discontinuous.

The presence of these intermediate states is completely unexpected, both at the classical level and, indeed, from previous numerical works for the $S=1/2$ KH model.
The fact that we need to go up to the $m\!=\!10$ truncation level to find two clearly separated termination points suggests that the intermediate phases may have a large unit cell, although this is by no means certain.
Moreover, the general tendency of the various LSUB$m$ data curves shown in Fig.~\ref{fig:Fig10} (and, indeed in {\it all\,} corresponding CCM calculations for other spin-lattice models) indicates that it is extremely unlikely that these intermediate phases will shrink and eventually disappear at some higher truncation level, $m\!>\!10$.

\setlength{\tabcolsep}{1.1em}
\renewcommand{\arraystretch}{1.4}
\begin{table*}[!t]
%\begin{ruledtabular}
\begin{tabular}{ [ c \thv c | c |c|c ]}
\thickhline
\diagbox[width=4.2cm,leftsep=0.2cm,innerleftsep=0.2cm,innerrightsep=2pt,linewidth=1pt,font=\normalsize]{Method}{Boundary region}
& 
  $\begin{array}{c}\text{N\'eel-zigzag}\\ (\text{QSL}_1) \end{array}$
& 
    $\begin{array}{c}\text{zigzag-FM}\\ \end{array}$ 
& 
    $\begin{array}{c}\text{FM-stripy}\\ (\text{QSL}_2) \end{array}$
& 
    $\begin{array}{c}\text{stripy-N\'eel}\\ \end{array}$
\\
\thickhline
$S=\infty$ & $90^\circ$ & $135^\circ$ & $270^\circ$ & $315^\circ$
\\
\thickhline
$\begin{array}{c}
%{\bf S=3/2,~\text{CCM}}\\
\vb*{S=3/2},~\text{CCM}\\
\text{\cdr{from duality}}
\end{array}$
&
$\begin{array}{c}
(89.46^\circ, 90.54^\circ) \\
\cdr{(89.45^\circ, 90.53^\circ)}
\end{array}$
&
$\begin{array}{c}
138.28^\circ \\
\cdr{138.29^\circ} 
\end{array}$
&
$\begin{array}{c}
(269.46^\circ, 270.54^\circ) \\
\cdr{(269.45^\circ, 270.53^\circ)}
\end{array}$
& 
$\begin{array}{c}
312.05^\circ\\
\cdr{312.06^\circ}
\end{array}$
\\
\hline
\text{fRG}~\cite{Fukui2022}
&
$(88.65^\circ, 91.35^\circ)$
& 
$137.92^\circ$
& 
$(266.84^\circ, 273.16^\circ)$
& 
$313.36^\circ$
\\
\thickhline
$\begin{array}{c}
%{\bf S=1,~\text{CCM}}\\
\vb*{S=1},~\text{CCM}\\
\text{{\cdr from duality}}
\end{array}$
&
$\begin{array}{c}
(89.1^\circ, 90.9^\circ) \\
\cdr{(89.07^\circ, 90.87^\circ)}
\end{array}$
& 
$\begin{array}{c}
140.04^\circ \\
\cdr{140.035^\circ} 
\end{array}$
&
$\begin{array}{c}
(269.1^\circ, 270.9^\circ) \\
\cdr{(269.07^\circ, 270.87^\circ)}
\end{array}$
& 
$\begin{array}{c}
310.716^\circ \\
\cdr{310.713^\circ}
\end{array}$ 
\\
\hline
\text{fRG}~\cite{Fukui2022}
&
$(87.75^\circ, 92.25^\circ)$
&
$137.92^\circ$
&
$(263.62^\circ, 274.99^\circ)$
&
$313.36^\circ$
\\
\text{iDMRG}~\cite{Dong2020PRB}
&
$(88.92^\circ,91.08^\circ)$
&
$156.6^\circ$
&
$(267.3^\circ,272.52^\circ)$
&
$329.4^\circ$
\\
\thickhline
$\begin{array}{c}
\vb*{S=1/2},~\text{CCM}\\
\text{{\cdr from duality}}
\end{array}$
&
$\begin{array}{c}
(87.48^\circ, 92.31^\circ) \\
\cdr{(87.45^\circ, 92.32^\circ)}
\end{array}$
&
$\begin{array}{c}
(138.96^\circ, 141.3^\circ)\\
\cdr{(138.88^\circ, 141.38^\circ)} 
\end{array}$
&
$\begin{array}{c}
(267.7^\circ, 272.16^\circ) \\ 
\cdr{(267.66^\circ, 272.1^\circ)}
\end{array}$
& 
$\begin{array}{c}
(309.78^\circ , 311.58^\circ)  \\
\cdr{(309.83^\circ, 311.52^\circ)}
\end{array}$
\\
\hline
\text{ED}~\cite{Gotfryd2017}
&
$(88.92^\circ, 91.08^\circ)$
&
$146.52^\circ$
& 
$(260.64^\circ, 277.02^\circ)$ 
&
$306.72^\circ$
\\
\text{CMFT}~\cite{Gotfryd2017}
&
$(89.28^\circ, 90.9^\circ)$
&
$148.5^\circ$
&
$(266.04^\circ, 273.42^\circ)$
&
$305.82^\circ$
\\
\text{fRG}~\cite{Fukui2022}
& 
$(86.84^\circ, 93.16^\circ)$
&
$142.92^\circ$
&
$(246.21^\circ, 283.76^\circ)$
& 
$311.24^\circ$
\\
\thickhline
\end{tabular}
%\end{ruledtabular}
\caption{{\bf Transition points between the various phases of the KH model.} Row 2 shows classical values. 
Rows 3, 6 and 10 show our CCM predictions (based on the given model states in each column) for $S\!=\!3/2$, $1$ and $1/2$, respecively, as found by CCM termination points or ground-state energy crossings.
Rows 4, 7 and 11 (in dark red) show results from the positions of the corresponding dual points, as found by the independent CCM calculations in other columns. 
Rows 5, 8, 9, and 12-14 show  values from fRG~(Ref.~\cite{Fukui2022} and private communication), iDMRG~\cite{Dong2020PRB}, and ED and CMFT calculations on the symmetric 24-site cluster~\cite{Gotfryd2017}.}\label{tab:PD}
\end{table*}

\subsection{CCM phase diagrams and duality benchmark}\label{sec:PD}
Collecting the extrapolated CCM data, particularly the positions of the various termination points and ground-state energy crossings, gives rise to the phase diagrams shown earlier in Fig.~\ref{fig:PD}, for $S\!=\!1/2$, $1$ and $3/2$. The narrow regions with question marks in the $S\!=\!1/2$ phase diagram designate the enigmatic intermediate phases discussed in Sec.~\ref{sec:intphases}.

The actual numerical values of the boundaries of the various phases are provided in rows 3-5 of Table~\ref{tab:PD}, with a precision within $0.001$~radians, which is the step size in our fine-tuning of the parameter $\phi$ near the transition points. For comparison, we also provide the corresponding transition points for the classical limit ($S\!=\!\infty$), and the ones obtained from fRG (Ref.~\cite{Fukui2022} and private communication), iDMRG~\cite{Dong2020PRB}, and ED and CMFT calculations on the symmetric 24-site cluster~\cite{Gotfryd2017}.

Importantly, we have checked whether the obtained values of the various boundaries are consistent with the duality transformation $\mc{T}_4$. To that end, we compare, in each entry of rows 3-5 of Table~\ref{tab:PD}, the boundaries found by CCM calculations based on the corresponding model states to the values resulting from those of the corresponding dual points, found by independent CCM calculations based on their dual model states. The agreement is, in most cases, well within our numerical precision. This provides a crucial benchmark of the method, as the CCM expansions around two states that are dual to each other (e.g., the zigzag and the N\'eel states), that come with very different magnetic unit cells, are completely independent.

\section{Conclusions}
\label{sec:Conclusions}
We have obtained the phase diagrams (Fig.~\ref{fig:PD}) of the spin-$S$ Kitaev-Heisenberg model on the honeycomb lattice, for $S\!=\!1/2$, $1$ and $3/2$, using extensive high-order CCM calculations based on the well-established SUB$m$--$m$ truncation scheme, with $m\leq 10$ for $S\!=\!1/2$ and $m\leq 8$ for $S\!=\!1,3/2$.
Besides the four collinear magnetic orders expected in the classical limit, the phase diagrams comprise the two spin liquid phases expected around the two Kitaev points~\cite{Kitaev2006,Chen2008,Mandal2009PRB,Chaloupka2010PRL,Chaloupka2013PRL,Obrien2016,Baskaran2008PRB,Rousochatzakis2018NC,Dong2020PRB,Lee2020PRR,Jin2022NC,Natori2023}, as well as two very narrow intermediate phases (one sandwiched between the zigzag and the FM phase, and the other between the N\'eel and the stripy phase) for $S\!=\!1/2$, whose precise nature calls for further dedicated investigations.  
The existence of the two Kitaev QSL phases and the two enigmatic phases for $S\!=\!1/2$ is manifested through the appearance of CCM termination points, which give direct evidence for quantum phase transitions. 

In addition to the accurate predictions for the various transition points (Table~\ref{tab:PD}), our CCM results provide clear insights on the order of the corresponding phase transitions, by an analysis of CCM termination/energy-crossing points, and the behavior of the corresponding (extrapolated) order parameter data in the vicinity of these transitions.

Besides the careful convergence analysis of the CCM results, and the comparison to published ED~\cite{Gotfryd2017}, CMFT~\cite{Gotfryd2017}, iDMRG~\cite{Dong2020PRB} and fRG~\cite{Fukui2022} data, we have benchmarked our results against the duality transformation $\mc{T}_4$ of the model and found excellent correspondence of the values of the transition points. This level of accuracy establishes the coupled cluster method as a versatile technique that can capture strong quantum fluctuations that are inherently present in highly frustrated models.

As far as we are aware, this is the first application of the coupled cluster method to a generalized Kitaev-like model with bond-dependent anisotropic interactions. As such, the present study serves as a testing ground for exploring the more realistic models of available materials, and for providing new insights for their complex phase diagrams and phenomenology, including recent $S\!>\!1/2$ models models with bilinear-biquadratic interactions~\cite{Pohle2023,Pohle2024}. 
Another exciting prospect is the potential to use dynamical extensions of the coupled cluster method to calculate dynamical response functions $\mc{S}({\bf q},\omega)$, whose accurate determination has been a great challenge for highly frustrated models.  We hope to be able to address this in a future paper.

\acknowledgments
IR and MG acknowledge the support of the Engineering and Physical Sciences Research Council (EPSRC, United Kingdom), Grant No. EP/V038281/1. RFB gratefully acknowledges the Leverhulme Trust (United Kingdom) for the award of an Emeritus Fellowship, Grant No. EM-2020-013.  DJJF, JR, and RFB acknowledge the support and 
hospitality of Loughborough University during fruitful visits concerning this research.
Finally, we would like to thank Kiyu Fukui %(University of Tokyo) 
for sharing their numerical fRG data for the zigzag--FM and stripy--N\'eel transition points (given in Table~\ref{tab:PD}), as well as Natalia Perkins, Joerg Schulenburg, Rico Pohle and Arnaud Ralko for useful discussions.
%\appendix

%\bibliography{references}

\begin{thebibliography}{130}%
\makeatletter
\providecommand \@ifxundefined [1]{%
 \@ifx{#1\undefined}
}%
\providecommand \@ifnum [1]{%
 \ifnum #1\expandafter \@firstoftwo
 \else \expandafter \@secondoftwo
 \fi
}%
\providecommand \@ifx [1]{%
 \ifx #1\expandafter \@firstoftwo
 \else \expandafter \@secondoftwo
 \fi
}%
\providecommand \natexlab [1]{#1}%
\providecommand \enquote  [1]{``#1''}%
\providecommand \bibnamefont  [1]{#1}%
\providecommand \bibfnamefont [1]{#1}%
\providecommand \citenamefont [1]{#1}%
\providecommand \href@noop [0]{\@secondoftwo}%
\providecommand \href [0]{\begingroup \@sanitize@url \@href}%
\providecommand \@href[1]{\@@startlink{#1}\@@href}%
\providecommand \@@href[1]{\endgroup#1\@@endlink}%
\providecommand \@sanitize@url [0]{\catcode `\\12\catcode `\$12\catcode
  `\&12\catcode `\#12\catcode `\^12\catcode `\_12\catcode `\%12\relax}%
\providecommand \@@startlink[1]{}%
\providecommand \@@endlink[0]{}%
\providecommand \url  [0]{\begingroup\@sanitize@url \@url }%
\providecommand \@url [1]{\endgroup\@href {#1}{\urlprefix }}%
\providecommand \urlprefix  [0]{URL }%
\providecommand \Eprint [0]{\href }%
\providecommand \doibase [0]{https://doi.org/}%
\providecommand \selectlanguage [0]{\@gobble}%
\providecommand \bibinfo  [0]{\@secondoftwo}%
\providecommand \bibfield  [0]{\@secondoftwo}%
\providecommand \translation [1]{[#1]}%
\providecommand \BibitemOpen [0]{}%
\providecommand \bibitemStop [0]{}%
\providecommand \bibitemNoStop [0]{.\EOS\space}%
\providecommand \EOS [0]{\spacefactor3000\relax}%
\providecommand \BibitemShut  [1]{\csname bibitem#1\endcsname}%
\let\auto@bib@innerbib\@empty
%</preamble>
\bibitem [{\citenamefont {Kitaev}(2006)}]{Kitaev2006}%
  \BibitemOpen
  \bibfield  {author} {\bibinfo {author} {\bibfnamefont {A.}~\bibnamefont
  {Kitaev}},\ }\bibfield  {title} {\bibinfo {title} {{Anyons in an exactly
  solved model and beyond}},\ }\href
  {https://doi.org/10.1016/j.aop.2005.10.005} {\bibfield  {journal} {\bibinfo
  {journal} {Ann.\ Phys. (N.Y.)}\ }\textbf {\bibinfo {volume} {321}},\ \bibinfo
  {pages} {2} (\bibinfo {year} {2006})}\BibitemShut {NoStop}%
\bibitem [{\citenamefont {Jackeli}\ and\ \citenamefont
  {Khaliullin}(2009)}]{Jackeli2009PRL}%
  \BibitemOpen
  \bibfield  {author} {\bibinfo {author} {\bibfnamefont {G.}~\bibnamefont
  {Jackeli}}\ and\ \bibinfo {author} {\bibfnamefont {G.}~\bibnamefont
  {Khaliullin}},\ }\bibfield  {title} {\bibinfo {title} {{Mott insulators in
  the strong spin-orbit coupling limit: From Heisenberg to a quantum compass
  and Kitaev models}},\ }\href {https://doi.org/10.1103/PhysRevLett.102.017205}
  {\bibfield  {journal} {\bibinfo  {journal} {Phys.\ Rev.\ Lett.}\ }\textbf
  {\bibinfo {volume} {102}},\ \bibinfo {pages} {017205} (\bibinfo {year}
  {2009})}\BibitemShut {NoStop}%
\bibitem [{\citenamefont {Chaloupka}\ \emph {et~al.}(2010)\citenamefont
  {Chaloupka}, \citenamefont {Jackeli},\ and\ \citenamefont
  {Khaliullin}}]{Chaloupka2010PRL}%
  \BibitemOpen
  \bibfield  {author} {\bibinfo {author} {\bibfnamefont {J.}~\bibnamefont
  {Chaloupka}}, \bibinfo {author} {\bibfnamefont {G.}~\bibnamefont {Jackeli}},\
  and\ \bibinfo {author} {\bibfnamefont {G.}~\bibnamefont {Khaliullin}},\
  }\bibfield  {title} {\bibinfo {title} {{Kitaev-Heisenberg model on a
  honeycomb lattice: Possible exotic phases in iridium oxides
  ${A}_{2}{\mathrm{IrO}}_{3}$}},\ }\href
  {https://doi.org/10.1103/PhysRevLett.105.027204} {\bibfield  {journal}
  {\bibinfo  {journal} {Phys.\ Rev.\ Lett.}\ }\textbf {\bibinfo {volume}
  {105}},\ \bibinfo {pages} {027204} (\bibinfo {year} {2010})}\BibitemShut
  {NoStop}%
\bibitem [{\citenamefont {Witczak-Krempa}\ \emph {et~al.}(2014)\citenamefont
  {Witczak-Krempa}, \citenamefont {Chen}, \citenamefont {Kim},\ and\
  \citenamefont {Balents}}]{Krempa2014ARCMP}%
  \BibitemOpen
  \bibfield  {author} {\bibinfo {author} {\bibfnamefont {W.}~\bibnamefont
  {Witczak-Krempa}}, \bibinfo {author} {\bibfnamefont {G.}~\bibnamefont
  {Chen}}, \bibinfo {author} {\bibfnamefont {Y.~B.}\ \bibnamefont {Kim}},\ and\
  \bibinfo {author} {\bibfnamefont {L.}~\bibnamefont {Balents}},\ }\bibfield
  {title} {\bibinfo {title} {Correlated quantum phenomena in the strong
  spin-orbit regime},\ }\href
  {https://doi.org/10.1146/annurev-conmatphys-020911-125138} {\bibfield
  {journal} {\bibinfo  {journal} {Annu.\ Rev.\ Condens.\ Matter Phys.}\
  }\textbf {\bibinfo {volume} {5}},\ \bibinfo {pages} {57} (\bibinfo {year}
  {2014})}\BibitemShut {NoStop}%
\bibitem [{\citenamefont {Rau}\ \emph {et~al.}(2016)\citenamefont {Rau},
  \citenamefont {Lee},\ and\ \citenamefont {Kee}}]{Rau2016ARCMP}%
  \BibitemOpen
  \bibfield  {author} {\bibinfo {author} {\bibfnamefont {J.~G.}\ \bibnamefont
  {Rau}}, \bibinfo {author} {\bibfnamefont {E.~K.-H.}\ \bibnamefont {Lee}},\
  and\ \bibinfo {author} {\bibfnamefont {H.-Y.}\ \bibnamefont {Kee}},\
  }\bibfield  {title} {\bibinfo {title} {Spin-orbit physics giving rise to
  novel phases in correlated systems: {I}ridates and related materials},\
  }\href {https://doi.org/10.1146/annurev-conmatphys-031115-011319} {\bibfield
  {journal} {\bibinfo  {journal} {Annu.\ Rev.\ Condens.\ Matter Phys.}\
  }\textbf {\bibinfo {volume} {7}},\ \bibinfo {pages} {195} (\bibinfo {year}
  {2016})}\BibitemShut {NoStop}%
\bibitem [{\citenamefont {Winter}\ \emph {et~al.}(2016)\citenamefont {Winter},
  \citenamefont {Li}, \citenamefont {Jeschke},\ and\ \citenamefont
  {Valent\'{\i}}}]{Winter2016PRB}%
  \BibitemOpen
  \bibfield  {author} {\bibinfo {author} {\bibfnamefont {S.~M.}\ \bibnamefont
  {Winter}}, \bibinfo {author} {\bibfnamefont {Y.}~\bibnamefont {Li}}, \bibinfo
  {author} {\bibfnamefont {H.~O.}\ \bibnamefont {Jeschke}},\ and\ \bibinfo
  {author} {\bibfnamefont {R.}~\bibnamefont {Valent\'{\i}}},\ }\bibfield
  {title} {\bibinfo {title} {{Challenges in design of Kitaev materials:
  Magnetic interactions from competing energy scales}},\ }\href
  {https://doi.org/10.1103/PhysRevB.93.214431} {\bibfield  {journal} {\bibinfo
  {journal} {Phys.\ Rev.\ B}\ }\textbf {\bibinfo {volume} {93}},\ \bibinfo
  {pages} {214431} (\bibinfo {year} {2016})}\BibitemShut {NoStop}%
\bibitem [{\citenamefont {Winter}\ \emph
  {et~al.}(2017{\natexlab{a}})\citenamefont {Winter}, \citenamefont {Tsirlin},
  \citenamefont {Daghofer}, \citenamefont {van~den Brink}, \citenamefont
  {Singh}, \citenamefont {Gegenwart},\ and\ \citenamefont
  {Valent\'{\i}}}]{Winter2017r}%
  \BibitemOpen
  \bibfield  {author} {\bibinfo {author} {\bibfnamefont {S.~M.}\ \bibnamefont
  {Winter}}, \bibinfo {author} {\bibfnamefont {A.~A.}\ \bibnamefont {Tsirlin}},
  \bibinfo {author} {\bibfnamefont {M.}~\bibnamefont {Daghofer}}, \bibinfo
  {author} {\bibfnamefont {J.}~\bibnamefont {van~den Brink}}, \bibinfo {author}
  {\bibfnamefont {Y.}~\bibnamefont {Singh}}, \bibinfo {author} {\bibfnamefont
  {P.}~\bibnamefont {Gegenwart}},\ and\ \bibinfo {author} {\bibfnamefont
  {R.}~\bibnamefont {Valent\'{\i}}},\ }\bibfield  {title} {\bibinfo {title}
  {Models and materials for generalized {K}itaev magnetism},\ }\href
  {https://doi.org/10.1088/1361-648X/aa8cf5} {\bibfield  {journal} {\bibinfo
  {journal} {J.\ Phys.: Condens.\ Matter}\ }\textbf {\bibinfo {volume} {29}},\
  \bibinfo {pages} {493002} (\bibinfo {year} {2017}{\natexlab{a}})}\BibitemShut
  {NoStop}%
\bibitem [{\citenamefont {Hermanns}\ \emph {et~al.}(2018)\citenamefont
  {Hermanns}, \citenamefont {Kimchi},\ and\ \citenamefont
  {Knolle}}]{Knolle2017ARCMP}%
  \BibitemOpen
  \bibfield  {author} {\bibinfo {author} {\bibfnamefont {M.}~\bibnamefont
  {Hermanns}}, \bibinfo {author} {\bibfnamefont {I.}~\bibnamefont {Kimchi}},\
  and\ \bibinfo {author} {\bibfnamefont {J.}~\bibnamefont {Knolle}},\
  }\bibfield  {title} {\bibinfo {title} {Physics of the {K}itaev model:
  {F}ractionalization, dynamic correlations, and material connections},\ }\href
  {https://doi.org/10.1146/annurev-conmatphys-033117-053934} {\bibfield
  {journal} {\bibinfo  {journal} {Annu.\ Rev.\ Condens.\ Matter Phys.}\
  }\textbf {\bibinfo {volume} {9}},\ \bibinfo {pages} {17} (\bibinfo {year}
  {2018})}\BibitemShut {NoStop}%
\bibitem [{\citenamefont {Takagi}\ \emph {et~al.}(2019)\citenamefont {Takagi},
  \citenamefont {Takayama}, \citenamefont {Jackeli}, \citenamefont
  {Khaliullin},\ and\ \citenamefont {Nagler}}]{Takagi2019NRP}%
  \BibitemOpen
  \bibfield  {author} {\bibinfo {author} {\bibfnamefont {H.}~\bibnamefont
  {Takagi}}, \bibinfo {author} {\bibfnamefont {T.}~\bibnamefont {Takayama}},
  \bibinfo {author} {\bibfnamefont {G.}~\bibnamefont {Jackeli}}, \bibinfo
  {author} {\bibfnamefont {G.}~\bibnamefont {Khaliullin}},\ and\ \bibinfo
  {author} {\bibfnamefont {S.~E.}\ \bibnamefont {Nagler}},\ }\bibfield  {title}
  {\bibinfo {title} {Concept and realization of {K}itaev quantum spin
  liquids},\ }\href {https://doi.org/10.1038/s42254-019-0038-2} {\bibfield
  {journal} {\bibinfo  {journal} {Nat.\ Rev.\ Phys.}\ }\textbf {\bibinfo
  {volume} {1}},\ \bibinfo {pages} {264} (\bibinfo {year} {2019})}\BibitemShut
  {NoStop}%
\bibitem [{\citenamefont {Janssen}\ and\ \citenamefont
  {Vojta}(2019)}]{Janssen2019}%
  \BibitemOpen
  \bibfield  {author} {\bibinfo {author} {\bibfnamefont {L.}~\bibnamefont
  {Janssen}}\ and\ \bibinfo {author} {\bibfnamefont {M.}~\bibnamefont
  {Vojta}},\ }\bibfield  {title} {\bibinfo {title} {{H}eisenberg-{K}itaev
  physics in magnetic fields},\ }\href
  {https://doi.org/10.1088/1361-648x/ab283e} {\bibfield  {journal} {\bibinfo
  {journal} {J.\ Phys.: Condens.\ Matter}\ }\textbf {\bibinfo {volume} {31}},\
  \bibinfo {pages} {423002} (\bibinfo {year} {2019})}\BibitemShut {NoStop}%
\bibitem [{\citenamefont {Motome}\ and\ \citenamefont
  {Nasu}(2020)}]{Motome2019JPSJ}%
  \BibitemOpen
  \bibfield  {author} {\bibinfo {author} {\bibfnamefont {Y.}~\bibnamefont
  {Motome}}\ and\ \bibinfo {author} {\bibfnamefont {J.}~\bibnamefont {Nasu}},\
  }\bibfield  {title} {\bibinfo {title} {Hunting {M}ajorana fermions in
  {K}itaev magnets},\ }\href {https://doi.org/10.7566/JPSJ.89.012002}
  {\bibfield  {journal} {\bibinfo  {journal} {J.\ Phys.\ Soc.\ Jpn.}\ }\textbf
  {\bibinfo {volume} {89}},\ \bibinfo {pages} {012002} (\bibinfo {year}
  {2020})}\BibitemShut {NoStop}%
\bibitem [{\citenamefont {Takayama}\ \emph {et~al.}(2021)\citenamefont
  {Takayama}, \citenamefont {Chaloupka}, \citenamefont {Smerald}, \citenamefont
  {Khaliullin},\ and\ \citenamefont {Takagi}}]{Takayama2021JPSJ}%
  \BibitemOpen
  \bibfield  {author} {\bibinfo {author} {\bibfnamefont {T.}~\bibnamefont
  {Takayama}}, \bibinfo {author} {\bibfnamefont {J.}~\bibnamefont {Chaloupka}},
  \bibinfo {author} {\bibfnamefont {A.}~\bibnamefont {Smerald}}, \bibinfo
  {author} {\bibfnamefont {G.}~\bibnamefont {Khaliullin}},\ and\ \bibinfo
  {author} {\bibfnamefont {H.}~\bibnamefont {Takagi}},\ }\bibfield  {title}
  {\bibinfo {title} {Spin–orbit-entangled electronic phases in {4\textit{d}}
  and {5\textit{d}} transition-metal compounds},\ }\href
  {https://doi.org/10.7566/JPSJ.90.062001} {\bibfield  {journal} {\bibinfo
  {journal} {J.\ Phys.\ Soc.\ Jpn.}\ }\textbf {\bibinfo {volume} {90}},\
  \bibinfo {pages} {062001} (\bibinfo {year} {2021})}\BibitemShut {NoStop}%
\bibitem [{\citenamefont {Trebst}\ and\ \citenamefont
  {Hickey}(2022)}]{Trebst2022}%
  \BibitemOpen
  \bibfield  {author} {\bibinfo {author} {\bibfnamefont {S.}~\bibnamefont
  {Trebst}}\ and\ \bibinfo {author} {\bibfnamefont {C.}~\bibnamefont
  {Hickey}},\ }\bibfield  {title} {\bibinfo {title} {{K}itaev materials},\
  }\href {https://doi.org/10.1016/j.physrep.2021.11.003} {\bibfield  {journal}
  {\bibinfo  {journal} {Phys.\ Rep.}\ }\textbf {\bibinfo {volume} {950}},\
  \bibinfo {pages} {1} (\bibinfo {year} {2022})}\BibitemShut {NoStop}%
\bibitem [{\citenamefont {Tsirlin}\ and\ \citenamefont
  {Gegenwart}(2022)}]{Tsirlin2022}%
  \BibitemOpen
  \bibfield  {author} {\bibinfo {author} {\bibfnamefont {A.~A.}\ \bibnamefont
  {Tsirlin}}\ and\ \bibinfo {author} {\bibfnamefont {P.}~\bibnamefont
  {Gegenwart}},\ }\bibfield  {title} {\bibinfo {title} {{K}itaev magnetism
  through the prism of lithium iridate},\ }\href
  {https://doi.org/10.1002/pssb.202100146} {\bibfield  {journal} {\bibinfo
  {journal} {Phys.\ Status Solidi B}\ }\textbf {\bibinfo {volume} {259}},\
  \bibinfo {pages} {2100146} (\bibinfo {year} {2022})}\BibitemShut {NoStop}%
\bibitem [{\citenamefont {Rousochatzakis}\ \emph {et~al.}(2024)\citenamefont
  {Rousochatzakis}, \citenamefont {Perkins}, \citenamefont {Luo},\ and\
  \citenamefont {Kee}}]{RousochatzakisRoPP2024}%
  \BibitemOpen
  \bibfield  {author} {\bibinfo {author} {\bibfnamefont {I.}~\bibnamefont
  {Rousochatzakis}}, \bibinfo {author} {\bibfnamefont {N.~B.}\ \bibnamefont
  {Perkins}}, \bibinfo {author} {\bibfnamefont {Q.}~\bibnamefont {Luo}},\ and\
  \bibinfo {author} {\bibfnamefont {H.-Y.}\ \bibnamefont {Kee}},\ }\bibfield
  {title} {\bibinfo {title} {Beyond {K}itaev physics in strong spin-orbit
  coupled magnets},\ }\href {https://doi.org/10.1088/1361-6633/ad208d}
  {\bibfield  {journal} {\bibinfo  {journal} {Rep.\ Prog.\ Phys.}\ }\textbf
  {\bibinfo {volume} {87}},\ \bibinfo {pages} {026502} (\bibinfo {year}
  {2024})}\BibitemShut {NoStop}%
\bibitem [{\citenamefont {Khaliullin}(2005)}]{Khaliullin2005}%
  \BibitemOpen
  \bibfield  {author} {\bibinfo {author} {\bibfnamefont {G.}~\bibnamefont
  {Khaliullin}},\ }\bibfield  {title} {\bibinfo {title} {Orbital order and
  fluctuations in {M}ott insulators},\ }\href
  {https://doi.org/10.1143/PTPS.160.155} {\bibfield  {journal} {\bibinfo
  {journal} {Prog.\ Theor.\ Phys.\ Suppl.}\ }\textbf {\bibinfo {volume}
  {160}},\ \bibinfo {pages} {155} (\bibinfo {year} {2005})}\BibitemShut
  {NoStop}%
\bibitem [{\citenamefont {Chen}\ and\ \citenamefont
  {Nussinov}(2008)}]{Chen2008}%
  \BibitemOpen
  \bibfield  {author} {\bibinfo {author} {\bibfnamefont {H.-D.}\ \bibnamefont
  {Chen}}\ and\ \bibinfo {author} {\bibfnamefont {Z.}~\bibnamefont
  {Nussinov}},\ }\bibfield  {title} {\bibinfo {title} {{Exact results of the
  Kitaev model on a hexagonal lattice: spin states, string and brane
  correlators, and anyonic excitations}},\ }\href
  {https://doi.org/10.1088/1751-8113/41/7/075001} {\bibfield  {journal}
  {\bibinfo  {journal} {J.\ Phys.\ A: Math.\ Theor.}\ }\textbf {\bibinfo
  {volume} {41}},\ \bibinfo {pages} {075001} (\bibinfo {year}
  {2008})}\BibitemShut {NoStop}%
\bibitem [{\citenamefont {Mandal}\ and\ \citenamefont
  {Surendran}(2009)}]{Mandal2009PRB}%
  \BibitemOpen
  \bibfield  {author} {\bibinfo {author} {\bibfnamefont {S.}~\bibnamefont
  {Mandal}}\ and\ \bibinfo {author} {\bibfnamefont {N.}~\bibnamefont
  {Surendran}},\ }\bibfield  {title} {\bibinfo {title} {{Exactly solvable
  Kitaev model in three dimensions}},\ }\href
  {https://doi.org/10.1103/PhysRevB.79.024426} {\bibfield  {journal} {\bibinfo
  {journal} {Phys.\ Rev.\ B}\ }\textbf {\bibinfo {volume} {79}},\ \bibinfo
  {pages} {024426} (\bibinfo {year} {2009})}\BibitemShut {NoStop}%
\bibitem [{\citenamefont {O'Brien}\ \emph {et~al.}(2016)\citenamefont
  {O'Brien}, \citenamefont {Hermanns},\ and\ \citenamefont
  {Trebst}}]{Obrien2016}%
  \BibitemOpen
  \bibfield  {author} {\bibinfo {author} {\bibfnamefont {K.}~\bibnamefont
  {O'Brien}}, \bibinfo {author} {\bibfnamefont {M.}~\bibnamefont {Hermanns}},\
  and\ \bibinfo {author} {\bibfnamefont {S.}~\bibnamefont {Trebst}},\
  }\bibfield  {title} {\bibinfo {title} {{Classification of gapless
  ${\mathbbm{Z}}_{2}$ spin liquids in three-dimensional Kitaev models}},\
  }\href {https://doi.org/10.1103/PhysRevB.93.085101} {\bibfield  {journal}
  {\bibinfo  {journal} {Phys.\ Rev.\ B}\ }\textbf {\bibinfo {volume} {93}},\
  \bibinfo {pages} {085101} (\bibinfo {year} {2016})}\BibitemShut {NoStop}%
\bibitem [{\citenamefont {Chaloupka}\ \emph {et~al.}(2013)\citenamefont
  {Chaloupka}, \citenamefont {Jackeli},\ and\ \citenamefont
  {Khaliullin}}]{Chaloupka2013PRL}%
  \BibitemOpen
  \bibfield  {author} {\bibinfo {author} {\bibfnamefont {J.}~\bibnamefont
  {Chaloupka}}, \bibinfo {author} {\bibfnamefont {G.}~\bibnamefont {Jackeli}},\
  and\ \bibinfo {author} {\bibfnamefont {G.}~\bibnamefont {Khaliullin}},\
  }\bibfield  {title} {\bibinfo {title} {{Zigzag magnetic order in the iridium
  oxide ${\mathrm{Na}}_{2}$IrO$_3$}},\ }\href
  {https://doi.org/10.1103/PhysRevLett.110.097204} {\bibfield  {journal}
  {\bibinfo  {journal} {Phys.\ Rev.\ Lett.}\ }\textbf {\bibinfo {volume}
  {110}},\ \bibinfo {pages} {097204} (\bibinfo {year} {2013})}\BibitemShut
  {NoStop}%
\bibitem [{\citenamefont {Rousochatzakis}\ \emph {et~al.}(2015)\citenamefont
  {Rousochatzakis}, \citenamefont {Reuther}, \citenamefont {Thomale},
  \citenamefont {Rachel},\ and\ \citenamefont
  {Perkins}}]{RousochatzakisPRX2015}%
  \BibitemOpen
  \bibfield  {author} {\bibinfo {author} {\bibfnamefont {I.}~\bibnamefont
  {Rousochatzakis}}, \bibinfo {author} {\bibfnamefont {J.}~\bibnamefont
  {Reuther}}, \bibinfo {author} {\bibfnamefont {R.}~\bibnamefont {Thomale}},
  \bibinfo {author} {\bibfnamefont {S.}~\bibnamefont {Rachel}},\ and\ \bibinfo
  {author} {\bibfnamefont {N.~B.}\ \bibnamefont {Perkins}},\ }\bibfield
  {title} {\bibinfo {title} {Phase diagram and quantum order by disorder in the
  {K}itaev ${K}_{1}\ensuremath{-}{K}_{2}$ honeycomb magnet},\ }\href
  {https://doi.org/10.1103/PhysRevX.5.041035} {\bibfield  {journal} {\bibinfo
  {journal} {Phys.\ Rev.\ X}\ }\textbf {\bibinfo {volume} {5}},\ \bibinfo
  {pages} {041035} (\bibinfo {year} {2015})}\BibitemShut {NoStop}%
\bibitem [{\citenamefont {Gotfryd}\ \emph {et~al.}(2017)\citenamefont
  {Gotfryd}, \citenamefont {Rusna\ifmmode~\check{c}\else \v{c}\fi{}ko},
  \citenamefont {Wohlfeld}, \citenamefont {Jackeli}, \citenamefont
  {Chaloupka},\ and\ \citenamefont {Ole\ifmmode~\acute{s}\else
  \'{s}\fi{}}}]{Gotfryd2017}%
  \BibitemOpen
  \bibfield  {author} {\bibinfo {author} {\bibfnamefont {D.}~\bibnamefont
  {Gotfryd}}, \bibinfo {author} {\bibfnamefont {J.}~\bibnamefont
  {Rusna\ifmmode~\check{c}\else \v{c}\fi{}ko}}, \bibinfo {author}
  {\bibfnamefont {K.}~\bibnamefont {Wohlfeld}}, \bibinfo {author}
  {\bibfnamefont {G.}~\bibnamefont {Jackeli}}, \bibinfo {author} {\bibfnamefont
  {J.}~\bibnamefont {Chaloupka}},\ and\ \bibinfo {author} {\bibfnamefont
  {A.~M.}\ \bibnamefont {Ole\ifmmode~\acute{s}\else \'{s}\fi{}}},\ }\bibfield
  {title} {\bibinfo {title} {Phase diagram and spin correlations of the
  {K}itaev-{H}eisenberg model: {I}mportance of quantum effects},\ }\href
  {https://doi.org/10.1103/PhysRevB.95.024426} {\bibfield  {journal} {\bibinfo
  {journal} {Phys.\ Rev.\ B}\ }\textbf {\bibinfo {volume} {95}},\ \bibinfo
  {pages} {024426} (\bibinfo {year} {2017})}\BibitemShut {NoStop}%
\bibitem [{\citenamefont {Catuneanu}\ \emph {et~al.}(2018)\citenamefont
  {Catuneanu}, \citenamefont {Yamaji}, \citenamefont {Wachtel}, \citenamefont
  {Kim},\ and\ \citenamefont {Kee}}]{Catuneanu2018npj}%
  \BibitemOpen
  \bibfield  {author} {\bibinfo {author} {\bibfnamefont {A.}~\bibnamefont
  {Catuneanu}}, \bibinfo {author} {\bibfnamefont {Y.}~\bibnamefont {Yamaji}},
  \bibinfo {author} {\bibfnamefont {G.}~\bibnamefont {Wachtel}}, \bibinfo
  {author} {\bibfnamefont {Y.~B.}\ \bibnamefont {Kim}},\ and\ \bibinfo {author}
  {\bibfnamefont {H.-Y.}\ \bibnamefont {Kee}},\ }\bibfield  {title} {\bibinfo
  {title} {Path to stable quantum spin liquids in spin-orbit coupled correlated
  materials},\ }\href {https://doi.org/10.1038/s41535-018-0095-2} {\bibfield
  {journal} {\bibinfo  {journal} {npj Quantum Mater.}\ }\textbf {\bibinfo
  {volume} {3}},\ \bibinfo {pages} {23} (\bibinfo {year} {2018})}\BibitemShut
  {NoStop}%
\bibitem [{\citenamefont {Hickey}\ \emph {et~al.}(2020)\citenamefont {Hickey},
  \citenamefont {Berke}, \citenamefont {Stavropoulos}, \citenamefont {Kee},\
  and\ \citenamefont {Trebst}}]{Hickey2020PRR}%
  \BibitemOpen
  \bibfield  {author} {\bibinfo {author} {\bibfnamefont {C.}~\bibnamefont
  {Hickey}}, \bibinfo {author} {\bibfnamefont {C.}~\bibnamefont {Berke}},
  \bibinfo {author} {\bibfnamefont {P.~P.}\ \bibnamefont {Stavropoulos}},
  \bibinfo {author} {\bibfnamefont {H.-Y.}\ \bibnamefont {Kee}},\ and\ \bibinfo
  {author} {\bibfnamefont {S.}~\bibnamefont {Trebst}},\ }\bibfield  {title}
  {\bibinfo {title} {Field-driven gapless spin liquid in the spin-$1$ {K}itaev
  honeycomb model},\ }\href {https://doi.org/10.1103/PhysRevResearch.2.023361}
  {\bibfield  {journal} {\bibinfo  {journal} {Phys.\ Rev.\ Res.}\ }\textbf
  {\bibinfo {volume} {2}},\ \bibinfo {pages} {023361} (\bibinfo {year}
  {2020})}\BibitemShut {NoStop}%
\bibitem [{\citenamefont {Rousochatzakis}\ \emph {et~al.}(2019)\citenamefont
  {Rousochatzakis}, \citenamefont {Kourtis}, \citenamefont {Knolle},
  \citenamefont {Moessner},\ and\ \citenamefont
  {Perkins}}]{Rousochatzakis2019PRB}%
  \BibitemOpen
  \bibfield  {author} {\bibinfo {author} {\bibfnamefont {I.}~\bibnamefont
  {Rousochatzakis}}, \bibinfo {author} {\bibfnamefont {S.}~\bibnamefont
  {Kourtis}}, \bibinfo {author} {\bibfnamefont {J.}~\bibnamefont {Knolle}},
  \bibinfo {author} {\bibfnamefont {R.}~\bibnamefont {Moessner}},\ and\
  \bibinfo {author} {\bibfnamefont {N.~B.}\ \bibnamefont {Perkins}},\
  }\bibfield  {title} {\bibinfo {title} {{Quantum spin liquid at finite
  temperature: Proximate dynamics and persistent typicality}},\ }\href
  {https://doi.org/10.1103/PhysRevB.100.045117} {\bibfield  {journal} {\bibinfo
   {journal} {Phys.\ Rev.\ B}\ }\textbf {\bibinfo {volume} {100}},\ \bibinfo
  {pages} {045117} (\bibinfo {year} {2019})}\BibitemShut {NoStop}%
\bibitem [{\citenamefont {Jiang}\ \emph {et~al.}(2011)\citenamefont {Jiang},
  \citenamefont {Gu}, \citenamefont {Qi},\ and\ \citenamefont
  {Trebst}}]{Jiang2011PRB}%
  \BibitemOpen
  \bibfield  {author} {\bibinfo {author} {\bibfnamefont {H.-C.}\ \bibnamefont
  {Jiang}}, \bibinfo {author} {\bibfnamefont {Z.-C.}\ \bibnamefont {Gu}},
  \bibinfo {author} {\bibfnamefont {X.-L.}\ \bibnamefont {Qi}},\ and\ \bibinfo
  {author} {\bibfnamefont {S.}~\bibnamefont {Trebst}},\ }\bibfield  {title}
  {\bibinfo {title} {{Possible proximity of the Mott insulating iridate
  Na${}_{2}$IrO${}_{3}$ to a topological phase: Phase diagram of the
  Heisenberg-Kitaev model in a magnetic field}},\ }\href
  {https://doi.org/10.1103/PhysRevB.83.245104} {\bibfield  {journal} {\bibinfo
  {journal} {Phys.\ Rev.\ B}\ }\textbf {\bibinfo {volume} {83}},\ \bibinfo
  {pages} {245104} (\bibinfo {year} {2011})}\BibitemShut {NoStop}%
\bibitem [{\citenamefont {Shinjo}\ \emph {et~al.}(2015)\citenamefont {Shinjo},
  \citenamefont {Sota},\ and\ \citenamefont {Tohyama}}]{Shinjo2015PRB}%
  \BibitemOpen
  \bibfield  {author} {\bibinfo {author} {\bibfnamefont {K.}~\bibnamefont
  {Shinjo}}, \bibinfo {author} {\bibfnamefont {S.}~\bibnamefont {Sota}},\ and\
  \bibinfo {author} {\bibfnamefont {T.}~\bibnamefont {Tohyama}},\ }\bibfield
  {title} {\bibinfo {title} {Density-matrix renormalization group study of the
  extended {K}itaev-{H}eisenberg model},\ }\href
  {https://doi.org/10.1103/PhysRevB.91.054401} {\bibfield  {journal} {\bibinfo
  {journal} {Phys.\ Rev.\ B}\ }\textbf {\bibinfo {volume} {91}},\ \bibinfo
  {pages} {054401} (\bibinfo {year} {2015})}\BibitemShut {NoStop}%
\bibitem [{\citenamefont {Gohlke}\ \emph {et~al.}(2017)\citenamefont {Gohlke},
  \citenamefont {Verresen}, \citenamefont {Moessner},\ and\ \citenamefont
  {Pollmann}}]{Gohlke2017PRL}%
  \BibitemOpen
  \bibfield  {author} {\bibinfo {author} {\bibfnamefont {M.}~\bibnamefont
  {Gohlke}}, \bibinfo {author} {\bibfnamefont {R.}~\bibnamefont {Verresen}},
  \bibinfo {author} {\bibfnamefont {R.}~\bibnamefont {Moessner}},\ and\
  \bibinfo {author} {\bibfnamefont {F.}~\bibnamefont {Pollmann}},\ }\bibfield
  {title} {\bibinfo {title} {{Dynamics of the Kitaev-Heisenberg model}},\
  }\href {https://doi.org/10.1103/PhysRevLett.119.157203} {\bibfield  {journal}
  {\bibinfo  {journal} {Phys.\ Rev.\ Lett.}\ }\textbf {\bibinfo {volume}
  {119}},\ \bibinfo {pages} {157203} (\bibinfo {year} {2017})}\BibitemShut
  {NoStop}%
\bibitem [{\citenamefont {Gohlke}\ \emph
  {et~al.}(2018{\natexlab{a}})\citenamefont {Gohlke}, \citenamefont {Wachtel},
  \citenamefont {Yamaji}, \citenamefont {Pollmann},\ and\ \citenamefont
  {Kim}}]{Gohlke2018PRBa}%
  \BibitemOpen
  \bibfield  {author} {\bibinfo {author} {\bibfnamefont {M.}~\bibnamefont
  {Gohlke}}, \bibinfo {author} {\bibfnamefont {G.}~\bibnamefont {Wachtel}},
  \bibinfo {author} {\bibfnamefont {Y.}~\bibnamefont {Yamaji}}, \bibinfo
  {author} {\bibfnamefont {F.}~\bibnamefont {Pollmann}},\ and\ \bibinfo
  {author} {\bibfnamefont {Y.~B.}\ \bibnamefont {Kim}},\ }\bibfield  {title}
  {\bibinfo {title} {{Quantum spin liquid signatures in Kitaev-like frustrated
  magnets}},\ }\href {https://doi.org/10.1103/PhysRevB.97.075126} {\bibfield
  {journal} {\bibinfo  {journal} {Phys.\ Rev.\ B}\ }\textbf {\bibinfo {volume}
  {97}},\ \bibinfo {pages} {075126} (\bibinfo {year}
  {2018}{\natexlab{a}})}\BibitemShut {NoStop}%
\bibitem [{\citenamefont {Gohlke}\ \emph
  {et~al.}(2018{\natexlab{b}})\citenamefont {Gohlke}, \citenamefont
  {Moessner},\ and\ \citenamefont {Pollmann}}]{Gohlke2018PRBb}%
  \BibitemOpen
  \bibfield  {author} {\bibinfo {author} {\bibfnamefont {M.}~\bibnamefont
  {Gohlke}}, \bibinfo {author} {\bibfnamefont {R.}~\bibnamefont {Moessner}},\
  and\ \bibinfo {author} {\bibfnamefont {F.}~\bibnamefont {Pollmann}},\
  }\bibfield  {title} {\bibinfo {title} {{Dynamical and topological properties
  of the Kitaev model in a [111] magnetic field}},\ }\href
  {https://doi.org/10.1103/PhysRevB.98.014418} {\bibfield  {journal} {\bibinfo
  {journal} {Phys.\ Rev.\ B}\ }\textbf {\bibinfo {volume} {98}},\ \bibinfo
  {pages} {014418} (\bibinfo {year} {2018}{\natexlab{b}})}\BibitemShut
  {NoStop}%
\bibitem [{\citenamefont {Gordon}\ \emph {et~al.}(2019)\citenamefont {Gordon},
  \citenamefont {Catuneanu}, \citenamefont {S{\o}rensen},\ and\ \citenamefont
  {Kee}}]{Gordon2019NC}%
  \BibitemOpen
  \bibfield  {author} {\bibinfo {author} {\bibfnamefont {J.~S.}\ \bibnamefont
  {Gordon}}, \bibinfo {author} {\bibfnamefont {A.}~\bibnamefont {Catuneanu}},
  \bibinfo {author} {\bibfnamefont {E.~S.}\ \bibnamefont {S{\o}rensen}},\ and\
  \bibinfo {author} {\bibfnamefont {H.-Y.}\ \bibnamefont {Kee}},\ }\bibfield
  {title} {\bibinfo {title} {{Theory of the field-revealed Kitaev spin
  liquid}},\ }\href {https://doi.org/10.1038/s41467-019-10405-8} {\bibfield
  {journal} {\bibinfo  {journal} {Nat.\ Commun.}\ }\textbf {\bibinfo {volume}
  {10}},\ \bibinfo {pages} {2470} (\bibinfo {year} {2019})}\BibitemShut
  {NoStop}%
\bibitem [{\citenamefont {Dong}\ and\ \citenamefont
  {Sheng}(2020)}]{Dong2020PRB}%
  \BibitemOpen
  \bibfield  {author} {\bibinfo {author} {\bibfnamefont {X.-Y.}\ \bibnamefont
  {Dong}}\ and\ \bibinfo {author} {\bibfnamefont {D.~N.}\ \bibnamefont
  {Sheng}},\ }\bibfield  {title} {\bibinfo {title} {Spin-1
  {K}itaev-{H}eisenberg model on a honeycomb lattice},\ }\href
  {https://doi.org/10.1103/PhysRevB.102.121102} {\bibfield  {journal} {\bibinfo
   {journal} {Phys.\ Rev.\ B}\ }\textbf {\bibinfo {volume} {102}},\ \bibinfo
  {pages} {121102} (\bibinfo {year} {2020})}\BibitemShut {NoStop}%
\bibitem [{\citenamefont {Gohlke}\ \emph {et~al.}(2020)\citenamefont {Gohlke},
  \citenamefont {Chern}, \citenamefont {Kee},\ and\ \citenamefont
  {Kim}}]{Gohlke2020PRR}%
  \BibitemOpen
  \bibfield  {author} {\bibinfo {author} {\bibfnamefont {M.}~\bibnamefont
  {Gohlke}}, \bibinfo {author} {\bibfnamefont {L.~E.}\ \bibnamefont {Chern}},
  \bibinfo {author} {\bibfnamefont {H.-Y.}\ \bibnamefont {Kee}},\ and\ \bibinfo
  {author} {\bibfnamefont {Y.~B.}\ \bibnamefont {Kim}},\ }\bibfield  {title}
  {\bibinfo {title} {{Emergence of nematic paramagnet via quantum
  order-by-disorder and pseudo-Goldstone modes in Kitaev magnets}},\ }\href
  {https://doi.org/10.1103/PhysRevResearch.2.043023} {\bibfield  {journal}
  {\bibinfo  {journal} {Phys.\ Rev.\ Res.}\ }\textbf {\bibinfo {volume} {2}},\
  \bibinfo {pages} {043023} (\bibinfo {year} {2020})}\BibitemShut {NoStop}%
\bibitem [{\citenamefont {Osorio~Iregui}\ \emph {et~al.}(2014)\citenamefont
  {Osorio~Iregui}, \citenamefont {Corboz},\ and\ \citenamefont
  {Troyer}}]{Osorio2014}%
  \BibitemOpen
  \bibfield  {author} {\bibinfo {author} {\bibfnamefont {J.}~\bibnamefont
  {Osorio~Iregui}}, \bibinfo {author} {\bibfnamefont {P.}~\bibnamefont
  {Corboz}},\ and\ \bibinfo {author} {\bibfnamefont {M.}~\bibnamefont
  {Troyer}},\ }\bibfield  {title} {\bibinfo {title} {Probing the stability of
  the spin-liquid phases in the {K}itaev-{H}eisenberg model using tensor
  network algorithms},\ }\href {https://doi.org/10.1103/PhysRevB.90.195102}
  {\bibfield  {journal} {\bibinfo  {journal} {Phys.\ Rev.\ B}\ }\textbf
  {\bibinfo {volume} {90}},\ \bibinfo {pages} {195102} (\bibinfo {year}
  {2014})}\BibitemShut {NoStop}%
\bibitem [{\citenamefont {Czarnik}\ \emph {et~al.}(2019)\citenamefont
  {Czarnik}, \citenamefont {Francuz},\ and\ \citenamefont
  {Dziarmaga}}]{Czarnik2019}%
  \BibitemOpen
  \bibfield  {author} {\bibinfo {author} {\bibfnamefont {P.}~\bibnamefont
  {Czarnik}}, \bibinfo {author} {\bibfnamefont {A.}~\bibnamefont {Francuz}},\
  and\ \bibinfo {author} {\bibfnamefont {J.}~\bibnamefont {Dziarmaga}},\
  }\bibfield  {title} {\bibinfo {title} {{Tensor network simulation of the
  Kitaev-Heisenberg model at finite temperature}},\ }\href
  {https://doi.org/10.1103/PhysRevB.100.165147} {\bibfield  {journal} {\bibinfo
   {journal} {Phys.\ Rev.\ B}\ }\textbf {\bibinfo {volume} {100}},\ \bibinfo
  {pages} {165147} (\bibinfo {year} {2019})}\BibitemShut {NoStop}%
\bibitem [{\citenamefont {Lee}\ \emph {et~al.}(2020{\natexlab{a}})\citenamefont
  {Lee}, \citenamefont {Kaneko}, \citenamefont {Chern}, \citenamefont {Okubo},
  \citenamefont {Yamaji}, \citenamefont {Kawashima},\ and\ \citenamefont
  {Kim}}]{Lee2020NC}%
  \BibitemOpen
  \bibfield  {author} {\bibinfo {author} {\bibfnamefont {H.-Y.}\ \bibnamefont
  {Lee}}, \bibinfo {author} {\bibfnamefont {R.}~\bibnamefont {Kaneko}},
  \bibinfo {author} {\bibfnamefont {L.~E.}\ \bibnamefont {Chern}}, \bibinfo
  {author} {\bibfnamefont {T.}~\bibnamefont {Okubo}}, \bibinfo {author}
  {\bibfnamefont {Y.}~\bibnamefont {Yamaji}}, \bibinfo {author} {\bibfnamefont
  {N.}~\bibnamefont {Kawashima}},\ and\ \bibinfo {author} {\bibfnamefont
  {Y.~B.}\ \bibnamefont {Kim}},\ }\bibfield  {title} {\bibinfo {title}
  {Magnetic field induced quantum phases in a tensor network study of {K}itaev
  magnets},\ }\href {https://doi.org/10.1038/s41467-020-15320-x} {\bibfield
  {journal} {\bibinfo  {journal} {Nat.\ Commun.}\ }\textbf {\bibinfo {volume}
  {11}},\ \bibinfo {pages} {1639} (\bibinfo {year}
  {2020}{\natexlab{a}})}\BibitemShut {NoStop}%
\bibitem [{\citenamefont {Reuther}\ \emph
  {et~al.}(2011{\natexlab{a}})\citenamefont {Reuther}, \citenamefont
  {Thomale},\ and\ \citenamefont {Trebst}}]{Reuther2011PRB}%
  \BibitemOpen
  \bibfield  {author} {\bibinfo {author} {\bibfnamefont {J.}~\bibnamefont
  {Reuther}}, \bibinfo {author} {\bibfnamefont {R.}~\bibnamefont {Thomale}},\
  and\ \bibinfo {author} {\bibfnamefont {S.}~\bibnamefont {Trebst}},\
  }\bibfield  {title} {\bibinfo {title} {{Finite-temperature phase diagram of
  the Heisenberg-Kitaev model}},\ }\href
  {https://doi.org/10.1103/PhysRevB.84.100406} {\bibfield  {journal} {\bibinfo
  {journal} {Phys.\ Rev.\ B}\ }\textbf {\bibinfo {volume} {84}},\ \bibinfo
  {pages} {100406} (\bibinfo {year} {2011}{\natexlab{a}})}\BibitemShut
  {NoStop}%
\bibitem [{\citenamefont {Buessen}\ and\ \citenamefont
  {Kim}(2021)}]{Buessen2021PRB}%
  \BibitemOpen
  \bibfield  {author} {\bibinfo {author} {\bibfnamefont {F.~L.}\ \bibnamefont
  {Buessen}}\ and\ \bibinfo {author} {\bibfnamefont {Y.~B.}\ \bibnamefont
  {Kim}},\ }\bibfield  {title} {\bibinfo {title} {{Functional renormalization
  group study of the Kitaev-$\mathrm{\ensuremath{\Gamma}}$ model on the
  honeycomb lattice and emergent incommensurate magnetic correlations}},\
  }\href {https://doi.org/10.1103/PhysRevB.103.184407} {\bibfield  {journal}
  {\bibinfo  {journal} {Phys.\ Rev.\ B}\ }\textbf {\bibinfo {volume} {103}},\
  \bibinfo {pages} {184407} (\bibinfo {year} {2021})}\BibitemShut {NoStop}%
\bibitem [{\citenamefont {Fukui}\ \emph {et~al.}(2022)\citenamefont {Fukui},
  \citenamefont {Kato}, \citenamefont {Nasu},\ and\ \citenamefont
  {Motome}}]{Fukui2022}%
  \BibitemOpen
  \bibfield  {author} {\bibinfo {author} {\bibfnamefont {K.}~\bibnamefont
  {Fukui}}, \bibinfo {author} {\bibfnamefont {Y.}~\bibnamefont {Kato}},
  \bibinfo {author} {\bibfnamefont {J.}~\bibnamefont {Nasu}},\ and\ \bibinfo
  {author} {\bibfnamefont {Y.}~\bibnamefont {Motome}},\ }\bibfield  {title}
  {\bibinfo {title} {{Ground-state phase diagram of spin-$S$ Kitaev-Heisenberg
  models}},\ }\href {https://doi.org/10.1103/PhysRevB.106.174416} {\bibfield
  {journal} {\bibinfo  {journal} {Phys. Rev. B}\ }\textbf {\bibinfo {volume}
  {106}},\ \bibinfo {pages} {174416} (\bibinfo {year} {2022})}\BibitemShut
  {NoStop}%
\bibitem [{\citenamefont {Burnell}\ and\ \citenamefont
  {Nayak}(2011)}]{Burnell2011PRB}%
  \BibitemOpen
  \bibfield  {author} {\bibinfo {author} {\bibfnamefont {F.~J.}\ \bibnamefont
  {Burnell}}\ and\ \bibinfo {author} {\bibfnamefont {C.}~\bibnamefont
  {Nayak}},\ }\bibfield  {title} {\bibinfo {title} {{SU}(2) slave fermion
  solution of the {K}itaev honeycomb lattice model},\ }\href
  {https://doi.org/10.1103/PhysRevB.84.125125} {\bibfield  {journal} {\bibinfo
  {journal} {Phys.\ Rev.\ B}\ }\textbf {\bibinfo {volume} {84}},\ \bibinfo
  {pages} {125125} (\bibinfo {year} {2011})}\BibitemShut {NoStop}%
\bibitem [{\citenamefont {Schaffer}\ \emph {et~al.}(2012)\citenamefont
  {Schaffer}, \citenamefont {Bhattacharjee},\ and\ \citenamefont
  {Kim}}]{Schaffer2012PRB}%
  \BibitemOpen
  \bibfield  {author} {\bibinfo {author} {\bibfnamefont {R.}~\bibnamefont
  {Schaffer}}, \bibinfo {author} {\bibfnamefont {S.}~\bibnamefont
  {Bhattacharjee}},\ and\ \bibinfo {author} {\bibfnamefont {Y.~B.}\
  \bibnamefont {Kim}},\ }\bibfield  {title} {\bibinfo {title} {{Quantum phase
  transition in Heisenberg-Kitaev model}},\ }\href
  {https://doi.org/10.1103/PhysRevB.86.224417} {\bibfield  {journal} {\bibinfo
  {journal} {Phys.\ Rev.\ B}\ }\textbf {\bibinfo {volume} {86}},\ \bibinfo
  {pages} {224417} (\bibinfo {year} {2012})}\BibitemShut {NoStop}%
\bibitem [{\citenamefont {Knolle}\ \emph {et~al.}(2018)\citenamefont {Knolle},
  \citenamefont {Bhattacharjee},\ and\ \citenamefont
  {Moessner}}]{Johannes2018PRB}%
  \BibitemOpen
  \bibfield  {author} {\bibinfo {author} {\bibfnamefont {J.}~\bibnamefont
  {Knolle}}, \bibinfo {author} {\bibfnamefont {S.}~\bibnamefont
  {Bhattacharjee}},\ and\ \bibinfo {author} {\bibfnamefont {R.}~\bibnamefont
  {Moessner}},\ }\bibfield  {title} {\bibinfo {title} {{Dynamics of a quantum
  spin liquid beyond integrability: The
  Kitaev-Heisenberg-$\mathrm{\ensuremath{\Gamma}}$ model in an augmented parton
  mean-field theory}},\ }\href {https://doi.org/10.1103/PhysRevB.97.134432}
  {\bibfield  {journal} {\bibinfo  {journal} {Phys.\ Rev.\ B}\ }\textbf
  {\bibinfo {volume} {97}},\ \bibinfo {pages} {134432} (\bibinfo {year}
  {2018})}\BibitemShut {NoStop}%
\bibitem [{\citenamefont {Ralko}\ and\ \citenamefont
  {Merino}(2024)}]{Ralko2024}%
  \BibitemOpen
  \bibfield  {author} {\bibinfo {author} {\bibfnamefont {A.}~\bibnamefont
  {Ralko}}\ and\ \bibinfo {author} {\bibfnamefont {J.}~\bibnamefont {Merino}},\
  }\href@noop {} {\bibinfo {title} {Chiral bosonic quantum spin liquid in the
  integer-spin {H}eisenberg-{K}itaev model}} (\bibinfo {year} {2024}),\ \Eprint
  {https://arxiv.org/abs/2405.10731} {arXiv:2405.10731 [cond-mat.str-el]}
  \BibitemShut {NoStop}%
\bibitem [{\citenamefont {Wang}\ \emph {et~al.}(2019)\citenamefont {Wang},
  \citenamefont {Normand},\ and\ \citenamefont {Liu}}]{Normand2019}%
  \BibitemOpen
  \bibfield  {author} {\bibinfo {author} {\bibfnamefont {J.}~\bibnamefont
  {Wang}}, \bibinfo {author} {\bibfnamefont {B.}~\bibnamefont {Normand}},\ and\
  \bibinfo {author} {\bibfnamefont {Z.-X.}\ \bibnamefont {Liu}},\ }\bibfield
  {title} {\bibinfo {title} {One proximate {K}itaev spin liquid in the
  ${K}\text{\ensuremath{-}}{J}\text{\ensuremath{-}}\mathrm{\ensuremath{\Gamma}}$
  model on the honeycomb lattice},\ }\href
  {https://doi.org/10.1103/PhysRevLett.123.197201} {\bibfield  {journal}
  {\bibinfo  {journal} {Phys.\ Rev.\ Lett.}\ }\textbf {\bibinfo {volume}
  {123}},\ \bibinfo {pages} {197201} (\bibinfo {year} {2019})}\BibitemShut
  {NoStop}%
\bibitem [{\citenamefont {Rao}\ \emph {et~al.}(2021)\citenamefont {Rao},
  \citenamefont {Liu}, \citenamefont {Machaczek},\ and\ \citenamefont
  {Pollet}}]{Rau2021PRR}%
  \BibitemOpen
  \bibfield  {author} {\bibinfo {author} {\bibfnamefont {N.}~\bibnamefont
  {Rao}}, \bibinfo {author} {\bibfnamefont {K.}~\bibnamefont {Liu}}, \bibinfo
  {author} {\bibfnamefont {M.}~\bibnamefont {Machaczek}},\ and\ \bibinfo
  {author} {\bibfnamefont {L.}~\bibnamefont {Pollet}},\ }\bibfield  {title}
  {\bibinfo {title} {Machine-learned phase diagrams of generalized {K}itaev
  honeycomb magnets},\ }\href
  {https://doi.org/10.1103/PhysRevResearch.3.033223} {\bibfield  {journal}
  {\bibinfo  {journal} {Phys.\ Rev.\ Res.}\ }\textbf {\bibinfo {volume} {3}},\
  \bibinfo {pages} {033223} (\bibinfo {year} {2021})}\BibitemShut {NoStop}%
\bibitem [{\citenamefont {Liu}\ \emph {et~al.}(2021)\citenamefont {Liu},
  \citenamefont {Sadoune}, \citenamefont {Rao}, \citenamefont {Greitemann},\
  and\ \citenamefont {Pollet}}]{Liu2021PRR}%
  \BibitemOpen
  \bibfield  {author} {\bibinfo {author} {\bibfnamefont {K.}~\bibnamefont
  {Liu}}, \bibinfo {author} {\bibfnamefont {N.}~\bibnamefont {Sadoune}},
  \bibinfo {author} {\bibfnamefont {N.}~\bibnamefont {Rao}}, \bibinfo {author}
  {\bibfnamefont {J.}~\bibnamefont {Greitemann}},\ and\ \bibinfo {author}
  {\bibfnamefont {L.}~\bibnamefont {Pollet}},\ }\bibfield  {title} {\bibinfo
  {title} {Revealing the phase diagram of {K}itaev materials by machine
  learning: {C}ooperation and competition between spin liquids},\ }\href
  {https://doi.org/10.1103/PhysRevResearch.3.023016} {\bibfield  {journal}
  {\bibinfo  {journal} {Phys.\ Rev.\ Res.}\ }\textbf {\bibinfo {volume} {3}},\
  \bibinfo {pages} {023016} (\bibinfo {year} {2021})}\BibitemShut {NoStop}%
\bibitem [{\citenamefont {Winter}\ \emph
  {et~al.}(2017{\natexlab{b}})\citenamefont {Winter}, \citenamefont {Riedl},
  \citenamefont {Maksimov}, \citenamefont {Chernyshev}, \citenamefont
  {Honecker},\ and\ \citenamefont {Valentí}}]{Winter2017}%
  \BibitemOpen
  \bibfield  {author} {\bibinfo {author} {\bibfnamefont {S.~M.}\ \bibnamefont
  {Winter}}, \bibinfo {author} {\bibfnamefont {K.}~\bibnamefont {Riedl}},
  \bibinfo {author} {\bibfnamefont {P.~A.}\ \bibnamefont {Maksimov}}, \bibinfo
  {author} {\bibfnamefont {A.~L.}\ \bibnamefont {Chernyshev}}, \bibinfo
  {author} {\bibfnamefont {A.}~\bibnamefont {Honecker}},\ and\ \bibinfo
  {author} {\bibfnamefont {R.}~\bibnamefont {Valentí}},\ }\bibfield  {title}
  {\bibinfo {title} {Breakdown of magnons in a strongly spin-orbital coupled
  magnet},\ }\href {https://doi.org/10.1038/s41467-017-01177-0} {\bibfield
  {journal} {\bibinfo  {journal} {Nat.\ Commun.}\ }\textbf {\bibinfo {volume}
  {8}},\ \bibinfo {pages} {1152} (\bibinfo {year}
  {2017}{\natexlab{b}})}\BibitemShut {NoStop}%
\bibitem [{\citenamefont {Smit}\ \emph {et~al.}(2020)\citenamefont {Smit},
  \citenamefont {Keupert}, \citenamefont {Tsyplyatyev}, \citenamefont
  {Maksimov}, \citenamefont {Chernyshev},\ and\ \citenamefont
  {Kopietz}}]{Smit2020}%
  \BibitemOpen
  \bibfield  {author} {\bibinfo {author} {\bibfnamefont {R.~L.}\ \bibnamefont
  {Smit}}, \bibinfo {author} {\bibfnamefont {S.}~\bibnamefont {Keupert}},
  \bibinfo {author} {\bibfnamefont {O.}~\bibnamefont {Tsyplyatyev}}, \bibinfo
  {author} {\bibfnamefont {P.~A.}\ \bibnamefont {Maksimov}}, \bibinfo {author}
  {\bibfnamefont {A.~L.}\ \bibnamefont {Chernyshev}},\ and\ \bibinfo {author}
  {\bibfnamefont {P.}~\bibnamefont {Kopietz}},\ }\bibfield  {title} {\bibinfo
  {title} {{Magnon damping in the zigzag phase of the
  Kitaev-Heisenberg-$\mathrm{\ensuremath{\Gamma}}$ model on a honeycomb
  lattice}},\ }\href {https://doi.org/10.1103/PhysRevB.101.054424} {\bibfield
  {journal} {\bibinfo  {journal} {Phys.\ Rev.\ B}\ }\textbf {\bibinfo {volume}
  {101}},\ \bibinfo {pages} {054424} (\bibinfo {year} {2020})}\BibitemShut
  {NoStop}%
\bibitem [{\citenamefont {Maksimov}\ and\ \citenamefont
  {Chernyshev}(2020)}]{Maksimov2020PRR}%
  \BibitemOpen
  \bibfield  {author} {\bibinfo {author} {\bibfnamefont {P.~A.}\ \bibnamefont
  {Maksimov}}\ and\ \bibinfo {author} {\bibfnamefont {A.~L.}\ \bibnamefont
  {Chernyshev}},\ }\bibfield  {title} {\bibinfo {title} {{Rethinking
  $\ensuremath{\alpha}\text{\ensuremath{-}}{\mathrm{RuCl}}_{3}$}},\ }\href
  {https://doi.org/10.1103/PhysRevResearch.2.033011} {\bibfield  {journal}
  {\bibinfo  {journal} {Phys.\ Rev.\ Res.}\ }\textbf {\bibinfo {volume} {2}},\
  \bibinfo {pages} {033011} (\bibinfo {year} {2020})}\BibitemShut {NoStop}%
\bibitem [{\citenamefont {K\"{u}mmel}\ \emph {et~al.}(1978)\citenamefont
  {K\"{u}mmel}, \citenamefont {L\"{u}hrmann},\ and\ \citenamefont
  {Zabolitzky}}]{Kummel_Luhr-Zab_1978}%
  \BibitemOpen
  \bibfield  {author} {\bibinfo {author} {\bibfnamefont {H.}~\bibnamefont
  {K\"{u}mmel}}, \bibinfo {author} {\bibfnamefont {K.~H.}\ \bibnamefont
  {L\"{u}hrmann}},\ and\ \bibinfo {author} {\bibfnamefont {J.~G.}\ \bibnamefont
  {Zabolitzky}},\ }\bibfield  {title} {\bibinfo {title} {Many-fermion theory in
  exp${S}$- (or coupled cluster) form},\ }\href
  {https://doi.org/10.1016/0370-1573(78)90081-9} {\bibfield  {journal}
  {\bibinfo  {journal} {Phys.\ Rep.}\ }\textbf {\bibinfo {volume} {36}},\
  \bibinfo {pages} {1} (\bibinfo {year} {1978})}\BibitemShut {NoStop}%
\bibitem [{\citenamefont {Bishop}\ and\ \citenamefont
  {L\"{u}hrmann}(1978)}]{Bishop-Luhrmann_1978}%
  \BibitemOpen
  \bibfield  {author} {\bibinfo {author} {\bibfnamefont {R.~F.}\ \bibnamefont
  {Bishop}}\ and\ \bibinfo {author} {\bibfnamefont {K.~H.}\ \bibnamefont
  {L\"{u}hrmann}},\ }\bibfield  {title} {\bibinfo {title} {Electron
  correlations: {I}. {G}round-state results in the high-density regime},\
  }\href {https://doi.org/10.1103/PhysRevB.17.3757} {\bibfield  {journal}
  {\bibinfo  {journal} {Phys.\ Rev.\ B}\ }\textbf {\bibinfo {volume} {17}},\
  \bibinfo {pages} {3757} (\bibinfo {year} {1978})}\BibitemShut {NoStop}%
\bibitem [{\citenamefont {Emrich}(1981{\natexlab{a}})}]{Emrich_1981a}%
  \BibitemOpen
  \bibfield  {author} {\bibinfo {author} {\bibfnamefont {K.}~\bibnamefont
  {Emrich}},\ }\bibfield  {title} {\bibinfo {title} {An extension of the
  coupled cluster formalism to excited states ({I})},\ }\href
  {https://doi.org/10.1016/0375-9474(81)90179-2} {\bibfield  {journal}
  {\bibinfo  {journal} {Nucl.\ Phys.\ A}\ }\textbf {\bibinfo {volume} {351}},\
  \bibinfo {pages} {379} (\bibinfo {year} {1981}{\natexlab{a}})}\BibitemShut
  {NoStop}%
\bibitem [{\citenamefont {Emrich}(1981{\natexlab{b}})}]{Emrich_1981b}%
  \BibitemOpen
  \bibfield  {author} {\bibinfo {author} {\bibfnamefont {K.}~\bibnamefont
  {Emrich}},\ }\bibfield  {title} {\bibinfo {title} {An extension of the
  coupled cluster formalism to excited states: ({II}). {A}pproximations and
  tests},\ }\href {https://doi.org/10.1016/0375-9474(81)90180-9} {\bibfield
  {journal} {\bibinfo  {journal} {Nucl.\ Phys.\ A}\ }\textbf {\bibinfo {volume}
  {351}},\ \bibinfo {pages} {397} (\bibinfo {year}
  {1981}{\natexlab{b}})}\BibitemShut {NoStop}%
\bibitem [{\citenamefont {Bishop}\ and\ \citenamefont
  {L\"{u}hrmann}(1982)}]{Bishop-Luhrmann_1982}%
  \BibitemOpen
  \bibfield  {author} {\bibinfo {author} {\bibfnamefont {R.~F.}\ \bibnamefont
  {Bishop}}\ and\ \bibinfo {author} {\bibfnamefont {K.~H.}\ \bibnamefont
  {L\"{u}hrmann}},\ }\bibfield  {title} {\bibinfo {title} {Electron
  correlations. {II}. {G}round-state results at low and metallic densities},\
  }\href {https://doi.org/10.1103/PhysRevB.26.5523} {\bibfield  {journal}
  {\bibinfo  {journal} {Phys.\ Rev.\ B}\ }\textbf {\bibinfo {volume} {26}},\
  \bibinfo {pages} {5523} (\bibinfo {year} {1982})}\BibitemShut {NoStop}%
\bibitem [{\citenamefont {Arponen}(1983)}]{Arponen_1983}%
  \BibitemOpen
  \bibfield  {author} {\bibinfo {author} {\bibfnamefont {J.}~\bibnamefont
  {Arponen}},\ }\bibfield  {title} {\bibinfo {title} {Variational principles
  and linked-cluster exp ${S\,}$ expansions for static and dynamic many-body
  problems},\ }\href {https://doi.org/10.1016/0003-4916(83)90284-1} {\bibfield
  {journal} {\bibinfo  {journal} {Ann.\ Phys.\ (N.Y.)}\ }\textbf {\bibinfo
  {volume} {151}},\ \bibinfo {pages} {311} (\bibinfo {year}
  {1983})}\BibitemShut {NoStop}%
\bibitem [{\citenamefont {Bishop}\ and\ \citenamefont
  {K\"{u}mmel}(1987)}]{Bishop-Kummel_1987}%
  \BibitemOpen
  \bibfield  {author} {\bibinfo {author} {\bibfnamefont {R.~F.}\ \bibnamefont
  {Bishop}}\ and\ \bibinfo {author} {\bibfnamefont {H.~G.}\ \bibnamefont
  {K\"{u}mmel}},\ }\bibfield  {title} {\bibinfo {title} {The coupled cluster
  method},\ }\href {https://doi.org/10.1063/1.881103} {\bibfield  {journal}
  {\bibinfo  {journal} {Phys.\ Today}\ }\textbf {\bibinfo {volume} {40(3)}},\
  \bibinfo {pages} {52} (\bibinfo {year} {1987})}\BibitemShut {NoStop}%
\bibitem [{\citenamefont {Arponen}\ \emph
  {et~al.}(1987{\natexlab{a}})\citenamefont {Arponen}, \citenamefont {Bishop},\
  and\ \citenamefont {Pajanne}}]{Arp-Bish-Paj_1987a}%
  \BibitemOpen
  \bibfield  {author} {\bibinfo {author} {\bibfnamefont {J.~S.}\ \bibnamefont
  {Arponen}}, \bibinfo {author} {\bibfnamefont {R.~F.}\ \bibnamefont
  {Bishop}},\ and\ \bibinfo {author} {\bibfnamefont {E.}~\bibnamefont
  {Pajanne}},\ }\bibfield  {title} {\bibinfo {title} {Extended coupled-cluster
  method. {I}. {G}eneralized coherent bosonization as a mapping of quantum
  theory into classical {H}amiltonian mechanics},\ }\href
  {https://doi.org/10.1103/PhysRevA.36.2519} {\bibfield  {journal} {\bibinfo
  {journal} {Phys.\ Rev.\ A}\ }\textbf {\bibinfo {volume} {36}},\ \bibinfo
  {pages} {2519} (\bibinfo {year} {1987}{\natexlab{a}})}\BibitemShut {NoStop}%
\bibitem [{\citenamefont {Arponen}\ \emph
  {et~al.}(1987{\natexlab{b}})\citenamefont {Arponen}, \citenamefont {Bishop},\
  and\ \citenamefont {Pajanne}}]{Arp-Bish-Paj_1987b}%
  \BibitemOpen
  \bibfield  {author} {\bibinfo {author} {\bibfnamefont {J.~S.}\ \bibnamefont
  {Arponen}}, \bibinfo {author} {\bibfnamefont {R.~F.}\ \bibnamefont
  {Bishop}},\ and\ \bibinfo {author} {\bibfnamefont {E.}~\bibnamefont
  {Pajanne}},\ }\bibfield  {title} {\bibinfo {title} {Extended coupled-cluster
  method. {II}. {E}xcited states and generalized random-phase approximation},\
  }\href {https://doi.org/10.1103/PhysRevA.36.2539} {\bibfield  {journal}
  {\bibinfo  {journal} {Phys.\ Rev.\ A}\ }\textbf {\bibinfo {volume} {36}},\
  \bibinfo {pages} {2539} (\bibinfo {year} {1987}{\natexlab{b}})}\BibitemShut
  {NoStop}%
\bibitem [{\citenamefont {Bishop}(1991)}]{Bishop_1991}%
  \BibitemOpen
  \bibfield  {author} {\bibinfo {author} {\bibfnamefont {R.~F.}\ \bibnamefont
  {Bishop}},\ }\bibfield  {title} {\bibinfo {title} {An overview of coupled
  cluster theory and its applications in physics},\ }\href
  {https://doi.org/10.1007/BF01119617} {\bibfield  {journal} {\bibinfo
  {journal} {Theor.\ Chim.\ Acta}\ }\textbf {\bibinfo {volume} {80}},\ \bibinfo
  {pages} {95} (\bibinfo {year} {1991})}\BibitemShut {NoStop}%
\bibitem [{\citenamefont {Arponen}\ and\ \citenamefont
  {Bishop}(1991)}]{Arponen-Bishop_1991}%
  \BibitemOpen
  \bibfield  {author} {\bibinfo {author} {\bibfnamefont {J.~S.}\ \bibnamefont
  {Arponen}}\ and\ \bibinfo {author} {\bibfnamefont {R.~F.}\ \bibnamefont
  {Bishop}},\ }\bibfield  {title} {\bibinfo {title} {Independent-cluster
  parametrizations of wave functions in model field theories. {I}.
  {I}ntroduction to their holomorphic representations},\ }\href
  {https://doi.org/10.1016/0003-4916(91)90183-9} {\bibfield  {journal}
  {\bibinfo  {journal} {Ann.\ Phys.\ (N.Y.)}\ }\textbf {\bibinfo {volume}
  {207}},\ \bibinfo {pages} {171} (\bibinfo {year} {1991})}\BibitemShut
  {NoStop}%
\bibitem [{\citenamefont {Arponen}\ and\ \citenamefont
  {Bishop}(1993)}]{Arponen-Bishop_1993a}%
  \BibitemOpen
  \bibfield  {author} {\bibinfo {author} {\bibfnamefont {J.~S.}\ \bibnamefont
  {Arponen}}\ and\ \bibinfo {author} {\bibfnamefont {R.~F.}\ \bibnamefont
  {Bishop}},\ }\bibfield  {title} {\bibinfo {title} {Independent-cluster
  parametrizations of wave functions in model field theories. {II}. {C}lassical
  mappings and their algebraic structure},\ }\href
  {https://doi.org/10.1006/aphy.1993.1082} {\bibfield  {journal} {\bibinfo
  {journal} {Ann.\ Phys.\ (N.Y.)}\ }\textbf {\bibinfo {volume} {227}},\
  \bibinfo {pages} {275} (\bibinfo {year} {1993})}\BibitemShut {NoStop}%
\bibitem [{\citenamefont {Bishop}(1998)}]{Bishop_1998}%
  \BibitemOpen
  \bibfield  {author} {\bibinfo {author} {\bibfnamefont {R.~F.}\ \bibnamefont
  {Bishop}},\ }\bibfield  {title} {\bibinfo {title} {The coupled cluster
  method},\ }in\ \href {https://doi.org/10.1007/BFb0104523} {\emph {\bibinfo
  {booktitle} {Microscopic Quantum Many-Body Theories and Their
  Applications}}},\ \bibinfo {series} {Lecture Notes in Physics}, Vol.~\bibinfo
  {volume} {{\bf 510}},\ \bibinfo {editor} {edited by\ \bibinfo {editor}
  {\bibfnamefont {J.}~\bibnamefont {Navarro}}\ and\ \bibinfo {editor}
  {\bibfnamefont {A.}~\bibnamefont {Polls}}}\ (\bibinfo  {publisher}
  {Springer-Verlag},\ \bibinfo {address} {Berlin},\ \bibinfo {year} {1998})\
  pp.\ \bibinfo {pages} {1--70}\BibitemShut {NoStop}%
\bibitem [{\citenamefont {Baskaran}\ \emph {et~al.}(2008)\citenamefont
  {Baskaran}, \citenamefont {Sen},\ and\ \citenamefont
  {Shankar}}]{Baskaran2008PRB}%
  \BibitemOpen
  \bibfield  {author} {\bibinfo {author} {\bibfnamefont {G.}~\bibnamefont
  {Baskaran}}, \bibinfo {author} {\bibfnamefont {D.}~\bibnamefont {Sen}},\ and\
  \bibinfo {author} {\bibfnamefont {R.}~\bibnamefont {Shankar}},\ }\bibfield
  {title} {\bibinfo {title} {Spin-${S}$ {K}itaev model: {C}lassical ground
  states, order from disorder, and exact correlation functions},\ }\href
  {https://doi.org/10.1103/PhysRevB.78.115116} {\bibfield  {journal} {\bibinfo
  {journal} {Phys.\ Rev.\ B}\ }\textbf {\bibinfo {volume} {78}},\ \bibinfo
  {pages} {115116} (\bibinfo {year} {2008})}\BibitemShut {NoStop}%
\bibitem [{\citenamefont {Chandra}\ \emph {et~al.}(2010)\citenamefont
  {Chandra}, \citenamefont {Ramola},\ and\ \citenamefont {Dhar}}]{Chandra2010}%
  \BibitemOpen
  \bibfield  {author} {\bibinfo {author} {\bibfnamefont {S.}~\bibnamefont
  {Chandra}}, \bibinfo {author} {\bibfnamefont {K.}~\bibnamefont {Ramola}},\
  and\ \bibinfo {author} {\bibfnamefont {D.}~\bibnamefont {Dhar}},\ }\bibfield
  {title} {\bibinfo {title} {Classical {H}eisenberg spins on a hexagonal
  lattice with {K}itaev couplings},\ }\href
  {https://doi.org/10.1103/PhysRevE.82.031113} {\bibfield  {journal} {\bibinfo
  {journal} {Phys.\ Rev.\ E}\ }\textbf {\bibinfo {volume} {82}},\ \bibinfo
  {pages} {031113} (\bibinfo {year} {2010})}\BibitemShut {NoStop}%
\bibitem [{\citenamefont {Rousochatzakis}\ \emph {et~al.}(2018)\citenamefont
  {Rousochatzakis}, \citenamefont {Sizyuk},\ and\ \citenamefont
  {Perkins}}]{Rousochatzakis2018NC}%
  \BibitemOpen
  \bibfield  {author} {\bibinfo {author} {\bibfnamefont {I.}~\bibnamefont
  {Rousochatzakis}}, \bibinfo {author} {\bibfnamefont {Y.}~\bibnamefont
  {Sizyuk}},\ and\ \bibinfo {author} {\bibfnamefont {N.~B.}\ \bibnamefont
  {Perkins}},\ }\bibfield  {title} {\bibinfo {title} {Quantum spin liquid in
  the semiclassical regime},\ }\href
  {https://doi.org/10.1038/s41467-018-03934-1} {\bibfield  {journal} {\bibinfo
  {journal} {Nat.\ Commun.}\ }\textbf {\bibinfo {volume} {9}},\ \bibinfo
  {pages} {1575} (\bibinfo {year} {2018})}\BibitemShut {NoStop}%
\bibitem [{\citenamefont {Koga}\ \emph {et~al.}(2018)\citenamefont {Koga},
  \citenamefont {Tomishige},\ and\ \citenamefont {Nasu}}]{Koga2018JPSJ}%
  \BibitemOpen
  \bibfield  {author} {\bibinfo {author} {\bibfnamefont {A.}~\bibnamefont
  {Koga}}, \bibinfo {author} {\bibfnamefont {H.}~\bibnamefont {Tomishige}},\
  and\ \bibinfo {author} {\bibfnamefont {J.}~\bibnamefont {Nasu}},\ }\bibfield
  {title} {\bibinfo {title} {Ground-state and thermodynamic properties of an
  {$S = 1$} {K}itaev model},\ }\href {https://doi.org/10.7566/JPSJ.87.063703}
  {\bibfield  {journal} {\bibinfo  {journal} {J.\ Phys.\ Soc.\ Jpn.}\ }\textbf
  {\bibinfo {volume} {87}},\ \bibinfo {pages} {063703} (\bibinfo {year}
  {2018})}\BibitemShut {NoStop}%
\bibitem [{\citenamefont {Oitmaa}\ \emph {et~al.}(2018)\citenamefont {Oitmaa},
  \citenamefont {Koga},\ and\ \citenamefont {Singh}}]{Oitmaa2018PRB}%
  \BibitemOpen
  \bibfield  {author} {\bibinfo {author} {\bibfnamefont {J.}~\bibnamefont
  {Oitmaa}}, \bibinfo {author} {\bibfnamefont {A.}~\bibnamefont {Koga}},\ and\
  \bibinfo {author} {\bibfnamefont {R.~R.~P.}\ \bibnamefont {Singh}},\
  }\bibfield  {title} {\bibinfo {title} {Incipient and well-developed entropy
  plateaus in spin-${S}\,$ {K}itaev models},\ }\href
  {https://doi.org/10.1103/PhysRevB.98.214404} {\bibfield  {journal} {\bibinfo
  {journal} {Phys.\ Rev.\ B}\ }\textbf {\bibinfo {volume} {98}},\ \bibinfo
  {pages} {214404} (\bibinfo {year} {2018})}\BibitemShut {NoStop}%
\bibitem [{\citenamefont {Minakawa}\ \emph {et~al.}(2019)\citenamefont
  {Minakawa}, \citenamefont {Nasu},\ and\ \citenamefont {Koga}}]{Minakawa2019}%
  \BibitemOpen
  \bibfield  {author} {\bibinfo {author} {\bibfnamefont {T.}~\bibnamefont
  {Minakawa}}, \bibinfo {author} {\bibfnamefont {J.}~\bibnamefont {Nasu}},\
  and\ \bibinfo {author} {\bibfnamefont {A.}~\bibnamefont {Koga}},\ }\bibfield
  {title} {\bibinfo {title} {{Quantum and classical behavior of spin-$S$ Kitaev
  models in the anisotropic limit}},\ }\href
  {https://doi.org/10.1103/PhysRevB.99.104408} {\bibfield  {journal} {\bibinfo
  {journal} {Phys.\ Rev.\ B}\ }\textbf {\bibinfo {volume} {99}},\ \bibinfo
  {pages} {104408} (\bibinfo {year} {2019})}\BibitemShut {NoStop}%
\bibitem [{\citenamefont {Zhu}\ \emph {et~al.}(2020)\citenamefont {Zhu},
  \citenamefont {Weng},\ and\ \citenamefont {Sheng}}]{Zhu2020PRR}%
  \BibitemOpen
  \bibfield  {author} {\bibinfo {author} {\bibfnamefont {Z.}~\bibnamefont
  {Zhu}}, \bibinfo {author} {\bibfnamefont {Z.-Y.}\ \bibnamefont {Weng}},\ and\
  \bibinfo {author} {\bibfnamefont {D.~N.}\ \bibnamefont {Sheng}},\ }\bibfield
  {title} {\bibinfo {title} {Magnetic field induced spin liquids in ${S}=1$
  {K}itaev honeycomb model},\ }\href
  {https://doi.org/10.1103/PhysRevResearch.2.022047} {\bibfield  {journal}
  {\bibinfo  {journal} {Phys.\ Rev.\ Res.}\ }\textbf {\bibinfo {volume} {2}},\
  \bibinfo {pages} {022047} (\bibinfo {year} {2020})}\BibitemShut {NoStop}%
\bibitem [{\citenamefont {Lee}\ \emph {et~al.}(2021)\citenamefont {Lee},
  \citenamefont {Suzuki}, \citenamefont {Kim},\ and\ \citenamefont
  {Kawashima}}]{LeePRB2021}%
  \BibitemOpen
  \bibfield  {author} {\bibinfo {author} {\bibfnamefont {H.-Y.}\ \bibnamefont
  {Lee}}, \bibinfo {author} {\bibfnamefont {T.}~\bibnamefont {Suzuki}},
  \bibinfo {author} {\bibfnamefont {Y.~B.}\ \bibnamefont {Kim}},\ and\ \bibinfo
  {author} {\bibfnamefont {N.}~\bibnamefont {Kawashima}},\ }\bibfield  {title}
  {\bibinfo {title} {Anisotropy as a diagnostic test for distinct
  tensor-network wave functions of integer- and half-integer-spin {K}itaev
  quantum spin liquids},\ }\href {https://doi.org/10.1103/PhysRevB.104.024417}
  {\bibfield  {journal} {\bibinfo  {journal} {Phys.\ Rev.\ B}\ }\textbf
  {\bibinfo {volume} {104}},\ \bibinfo {pages} {024417} (\bibinfo {year}
  {2021})}\BibitemShut {NoStop}%
\bibitem [{\citenamefont {Khait}\ \emph {et~al.}(2021)\citenamefont {Khait},
  \citenamefont {Stavropoulos}, \citenamefont {Kee},\ and\ \citenamefont
  {Kim}}]{Khait2021PRR}%
  \BibitemOpen
  \bibfield  {author} {\bibinfo {author} {\bibfnamefont {I.}~\bibnamefont
  {Khait}}, \bibinfo {author} {\bibfnamefont {P.~P.}\ \bibnamefont
  {Stavropoulos}}, \bibinfo {author} {\bibfnamefont {H.-Y.}\ \bibnamefont
  {Kee}},\ and\ \bibinfo {author} {\bibfnamefont {Y.~B.}\ \bibnamefont {Kim}},\
  }\bibfield  {title} {\bibinfo {title} {{Characterizing spin-one Kitaev
  quantum spin liquids}},\ }\href
  {https://doi.org/10.1103/PhysRevResearch.3.013160} {\bibfield  {journal}
  {\bibinfo  {journal} {Phys.\ Rev.\ Res.}\ }\textbf {\bibinfo {volume} {3}},\
  \bibinfo {pages} {013160} (\bibinfo {year} {2021})}\BibitemShut {NoStop}%
\bibitem [{\citenamefont {Jin}\ \emph {et~al.}(2022)\citenamefont {Jin},
  \citenamefont {Natori}, \citenamefont {Pollmann},\ and\ \citenamefont
  {Knolle}}]{Jin2022NC}%
  \BibitemOpen
  \bibfield  {author} {\bibinfo {author} {\bibfnamefont {H.-K.}\ \bibnamefont
  {Jin}}, \bibinfo {author} {\bibfnamefont {W.~M.~H.}\ \bibnamefont {Natori}},
  \bibinfo {author} {\bibfnamefont {F.}~\bibnamefont {Pollmann}},\ and\
  \bibinfo {author} {\bibfnamefont {J.}~\bibnamefont {Knolle}},\ }\bibfield
  {title} {\bibinfo {title} {{Unveiling the S=3/2 Kitaev honeycomb spin
  liquids}},\ }\href {https://doi.org/10.1038/s41467-022-31503-0} {\bibfield
  {journal} {\bibinfo  {journal} {Nat.\ Commun.}\ }\textbf {\bibinfo {volume}
  {13}},\ \bibinfo {pages} {3813} (\bibinfo {year} {2022})}\BibitemShut
  {NoStop}%
\bibitem [{\citenamefont {Chen}\ \emph {et~al.}(2022)\citenamefont {Chen},
  \citenamefont {Genzor}, \citenamefont {Kim},\ and\ \citenamefont
  {Kao}}]{ChenPRB2022}%
  \BibitemOpen
  \bibfield  {author} {\bibinfo {author} {\bibfnamefont {Y.-H.}\ \bibnamefont
  {Chen}}, \bibinfo {author} {\bibfnamefont {J.}~\bibnamefont {Genzor}},
  \bibinfo {author} {\bibfnamefont {Y.~B.}\ \bibnamefont {Kim}},\ and\ \bibinfo
  {author} {\bibfnamefont {Y.-J.}\ \bibnamefont {Kao}},\ }\bibfield  {title}
  {\bibinfo {title} {Excitation spectrum of spin-1 {K}itaev spin liquids},\
  }\href {https://doi.org/10.1103/PhysRevB.105.L060403} {\bibfield  {journal}
  {\bibinfo  {journal} {Phys.\ Rev.\ B}\ }\textbf {\bibinfo {volume} {105}},\
  \bibinfo {pages} {L060403} (\bibinfo {year} {2022})}\BibitemShut {NoStop}%
\bibitem [{\citenamefont {Bradley}\ and\ \citenamefont
  {Singh}(2022)}]{Bradley2022}%
  \BibitemOpen
  \bibfield  {author} {\bibinfo {author} {\bibfnamefont {O.}~\bibnamefont
  {Bradley}}\ and\ \bibinfo {author} {\bibfnamefont {R.~R.~P.}\ \bibnamefont
  {Singh}},\ }\bibfield  {title} {\bibinfo {title} {{Instabilities of spin-1
  Kitaev spin liquid phase in presence of single-ion anisotropies}},\ }\href
  {https://doi.org/10.1103/PhysRevB.105.L060405} {\bibfield  {journal}
  {\bibinfo  {journal} {Phys.\ Rev.\ B}\ }\textbf {\bibinfo {volume} {105}},\
  \bibinfo {pages} {L060405} (\bibinfo {year} {2022})}\BibitemShut {NoStop}%
\bibitem [{\citenamefont {Gordon}\ and\ \citenamefont
  {Kee}(2022)}]{GordonPRR2022}%
  \BibitemOpen
  \bibfield  {author} {\bibinfo {author} {\bibfnamefont {J.~S.}\ \bibnamefont
  {Gordon}}\ and\ \bibinfo {author} {\bibfnamefont {H.-Y.}\ \bibnamefont
  {Kee}},\ }\bibfield  {title} {\bibinfo {title} {Insights into the anisotropic
  spin-${S}$ {K}itaev chain},\ }\href
  {https://doi.org/10.1103/PhysRevResearch.4.013205} {\bibfield  {journal}
  {\bibinfo  {journal} {Phys.\ Rev.\ Res.}\ }\textbf {\bibinfo {volume} {4}},\
  \bibinfo {pages} {013205} (\bibinfo {year} {2022})}\BibitemShut {NoStop}%
\bibitem [{\citenamefont {Ma}(2023)}]{MaPRL2023}%
  \BibitemOpen
  \bibfield  {author} {\bibinfo {author} {\bibfnamefont {H.}~\bibnamefont
  {Ma}},\ }\bibfield  {title} {\bibinfo {title} {{$\mathbbm{Z}_2$} spin liquids
  in the higher spin-${S}$ {K}itaev honeycomb model: {A}n exact deconfined
  {$\mathbbm{Z}_2$} gauge structure in a nonintegrable model},\ }\href
  {https://doi.org/10.1103/PhysRevLett.130.156701} {\bibfield  {journal}
  {\bibinfo  {journal} {Phys.\ Rev.\ Lett.}\ }\textbf {\bibinfo {volume}
  {130}},\ \bibinfo {pages} {156701} (\bibinfo {year} {2023})}\BibitemShut
  {NoStop}%
\bibitem [{\citenamefont {Cen}\ and\ \citenamefont {Kee}(2023)}]{Cen2023}%
  \BibitemOpen
  \bibfield  {author} {\bibinfo {author} {\bibfnamefont {J.}~\bibnamefont
  {Cen}}\ and\ \bibinfo {author} {\bibfnamefont {H.-Y.}\ \bibnamefont {Kee}},\
  }\bibfield  {title} {\bibinfo {title} {Determining {K}itaev interaction in
  spin-${S}$ honeycomb {M}ott insulators},\ }\href
  {https://doi.org/10.1103/PhysRevB.107.014411} {\bibfield  {journal} {\bibinfo
   {journal} {Phys.\ Rev.\ B}\ }\textbf {\bibinfo {volume} {107}},\ \bibinfo
  {pages} {014411} (\bibinfo {year} {2023})}\BibitemShut {NoStop}%
\bibitem [{\citenamefont {Liu}\ \emph {et~al.}(2024)\citenamefont {Liu},
  \citenamefont {Lam}, \citenamefont {Ma},\ and\ \citenamefont
  {Zou}}]{liu2023symmetries}%
  \BibitemOpen
  \bibfield  {author} {\bibinfo {author} {\bibfnamefont {R.}~\bibnamefont
  {Liu}}, \bibinfo {author} {\bibfnamefont {H.~T.}\ \bibnamefont {Lam}},
  \bibinfo {author} {\bibfnamefont {H.}~\bibnamefont {Ma}},\ and\ \bibinfo
  {author} {\bibfnamefont {L.}~\bibnamefont {Zou}},\ }\bibfield  {title}
  {\bibinfo {title} {{Symmetries and anomalies of Kitaev spin-S models:
  Identifying symmetry-enforced exotic quantum matter}},\ }\href
  {https://doi.org/10.21468/SciPostPhys.16.4.100} {\bibfield  {journal}
  {\bibinfo  {journal} {SciPost Phys.}\ }\textbf {\bibinfo {volume} {16}},\
  \bibinfo {pages} {100} (\bibinfo {year} {2024})}\BibitemShut {NoStop}%
\bibitem [{\citenamefont {Yamada}\ \emph {et~al.}(2018)\citenamefont {Yamada},
  \citenamefont {Oshikawa},\ and\ \citenamefont {Jackeli}}]{Yamada2018}%
  \BibitemOpen
  \bibfield  {author} {\bibinfo {author} {\bibfnamefont {M.~G.}\ \bibnamefont
  {Yamada}}, \bibinfo {author} {\bibfnamefont {M.}~\bibnamefont {Oshikawa}},\
  and\ \bibinfo {author} {\bibfnamefont {G.}~\bibnamefont {Jackeli}},\
  }\bibfield  {title} {\bibinfo {title} {Emergent {$\mathrm{SU}(4)$} symmetry
  in {$\ensuremath{\alpha}\text{\ensuremath{-}}{\mathrm{ZrCl}}_{3}$} and
  crystalline spin-orbital iquids},\ }\href
  {https://doi.org/10.1103/PhysRevLett.121.097201} {\bibfield  {journal}
  {\bibinfo  {journal} {Phys.\ Rev.\ Lett.}\ }\textbf {\bibinfo {volume}
  {121}},\ \bibinfo {pages} {097201} (\bibinfo {year} {2018})}\BibitemShut
  {NoStop}%
\bibitem [{\citenamefont {Lee}\ \emph {et~al.}(2020{\natexlab{b}})\citenamefont
  {Lee}, \citenamefont {Utermohlen}, \citenamefont {Weber}, \citenamefont
  {Hwang}, \citenamefont {Zhang}, \citenamefont {van Tol}, \citenamefont
  {Goldberger}, \citenamefont {Trivedi},\ and\ \citenamefont
  {Hammel}}]{Lee20220PRL}%
  \BibitemOpen
  \bibfield  {author} {\bibinfo {author} {\bibfnamefont {I.}~\bibnamefont
  {Lee}}, \bibinfo {author} {\bibfnamefont {F.~G.}\ \bibnamefont {Utermohlen}},
  \bibinfo {author} {\bibfnamefont {D.}~\bibnamefont {Weber}}, \bibinfo
  {author} {\bibfnamefont {K.}~\bibnamefont {Hwang}}, \bibinfo {author}
  {\bibfnamefont {C.}~\bibnamefont {Zhang}}, \bibinfo {author} {\bibfnamefont
  {J.}~\bibnamefont {van Tol}}, \bibinfo {author} {\bibfnamefont {J.~E.}\
  \bibnamefont {Goldberger}}, \bibinfo {author} {\bibfnamefont
  {N.}~\bibnamefont {Trivedi}},\ and\ \bibinfo {author} {\bibfnamefont {P.~C.}\
  \bibnamefont {Hammel}},\ }\bibfield  {title} {\bibinfo {title} {Fundamental
  spin interactions underlying the magnetic anisotropy in the {K}itaev
  ferromagnet {${\mathrm{CrI}}_{3}$}},\ }\href
  {https://doi.org/10.1103/PhysRevLett.124.017201} {\bibfield  {journal}
  {\bibinfo  {journal} {Phys.\ Rev.\ Lett.}\ }\textbf {\bibinfo {volume}
  {124}},\ \bibinfo {pages} {017201} (\bibinfo {year}
  {2020}{\natexlab{b}})}\BibitemShut {NoStop}%
\bibitem [{\citenamefont {Xu}\ \emph {et~al.}(2020)\citenamefont {Xu},
  \citenamefont {Feng}, \citenamefont {Kawamura}, \citenamefont {Yamaji},
  \citenamefont {Nahas}, \citenamefont {Prokhorenko}, \citenamefont {Qi},
  \citenamefont {Xiang},\ and\ \citenamefont {Bellaiche}}]{Xu2020PRL}%
  \BibitemOpen
  \bibfield  {author} {\bibinfo {author} {\bibfnamefont {C.}~\bibnamefont
  {Xu}}, \bibinfo {author} {\bibfnamefont {J.}~\bibnamefont {Feng}}, \bibinfo
  {author} {\bibfnamefont {M.}~\bibnamefont {Kawamura}}, \bibinfo {author}
  {\bibfnamefont {Y.}~\bibnamefont {Yamaji}}, \bibinfo {author} {\bibfnamefont
  {Y.}~\bibnamefont {Nahas}}, \bibinfo {author} {\bibfnamefont
  {S.}~\bibnamefont {Prokhorenko}}, \bibinfo {author} {\bibfnamefont
  {Y.}~\bibnamefont {Qi}}, \bibinfo {author} {\bibfnamefont {H.}~\bibnamefont
  {Xiang}},\ and\ \bibinfo {author} {\bibfnamefont {L.}~\bibnamefont
  {Bellaiche}},\ }\bibfield  {title} {\bibinfo {title} {Possible {K}itaev
  quantum spin liquid state in 2{D} materials with $s=3/2$},\ }\href
  {https://doi.org/10.1103/PhysRevLett.124.087205} {\bibfield  {journal}
  {\bibinfo  {journal} {Phys.\ Rev.\ Lett.}\ }\textbf {\bibinfo {volume}
  {124}},\ \bibinfo {pages} {087205} (\bibinfo {year} {2020})}\BibitemShut
  {NoStop}%
\bibitem [{\citenamefont {Stavropoulos}\ \emph {et~al.}(2019)\citenamefont
  {Stavropoulos}, \citenamefont {Pereira},\ and\ \citenamefont
  {Kee}}]{Stavropoulos2019PRL}%
  \BibitemOpen
  \bibfield  {author} {\bibinfo {author} {\bibfnamefont {P.~P.}\ \bibnamefont
  {Stavropoulos}}, \bibinfo {author} {\bibfnamefont {D.}~\bibnamefont
  {Pereira}},\ and\ \bibinfo {author} {\bibfnamefont {H.-Y.}\ \bibnamefont
  {Kee}},\ }\bibfield  {title} {\bibinfo {title} {Microscopic mechanism for a
  higher-spin {K}itaev model},\ }\href
  {https://doi.org/10.1103/PhysRevLett.123.037203} {\bibfield  {journal}
  {\bibinfo  {journal} {Phys.\ Rev.\ Lett.}\ }\textbf {\bibinfo {volume}
  {123}},\ \bibinfo {pages} {037203} (\bibinfo {year} {2019})}\BibitemShut
  {NoStop}%
\bibitem [{\citenamefont {Stavropoulos}\ \emph {et~al.}(2021)\citenamefont
  {Stavropoulos}, \citenamefont {Liu},\ and\ \citenamefont
  {Kee}}]{Stavropoulos2021PRR}%
  \BibitemOpen
  \bibfield  {author} {\bibinfo {author} {\bibfnamefont {P.~P.}\ \bibnamefont
  {Stavropoulos}}, \bibinfo {author} {\bibfnamefont {X.}~\bibnamefont {Liu}},\
  and\ \bibinfo {author} {\bibfnamefont {H.-Y.}\ \bibnamefont {Kee}},\
  }\bibfield  {title} {\bibinfo {title} {Magnetic anisotropy in spin-3/2 with
  heavy ligand in honeycomb {M}ott insulators: {A}pplication to
  {${\mathrm{CrI}}_{3}$}},\ }\href
  {https://doi.org/10.1103/PhysRevResearch.3.013216} {\bibfield  {journal}
  {\bibinfo  {journal} {Phys.\ Rev.\ Res}\ }\textbf {\bibinfo {volume} {3}},\
  \bibinfo {pages} {013216} (\bibinfo {year} {2021})}\BibitemShut {NoStop}%
\bibitem [{\citenamefont {Samarakoon}\ \emph {et~al.}(2021)\citenamefont
  {Samarakoon}, \citenamefont {Chen}, \citenamefont {Zhou},\ and\ \citenamefont
  {Garlea}}]{Samarakoon2021}%
  \BibitemOpen
  \bibfield  {author} {\bibinfo {author} {\bibfnamefont {A.~M.}\ \bibnamefont
  {Samarakoon}}, \bibinfo {author} {\bibfnamefont {Q.}~\bibnamefont {Chen}},
  \bibinfo {author} {\bibfnamefont {H.}~\bibnamefont {Zhou}},\ and\ \bibinfo
  {author} {\bibfnamefont {V.~O.}\ \bibnamefont {Garlea}},\ }\bibfield  {title}
  {\bibinfo {title} {{Static and dynamic magnetic properties of honeycomb
  lattice antiferromagnets ${\mathrm{Na}}_{2}{M}_{2}{\mathrm{TeO}}_{6}$,
  $M=\mathrm{Co}$ and Ni}},\ }\href
  {https://doi.org/10.1103/PhysRevB.104.184415} {\bibfield  {journal} {\bibinfo
   {journal} {Phys.\ Rev.\ B}\ }\textbf {\bibinfo {volume} {104}},\ \bibinfo
  {pages} {184415} (\bibinfo {year} {2021})}\BibitemShut {NoStop}%
\bibitem [{\citenamefont {Chaloupka}\ and\ \citenamefont
  {Khaliullin}(2015)}]{Chaloupka2015PRB}%
  \BibitemOpen
  \bibfield  {author} {\bibinfo {author} {\bibfnamefont {J.}~\bibnamefont
  {Chaloupka}}\ and\ \bibinfo {author} {\bibfnamefont {G.}~\bibnamefont
  {Khaliullin}},\ }\bibfield  {title} {\bibinfo {title} {{Hidden symmetries of
  the extended Kitaev-Heisenberg model: Implications for the honeycomb-lattice
  iridates ${A}_{2}{\mathrm{IrO}}_{3}$}},\ }\href
  {https://doi.org/10.1103/PhysRevB.92.024413} {\bibfield  {journal} {\bibinfo
  {journal} {Phys.\ Rev.\ B}\ }\textbf {\bibinfo {volume} {92}},\ \bibinfo
  {pages} {024413} (\bibinfo {year} {2015})}\BibitemShut {NoStop}%
\bibitem [{\citenamefont {Elitzur}(1975)}]{Elitzur1975}%
  \BibitemOpen
  \bibfield  {author} {\bibinfo {author} {\bibfnamefont {S.}~\bibnamefont
  {Elitzur}},\ }\bibfield  {title} {\bibinfo {title} {Impossibility of
  spontaneously breaking local symmetries},\ }\href
  {https://doi.org/10.1103/PhysRevD.12.3978} {\bibfield  {journal} {\bibinfo
  {journal} {Phys.\ Rev.\ D}\ }\textbf {\bibinfo {volume} {12}},\ \bibinfo
  {pages} {3978} (\bibinfo {year} {1975})}\BibitemShut {NoStop}%
\bibitem [{\citenamefont {Lee}\ \emph {et~al.}(2020{\natexlab{c}})\citenamefont
  {Lee}, \citenamefont {Kawashima},\ and\ \citenamefont {Kim}}]{Lee2020PRR}%
  \BibitemOpen
  \bibfield  {author} {\bibinfo {author} {\bibfnamefont {H.-Y.}\ \bibnamefont
  {Lee}}, \bibinfo {author} {\bibfnamefont {N.}~\bibnamefont {Kawashima}},\
  and\ \bibinfo {author} {\bibfnamefont {Y.~B.}\ \bibnamefont {Kim}},\
  }\bibfield  {title} {\bibinfo {title} {{Tensor network wave function of $S=1$
  Kitaev spin liquids}},\ }\href
  {https://doi.org/10.1103/PhysRevResearch.2.033318} {\bibfield  {journal}
  {\bibinfo  {journal} {Phys.\ Rev.\ Res.}\ }\textbf {\bibinfo {volume} {2}},\
  \bibinfo {pages} {033318} (\bibinfo {year} {2020}{\natexlab{c}})}\BibitemShut
  {NoStop}%
\bibitem [{\citenamefont {Natori}\ \emph {et~al.}(2023)\citenamefont {Natori},
  \citenamefont {Jin},\ and\ \citenamefont {Knolle}}]{Natori2023}%
  \BibitemOpen
  \bibfield  {author} {\bibinfo {author} {\bibfnamefont {W.~M.~H.}\
  \bibnamefont {Natori}}, \bibinfo {author} {\bibfnamefont {H.-K.}\
  \bibnamefont {Jin}},\ and\ \bibinfo {author} {\bibfnamefont {J.}~\bibnamefont
  {Knolle}},\ }\bibfield  {title} {\bibinfo {title} {Quantum liquids of the
  ${S}=\frac{3}{2}$ {K}itaev honeycomb and related {K}ugel-{K}homskii models},\
  }\href {https://doi.org/10.1103/PhysRevB.108.075111} {\bibfield  {journal}
  {\bibinfo  {journal} {Phys.\ Rev.\ B}\ }\textbf {\bibinfo {volume} {108}},\
  \bibinfo {pages} {075111} (\bibinfo {year} {2023})}\BibitemShut {NoStop}%
\bibitem [{\citenamefont {Sizyuk}\ \emph {et~al.}(2016)\citenamefont {Sizyuk},
  \citenamefont {W\"olfle},\ and\ \citenamefont {Perkins}}]{Sizyuk2016}%
  \BibitemOpen
  \bibfield  {author} {\bibinfo {author} {\bibfnamefont {Y.}~\bibnamefont
  {Sizyuk}}, \bibinfo {author} {\bibfnamefont {P.}~\bibnamefont {W\"olfle}},\
  and\ \bibinfo {author} {\bibfnamefont {N.~B.}\ \bibnamefont {Perkins}},\
  }\bibfield  {title} {\bibinfo {title} {{Selection of direction of the ordered
  moments in ${\text{Na}}_{2}{\text{IrO}}_{3}$ and
  $\ensuremath{\alpha}\ensuremath{-}{\text{RuCl}}_{3}$}},\ }\href
  {https://doi.org/10.1103/PhysRevB.94.085109} {\bibfield  {journal} {\bibinfo
  {journal} {Phys.\ Rev.\ B}\ }\textbf {\bibinfo {volume} {94}},\ \bibinfo
  {pages} {085109} (\bibinfo {year} {2016})}\BibitemShut {NoStop}%
\bibitem [{\citenamefont {W\"olfle}\ \emph {et~al.}(2017)\citenamefont
  {W\"olfle}, \citenamefont {Perkins},\ and\ \citenamefont
  {Sizyuk}}]{Peter2017}%
  \BibitemOpen
  \bibfield  {author} {\bibinfo {author} {\bibfnamefont {P.}~\bibnamefont
  {W\"olfle}}, \bibinfo {author} {\bibfnamefont {N.~B.}\ \bibnamefont
  {Perkins}},\ and\ \bibinfo {author} {\bibfnamefont {Y.}~\bibnamefont
  {Sizyuk}},\ }\bibfield  {title} {\bibinfo {title} {{Free energy of quantum
  spin systems: Functional integral representation}},\ }\href
  {https://doi.org/10.1103/PhysRevB.95.184408} {\bibfield  {journal} {\bibinfo
  {journal} {Phys.\ Rev.\ B}\ }\textbf {\bibinfo {volume} {95}},\ \bibinfo
  {pages} {184408} (\bibinfo {year} {2017})}\BibitemShut {NoStop}%
\bibitem [{\citenamefont {Sela}\ \emph {et~al.}(2014)\citenamefont {Sela},
  \citenamefont {Jiang}, \citenamefont {Gerlach},\ and\ \citenamefont
  {Trebst}}]{Sela2014PRB}%
  \BibitemOpen
  \bibfield  {author} {\bibinfo {author} {\bibfnamefont {E.}~\bibnamefont
  {Sela}}, \bibinfo {author} {\bibfnamefont {H.-C.}\ \bibnamefont {Jiang}},
  \bibinfo {author} {\bibfnamefont {M.~H.}\ \bibnamefont {Gerlach}},\ and\
  \bibinfo {author} {\bibfnamefont {S.}~\bibnamefont {Trebst}},\ }\bibfield
  {title} {\bibinfo {title} {{Order-by-disorder and spin-orbital liquids in a
  distorted Heisenberg-Kitaev model}},\ }\href
  {https://doi.org/10.1103/PhysRevB.90.035113} {\bibfield  {journal} {\bibinfo
  {journal} {Phys.\ Rev.\ B}\ }\textbf {\bibinfo {volume} {90}},\ \bibinfo
  {pages} {035113} (\bibinfo {year} {2014})}\BibitemShut {NoStop}%
\bibitem [{\citenamefont {Chaloupka}\ and\ \citenamefont
  {Khaliullin}(2016)}]{Chaloupka2016}%
  \BibitemOpen
  \bibfield  {author} {\bibinfo {author} {\bibfnamefont {J.}~\bibnamefont
  {Chaloupka}}\ and\ \bibinfo {author} {\bibfnamefont {G.}~\bibnamefont
  {Khaliullin}},\ }\bibfield  {title} {\bibinfo {title} {{Magnetic anisotropy
  in the Kitaev model systems ${\mathrm{Na}}_{2}{\mathrm{IrO}}_{3}$ and
  ${\mathrm{RuCl}}_{3}$}},\ }\href {https://doi.org/10.1103/PhysRevB.94.064435}
  {\bibfield  {journal} {\bibinfo  {journal} {Phys.\ Rev.\ B}\ }\textbf
  {\bibinfo {volume} {94}},\ \bibinfo {pages} {064435} (\bibinfo {year}
  {2016})}\BibitemShut {NoStop}%
\bibitem [{\citenamefont {Janssen}\ \emph {et~al.}(2016)\citenamefont
  {Janssen}, \citenamefont {Andrade},\ and\ \citenamefont
  {Vojta}}]{Janssen2016}%
  \BibitemOpen
  \bibfield  {author} {\bibinfo {author} {\bibfnamefont {L.}~\bibnamefont
  {Janssen}}, \bibinfo {author} {\bibfnamefont {E.~C.}\ \bibnamefont
  {Andrade}},\ and\ \bibinfo {author} {\bibfnamefont {M.}~\bibnamefont
  {Vojta}},\ }\bibfield  {title} {\bibinfo {title} {{Honeycomb-lattice
  Heisenberg-Kitaev Model in a magnetic field: Spin canting, metamagnetism, and
  vortex crystals}},\ }\href {https://doi.org/10.1103/PhysRevLett.117.277202}
  {\bibfield  {journal} {\bibinfo  {journal} {Phys.\ Rev.\ Lett.}\ }\textbf
  {\bibinfo {volume} {117}},\ \bibinfo {pages} {277202} (\bibinfo {year}
  {2016})}\BibitemShut {NoStop}%
\bibitem [{\citenamefont {Chern}\ \emph {et~al.}(2017)\citenamefont {Chern},
  \citenamefont {Sizyuk}, \citenamefont {Price},\ and\ \citenamefont
  {Perkins}}]{Chern2017}%
  \BibitemOpen
  \bibfield  {author} {\bibinfo {author} {\bibfnamefont {G.-W.}\ \bibnamefont
  {Chern}}, \bibinfo {author} {\bibfnamefont {Y.}~\bibnamefont {Sizyuk}},
  \bibinfo {author} {\bibfnamefont {C.}~\bibnamefont {Price}},\ and\ \bibinfo
  {author} {\bibfnamefont {N.~B.}\ \bibnamefont {Perkins}},\ }\bibfield
  {title} {\bibinfo {title} {{Kitaev-Heisenberg model in a magnetic field:
  Order-by-disorder and commensurate-incommensurate transitions}},\ }\href
  {https://doi.org/10.1103/PhysRevB.95.144427} {\bibfield  {journal} {\bibinfo
  {journal} {Phys.\ Rev.\ B}\ }\textbf {\bibinfo {volume} {95}},\ \bibinfo
  {pages} {144427} (\bibinfo {year} {2017})}\BibitemShut {NoStop}%
\bibitem [{\citenamefont {Bartlett}\ and\ \citenamefont
  {Musia\l{}}(2007)}]{Bartlett-Musial_2007}%
  \BibitemOpen
  \bibfield  {author} {\bibinfo {author} {\bibfnamefont {R.~J.}\ \bibnamefont
  {Bartlett}}\ and\ \bibinfo {author} {\bibfnamefont {M.}~\bibnamefont
  {Musia\l{}}},\ }\bibfield  {title} {\bibinfo {title} {Coupled-cluster theory
  in quantum chemistry},\ }\href {https://doi.org/10.1103/RevModPhys.79.291}
  {\bibfield  {journal} {\bibinfo  {journal} {Rev.\ Mod.\ Phys.}\ }\textbf
  {\bibinfo {volume} {79}},\ \bibinfo {pages} {291} (\bibinfo {year}
  {2007})}\BibitemShut {NoStop}%
\bibitem [{\citenamefont {Hagen}\ \emph {et~al.}(2014)\citenamefont {Hagen},
  \citenamefont {Papenbrock}, \citenamefont {Hjorth-Jensen},\ and\
  \citenamefont {Dean}}]{Hagen_2014}%
  \BibitemOpen
  \bibfield  {author} {\bibinfo {author} {\bibfnamefont {G.}~\bibnamefont
  {Hagen}}, \bibinfo {author} {\bibfnamefont {T.}~\bibnamefont {Papenbrock}},
  \bibinfo {author} {\bibfnamefont {M.}~\bibnamefont {Hjorth-Jensen}},\ and\
  \bibinfo {author} {\bibfnamefont {D.~J.}\ \bibnamefont {Dean}},\ }\bibfield
  {title} {\bibinfo {title} {Coupled-cluster computations of atomic nuclei},\
  }\href {https://doi.org/10.1088/0034-4885/77/9/096302} {\bibfield  {journal}
  {\bibinfo  {journal} {Rep.\ Prog.\ Phys.}\ }\textbf {\bibinfo {volume}
  {77}},\ \bibinfo {pages} {096302} (\bibinfo {year} {2014})}\BibitemShut
  {NoStop}%
\bibitem [{\citenamefont {K\"ummel}(1983)}]{Kuemmel_1983}%
  \BibitemOpen
  \bibfield  {author} {\bibinfo {author} {\bibfnamefont {H.}~\bibnamefont
  {K\"ummel}},\ }\bibfield  {title} {\bibinfo {title} {Effective operators in
  the relativistic meson-nucleon system},\ }\href
  {https://doi.org/10.1103/PhysRevC.27.765} {\bibfield  {journal} {\bibinfo
  {journal} {Phys.\ Rev.\ C}\ }\textbf {\bibinfo {volume} {27}},\ \bibinfo
  {pages} {765} (\bibinfo {year} {1983})}\BibitemShut {NoStop}%
\bibitem [{\citenamefont {Hasberg}\ and\ \citenamefont
  {K\"ummel}(1986)}]{Hasberg-Kuemmel_1986}%
  \BibitemOpen
  \bibfield  {author} {\bibinfo {author} {\bibfnamefont {G.}~\bibnamefont
  {Hasberg}}\ and\ \bibinfo {author} {\bibfnamefont {H.}~\bibnamefont
  {K\"ummel}},\ }\bibfield  {title} {\bibinfo {title} {Coupled cluster
  description of pion-nucleon systems},\ }\href
  {https://doi.org/10.1103/PhysRevC.33.1367} {\bibfield  {journal} {\bibinfo
  {journal} {Phys.\ Rev.\ C}\ }\textbf {\bibinfo {volume} {33}},\ \bibinfo
  {pages} {1367} (\bibinfo {year} {1986})}\BibitemShut {NoStop}%
\bibitem [{\citenamefont {Kaulfuss}(1985)}]{Kaulfuss_1985}%
  \BibitemOpen
  \bibfield  {author} {\bibinfo {author} {\bibfnamefont {U.}~\bibnamefont
  {Kaulfuss}},\ }\bibfield  {title} {\bibinfo {title} {Renormalization of the
  coupled cluster equations in three-dimensional ${\ensuremath{\varphi}}^{4}$
  quantum field theory},\ }\href {https://doi.org/10.1103/PhysRevD.32.1421}
  {\bibfield  {journal} {\bibinfo  {journal} {Phys.\ Rev.\ D}\ }\textbf
  {\bibinfo {volume} {32}},\ \bibinfo {pages} {1421} (\bibinfo {year}
  {1985})}\BibitemShut {NoStop}%
\bibitem [{\citenamefont {Altenbokum}\ and\ \citenamefont
  {K\"ummel}(1985)}]{Altenokum-Kuemmel_1985}%
  \BibitemOpen
  \bibfield  {author} {\bibinfo {author} {\bibfnamefont {M.}~\bibnamefont
  {Altenbokum}}\ and\ \bibinfo {author} {\bibfnamefont {H.}~\bibnamefont
  {K\"ummel}},\ }\bibfield  {title} {\bibinfo {title} {Soliton sector of the
  (${\ensuremath{\varphi}}^{4}$${)}_{2}$ quantum field theory in the {H}artree
  approximation},\ }\href {https://doi.org/10.1103/PhysRevD.32.2014} {\bibfield
   {journal} {\bibinfo  {journal} {Phys.\ Rev.\ D}\ }\textbf {\bibinfo {volume}
  {32}},\ \bibinfo {pages} {2014} (\bibinfo {year} {1985})}\BibitemShut
  {NoStop}%
\bibitem [{\citenamefont {Baker}\ \emph {et~al.}(1996)\citenamefont {Baker},
  \citenamefont {Bishop},\ and\ \citenamefont
  {Davidson}}]{Baker-Bishop-Davidson_1996}%
  \BibitemOpen
  \bibfield  {author} {\bibinfo {author} {\bibfnamefont {S.~J.}\ \bibnamefont
  {Baker}}, \bibinfo {author} {\bibfnamefont {R.~F.}\ \bibnamefont {Bishop}},\
  and\ \bibinfo {author} {\bibfnamefont {N.~J.}\ \bibnamefont {Davidson}},\
  }\bibfield  {title} {\bibinfo {title} {Coupled cluster analysis of the {U}(1)
  lattice gauge model using a correlated "mean-field" reference state},\ }\href
  {https://doi.org/10.1103/PhysRevD.53.2610} {\bibfield  {journal} {\bibinfo
  {journal} {Phys.\ Rev.\ D}\ }\textbf {\bibinfo {volume} {53}},\ \bibinfo
  {pages} {2610} (\bibinfo {year} {1996})}\BibitemShut {NoStop}%
\bibitem [{\citenamefont {Ligterink}\ \emph {et~al.}(1998)\citenamefont
  {Ligterink}, \citenamefont {Walet},\ and\ \citenamefont
  {Bishop}}]{Ligterink-Walet-Bishop_1998}%
  \BibitemOpen
  \bibfield  {author} {\bibinfo {author} {\bibfnamefont {N.~E.}\ \bibnamefont
  {Ligterink}}, \bibinfo {author} {\bibfnamefont {N.~R.}\ \bibnamefont
  {Walet}},\ and\ \bibinfo {author} {\bibfnamefont {R.~F.}\ \bibnamefont
  {Bishop}},\ }\bibfield  {title} {\bibinfo {title} {A coupled-cluster
  formulation of {H}amiltonian lattice field theory: {T}he nonlinear sigma
  model},\ }\href {https://doi.org/10.1006/aphy.1998.5812} {\bibfield
  {journal} {\bibinfo  {journal} {Ann.\ Phys.\ (N.Y.)}\ }\textbf {\bibinfo
  {volume} {267}},\ \bibinfo {pages} {97} (\bibinfo {year} {1998})}\BibitemShut
  {NoStop}%
\bibitem [{\citenamefont {Ligterink}\ \emph {et~al.}(2000)\citenamefont
  {Ligterink}, \citenamefont {Walet},\ and\ \citenamefont
  {Bishop}}]{Ligterink-Walet-Bishop_2000}%
  \BibitemOpen
  \bibfield  {author} {\bibinfo {author} {\bibfnamefont {N.~E.}\ \bibnamefont
  {Ligterink}}, \bibinfo {author} {\bibfnamefont {N.~R.}\ \bibnamefont
  {Walet}},\ and\ \bibinfo {author} {\bibfnamefont {R.~F.}\ \bibnamefont
  {Bishop}},\ }\bibfield  {title} {\bibinfo {title} {Toward a many-body
  treatment of {H}amiltonian lattice {$SU(N)$} gauge theory},\ }\href
  {https://doi.org/10.1006/aphy.2000.6070} {\bibfield  {journal} {\bibinfo
  {journal} {Ann.\ Phys.\ (N.Y.)}\ }\textbf {\bibinfo {volume} {284}},\
  \bibinfo {pages} {215} (\bibinfo {year} {2000})}\BibitemShut {NoStop}%
\bibitem [{\citenamefont {Zeng}\ \emph {et~al.}(1998)\citenamefont {Zeng},
  \citenamefont {Farnell},\ and\ \citenamefont {Bishop}}]{Zeng_et-al_1998}%
  \BibitemOpen
  \bibfield  {author} {\bibinfo {author} {\bibfnamefont {C.}~\bibnamefont
  {Zeng}}, \bibinfo {author} {\bibfnamefont {D.~J.~J.}\ \bibnamefont
  {Farnell}},\ and\ \bibinfo {author} {\bibfnamefont {R.~F.}\ \bibnamefont
  {Bishop}},\ }\bibfield  {title} {\bibinfo {title} {An efficient
  implementation of high-order coupled-cluster techniques applied to quantum
  magnets},\ }\href {https://doi.org/10.1023/A:1023220222019} {\bibfield
  {journal} {\bibinfo  {journal} {J.\ Stat.\ Phys.}\ }\textbf {\bibinfo
  {volume} {90}},\ \bibinfo {pages} {327} (\bibinfo {year} {1998})}\BibitemShut
  {NoStop}%
\bibitem [{\citenamefont {Bishop}\ \emph {et~al.}(2000)\citenamefont {Bishop},
  \citenamefont {Farnell}, \citenamefont {Kr\"uger}, \citenamefont {Parkinson},
  \citenamefont {Richter},\ and\ \citenamefont {Zeng}}]{Bishop2000}%
  \BibitemOpen
  \bibfield  {author} {\bibinfo {author} {\bibfnamefont {R.~F.}\ \bibnamefont
  {Bishop}}, \bibinfo {author} {\bibfnamefont {D.~J.~J.}\ \bibnamefont
  {Farnell}}, \bibinfo {author} {\bibfnamefont {S.~E.}\ \bibnamefont
  {Kr\"uger}}, \bibinfo {author} {\bibfnamefont {J.~B.}\ \bibnamefont
  {Parkinson}}, \bibinfo {author} {\bibfnamefont {J.}~\bibnamefont {Richter}},\
  and\ \bibinfo {author} {\bibfnamefont {C.}~\bibnamefont {Zeng}},\ }\bibfield
  {title} {\bibinfo {title} {High-order coupled cluster method calculations for
  the ground- and excited-state properties of the spin-half {$XXZ$} model},\
  }\href {https://doi.org/10.1088/0953-8984/12/30/317} {\bibfield  {journal}
  {\bibinfo  {journal} {J.\ Phys.: Condens.\ Matter}\ }\textbf {\bibinfo
  {volume} {12}},\ \bibinfo {pages} {6887} (\bibinfo {year}
  {2000})}\BibitemShut {NoStop}%
\bibitem [{\citenamefont {Kr\"uger}\ \emph {et~al.}(2000)\citenamefont
  {Kr\"uger}, \citenamefont {Richter}, \citenamefont {Schulenburg},
  \citenamefont {Farnell},\ and\ \citenamefont {Bishop}}]{Kruger2000}%
  \BibitemOpen
  \bibfield  {author} {\bibinfo {author} {\bibfnamefont {S.~E.}\ \bibnamefont
  {Kr\"uger}}, \bibinfo {author} {\bibfnamefont {J.}~\bibnamefont {Richter}},
  \bibinfo {author} {\bibfnamefont {J.}~\bibnamefont {Schulenburg}}, \bibinfo
  {author} {\bibfnamefont {D.~J.~J.}\ \bibnamefont {Farnell}},\ and\ \bibinfo
  {author} {\bibfnamefont {R.~F.}\ \bibnamefont {Bishop}},\ }\bibfield  {title}
  {\bibinfo {title} {Quantum phase transitions of a square-lattice {H}eisenberg
  antiferromagnet with two kinds of nearest-neighbor bonds: {A} high-order
  coupled-cluster treatment},\ }\href
  {https://doi.org/10.1103/PhysRevB.61.14607} {\bibfield  {journal} {\bibinfo
  {journal} {Phys.\ Rev.\ B}\ }\textbf {\bibinfo {volume} {61}},\ \bibinfo
  {pages} {14607} (\bibinfo {year} {2000})}\BibitemShut {NoStop}%
\bibitem [{\citenamefont {Farnell}\ \emph {et~al.}(2001)\citenamefont
  {Farnell}, \citenamefont {Gernoth},\ and\ \citenamefont
  {Bishop}}]{Farnell-et-al_2001}%
  \BibitemOpen
  \bibfield  {author} {\bibinfo {author} {\bibfnamefont {D.~J.~J.}\
  \bibnamefont {Farnell}}, \bibinfo {author} {\bibfnamefont {K.~A.}\
  \bibnamefont {Gernoth}},\ and\ \bibinfo {author} {\bibfnamefont {R.~F.}\
  \bibnamefont {Bishop}},\ }\bibfield  {title} {\bibinfo {title} {High-order
  coupled-cluster method for general spin-lattice problems: {A}n illustration
  via the anisotropic {H}eisenberg model},\ }\href
  {https://doi.org/10.1103/PhysRevB.64.172409} {\bibfield  {journal} {\bibinfo
  {journal} {Phys.\ Rev.\ B}\ }\textbf {\bibinfo {volume} {64}},\ \bibinfo
  {pages} {172409} (\bibinfo {year} {2001})}\BibitemShut {NoStop}%
\bibitem [{\citenamefont {Farnell}\ and\ \citenamefont
  {Bishop}(2004)}]{Farnell-Bishop_2004}%
  \BibitemOpen
  \bibfield  {author} {\bibinfo {author} {\bibfnamefont {D.~J.~J.}\
  \bibnamefont {Farnell}}\ and\ \bibinfo {author} {\bibfnamefont {R.~F.}\
  \bibnamefont {Bishop}},\ }\bibfield  {title} {\bibinfo {title} {The coupled
  cluster method applied to quantum magnetism},\ }in\ \href
  {https://doi.org/10.1007/BFb0119597} {\emph {\bibinfo {booktitle} {Quantum
  Magnetism}}},\ \bibinfo {series} {Lecture Notes in Physics}, Vol.~\bibinfo
  {volume} {{\bf 645}},\ \bibinfo {editor} {edited by\ \bibinfo {editor}
  {\bibfnamefont {U.}~\bibnamefont {Schollw{\"o}ck}}, \bibinfo {editor}
  {\bibfnamefont {J.}~\bibnamefont {Richter}}, \bibinfo {editor} {\bibfnamefont
  {D.~J.~J.}\ \bibnamefont {Farnell}},\ and\ \bibinfo {editor} {\bibfnamefont
  {R.~F.}\ \bibnamefont {Bishop}}}\ (\bibinfo  {publisher} {Springer-Verlag},\
  \bibinfo {address} {Berlin},\ \bibinfo {year} {2004})\ pp.\ \bibinfo {pages}
  {307--348}\BibitemShut {NoStop}%
\bibitem [{\citenamefont {Darradi}\ \emph {et~al.}(2005)\citenamefont
  {Darradi}, \citenamefont {Richter},\ and\ \citenamefont
  {Farnell}}]{Darradi2005}%
  \BibitemOpen
  \bibfield  {author} {\bibinfo {author} {\bibfnamefont {R.}~\bibnamefont
  {Darradi}}, \bibinfo {author} {\bibfnamefont {J.}~\bibnamefont {Richter}},\
  and\ \bibinfo {author} {\bibfnamefont {D.~J.~J.}\ \bibnamefont {Farnell}},\
  }\bibfield  {title} {\bibinfo {title} {Coupled cluster treatment of the
  {S}hastry-{S}utherland antiferromagnet},\ }\href
  {https://doi.org/10.1103/PhysRevB.72.104425} {\bibfield  {journal} {\bibinfo
  {journal} {Phys.\ Rev.\ B}\ }\textbf {\bibinfo {volume} {72}},\ \bibinfo
  {pages} {104425} (\bibinfo {year} {2005})}\BibitemShut {NoStop}%
\bibitem [{\citenamefont {Bishop}\ \emph
  {et~al.}(2008{\natexlab{a}})\citenamefont {Bishop}, \citenamefont {Li},
  \citenamefont {Darradi},\ and\ \citenamefont {Richter}}]{Bishop2008}%
  \BibitemOpen
  \bibfield  {author} {\bibinfo {author} {\bibfnamefont {R.~F.}\ \bibnamefont
  {Bishop}}, \bibinfo {author} {\bibfnamefont {P.~H.~Y.}\ \bibnamefont {Li}},
  \bibinfo {author} {\bibfnamefont {R.}~\bibnamefont {Darradi}},\ and\ \bibinfo
  {author} {\bibfnamefont {J.}~\bibnamefont {Richter}},\ }\bibfield  {title}
  {\bibinfo {title} {The quantum {$J_1$--$J{_1}^{\ensuremath{\prime}}$--$J_2$}
  spin-$1/2$ {H}eisenberg model: influence of the interchain coupling on the
  ground-state magnetic ordering in two dimensions},\ }\href
  {https://doi.org/10.1088/0953-8984/20/25/255251} {\bibfield  {journal}
  {\bibinfo  {journal} {J.\ Phys.: Condens.\ Matter}\ }\textbf {\bibinfo
  {volume} {20}},\ \bibinfo {pages} {255251} (\bibinfo {year}
  {2008}{\natexlab{a}})}\BibitemShut {NoStop}%
\bibitem [{\citenamefont {Bishop}\ \emph
  {et~al.}(2008{\natexlab{b}})\citenamefont {Bishop}, \citenamefont {Li},
  \citenamefont {Darradi}, \citenamefont {Schulenburg},\ and\ \citenamefont
  {Richter}}]{Bishop-et-al_2008}%
  \BibitemOpen
  \bibfield  {author} {\bibinfo {author} {\bibfnamefont {R.~F.}\ \bibnamefont
  {Bishop}}, \bibinfo {author} {\bibfnamefont {P.~H.~Y.}\ \bibnamefont {Li}},
  \bibinfo {author} {\bibfnamefont {R.}~\bibnamefont {Darradi}}, \bibinfo
  {author} {\bibfnamefont {J.}~\bibnamefont {Schulenburg}},\ and\ \bibinfo
  {author} {\bibfnamefont {J.}~\bibnamefont {Richter}},\ }\bibfield  {title}
  {\bibinfo {title} {Effect of anisotropy on the ground-state magnetic ordering
  of the spin-half quantum ${J}_{1}^{XXZ}$--${J}_{2}^{XXZ}$ model on the square
  lattice},\ }\href {https://doi.org/10.1103/PhysRevB.78.054412} {\bibfield
  {journal} {\bibinfo  {journal} {Phys.\ Rev.\ B}\ }\textbf {\bibinfo {volume}
  {78}},\ \bibinfo {pages} {054412} (\bibinfo {year}
  {2008}{\natexlab{b}})}\BibitemShut {NoStop}%
\bibitem [{\citenamefont {Darradi}\ \emph {et~al.}(2008)\citenamefont
  {Darradi}, \citenamefont {Derzhko}, \citenamefont {Zinke}, \citenamefont
  {Schulenburg}, \citenamefont {Kr\"uger},\ and\ \citenamefont
  {Richter}}]{Darradi-et-al_2008}%
  \BibitemOpen
  \bibfield  {author} {\bibinfo {author} {\bibfnamefont {R.}~\bibnamefont
  {Darradi}}, \bibinfo {author} {\bibfnamefont {O.}~\bibnamefont {Derzhko}},
  \bibinfo {author} {\bibfnamefont {R.}~\bibnamefont {Zinke}}, \bibinfo
  {author} {\bibfnamefont {J.}~\bibnamefont {Schulenburg}}, \bibinfo {author}
  {\bibfnamefont {S.~E.}\ \bibnamefont {Kr\"uger}},\ and\ \bibinfo {author}
  {\bibfnamefont {J.}~\bibnamefont {Richter}},\ }\bibfield  {title} {\bibinfo
  {title} {Ground state phases of the spin-1/2 ${J}_{1}$--${J}_{2}$
  {H}eisenberg antiferromagnet on the square lattice: {A} high-order coupled
  cluster treatment},\ }\href {https://doi.org/10.1103/PhysRevB.78.214415}
  {\bibfield  {journal} {\bibinfo  {journal} {Phys.\ Rev.\ B}\ }\textbf
  {\bibinfo {volume} {78}},\ \bibinfo {pages} {214415} (\bibinfo {year}
  {2008})}\BibitemShut {NoStop}%
\bibitem [{\citenamefont {Bishop}\ and\ \citenamefont {Li}(2011)}]{Bishop2011}%
  \BibitemOpen
  \bibfield  {author} {\bibinfo {author} {\bibfnamefont {R.~F.}\ \bibnamefont
  {Bishop}}\ and\ \bibinfo {author} {\bibfnamefont {P.~H.~Y.}\ \bibnamefont
  {Li}},\ }\bibfield  {title} {\bibinfo {title} {Coupled-cluster method: {A}
  lattice-path-based subsystem approximation scheme for quantum lattice
  models},\ }\href {https://doi.org/10.1103/PhysRevA.83.042111} {\bibfield
  {journal} {\bibinfo  {journal} {Phys.\ Rev.\ A}\ }\textbf {\bibinfo {volume}
  {83}},\ \bibinfo {pages} {042111} (\bibinfo {year} {2011})}\BibitemShut
  {NoStop}%
\bibitem [{\citenamefont {Farnell}\ \emph {et~al.}(2011)\citenamefont
  {Farnell}, \citenamefont {Darradi}, \citenamefont {Schmidt},\ and\
  \citenamefont {Richter}}]{Farnell2011}%
  \BibitemOpen
  \bibfield  {author} {\bibinfo {author} {\bibfnamefont {D.~J.~J.}\
  \bibnamefont {Farnell}}, \bibinfo {author} {\bibfnamefont {R.}~\bibnamefont
  {Darradi}}, \bibinfo {author} {\bibfnamefont {R.}~\bibnamefont {Schmidt}},\
  and\ \bibinfo {author} {\bibfnamefont {J.}~\bibnamefont {Richter}},\
  }\bibfield  {title} {\bibinfo {title} {Spin-half {H}eisenberg antiferromagnet
  on two archimedian lattices: {F}rom the bounce lattice to the maple-leaf
  lattice and beyond},\ }\href {https://doi.org/10.1103/PhysRevB.84.104406}
  {\bibfield  {journal} {\bibinfo  {journal} {Phys.\ Rev.\ B}\ }\textbf
  {\bibinfo {volume} {84}},\ \bibinfo {pages} {104406} (\bibinfo {year}
  {2011})}\BibitemShut {NoStop}%
\bibitem [{\citenamefont {Reuther}\ \emph
  {et~al.}(2011{\natexlab{b}})\citenamefont {Reuther}, \citenamefont
  {W\"olfle}, \citenamefont {Darradi}, \citenamefont {Brenig}, \citenamefont
  {Arlego},\ and\ \citenamefont {Richter}}]{Reuther-et-al_2011}%
  \BibitemOpen
  \bibfield  {author} {\bibinfo {author} {\bibfnamefont {J.}~\bibnamefont
  {Reuther}}, \bibinfo {author} {\bibfnamefont {P.}~\bibnamefont {W\"olfle}},
  \bibinfo {author} {\bibfnamefont {R.}~\bibnamefont {Darradi}}, \bibinfo
  {author} {\bibfnamefont {W.}~\bibnamefont {Brenig}}, \bibinfo {author}
  {\bibfnamefont {M.}~\bibnamefont {Arlego}},\ and\ \bibinfo {author}
  {\bibfnamefont {J.}~\bibnamefont {Richter}},\ }\bibfield  {title} {\bibinfo
  {title} {Quantum phases of the planar antiferromagnetic
  ${J}_{1}\ensuremath{-}{J}_{2}\ensuremath{-}{J}_{3}$ {H}eisenberg model},\
  }\href {https://doi.org/10.1103/PhysRevB.83.064416} {\bibfield  {journal}
  {\bibinfo  {journal} {Phys.\ Rev.\ B}\ }\textbf {\bibinfo {volume} {83}},\
  \bibinfo {pages} {064416} (\bibinfo {year} {2011}{\natexlab{b}})}\BibitemShut
  {NoStop}%
\bibitem [{\citenamefont {G\"otze}\ \emph {et~al.}(2011)\citenamefont
  {G\"otze}, \citenamefont {Farnell}, \citenamefont {Bishop}, \citenamefont
  {Li},\ and\ \citenamefont {Richter}}]{Gotze2011}%
  \BibitemOpen
  \bibfield  {author} {\bibinfo {author} {\bibfnamefont {O.}~\bibnamefont
  {G\"otze}}, \bibinfo {author} {\bibfnamefont {D.~J.~J.}\ \bibnamefont
  {Farnell}}, \bibinfo {author} {\bibfnamefont {R.~F.}\ \bibnamefont {Bishop}},
  \bibinfo {author} {\bibfnamefont {P.~H.~Y.}\ \bibnamefont {Li}},\ and\
  \bibinfo {author} {\bibfnamefont {J.}~\bibnamefont {Richter}},\ }\bibfield
  {title} {\bibinfo {title} {Heisenberg antiferromagnet on the kagome lattice
  with arbitrary spin: {A} higher-order coupled cluster treatment},\ }\href
  {https://doi.org/10.1103/PhysRevB.84.224428} {\bibfield  {journal} {\bibinfo
  {journal} {Phys.\ Rev.\ B}\ }\textbf {\bibinfo {volume} {84}},\ \bibinfo
  {pages} {224428} (\bibinfo {year} {2011})}\BibitemShut {NoStop}%
\bibitem [{\citenamefont {G\"otze}\ \emph {et~al.}(2012)\citenamefont
  {G\"otze}, \citenamefont {Kr\"uger}, \citenamefont {Fleck}, \citenamefont
  {Schulenburg},\ and\ \citenamefont {Richter}}]{Gotze-et-al_2012}%
  \BibitemOpen
  \bibfield  {author} {\bibinfo {author} {\bibfnamefont {O.}~\bibnamefont
  {G\"otze}}, \bibinfo {author} {\bibfnamefont {S.~E.}\ \bibnamefont
  {Kr\"uger}}, \bibinfo {author} {\bibfnamefont {F.}~\bibnamefont {Fleck}},
  \bibinfo {author} {\bibfnamefont {J.}~\bibnamefont {Schulenburg}},\ and\
  \bibinfo {author} {\bibfnamefont {J.}~\bibnamefont {Richter}},\ }\bibfield
  {title} {\bibinfo {title} {Ground-state phase diagram of the
  spin-$\frac{1}{2}$ square- lattice ${J}_{1}$-${J}_{2}$ model with plaquette
  structure},\ }\href {https://doi.org/10.1103/PhysRevB.85.224424} {\bibfield
  {journal} {\bibinfo  {journal} {Phys.\ Rev.\ B}\ }\textbf {\bibinfo {volume}
  {85}},\ \bibinfo {pages} {224424h} (\bibinfo {year} {2012})}\BibitemShut
  {NoStop}%
\bibitem [{\citenamefont {Bishop}\ \emph {et~al.}(2014)\citenamefont {Bishop},
  \citenamefont {Li},\ and\ \citenamefont
  {Campbell}}]{Bishop-Li-Campbell_2014}%
  \BibitemOpen
  \bibfield  {author} {\bibinfo {author} {\bibfnamefont {R.~F.}\ \bibnamefont
  {Bishop}}, \bibinfo {author} {\bibfnamefont {P.~H.~Y.}\ \bibnamefont {Li}},\
  and\ \bibinfo {author} {\bibfnamefont {C.~E.}\ \bibnamefont {Campbell}},\
  }\bibfield  {title} {\bibinfo {title} {Highly frustrated spin-lattice models
  of magnetism and their quantum phase transitions: {A} microscopic treatment
  via the coupled cluster method},\ }\href {https://doi.org/10.1063/1.4899216}
  {\bibfield  {journal} {\bibinfo  {journal} {AIP Conf.\ Proc.}\ }\textbf
  {\bibinfo {volume} {1619}},\ \bibinfo {pages} {40} (\bibinfo {year}
  {2014})}\BibitemShut {NoStop}%
\bibitem [{\citenamefont {Farnell}\ \emph {et~al.}(2014)\citenamefont
  {Farnell}, \citenamefont {G\"otze}, \citenamefont {Richter}, \citenamefont
  {Bishop},\ and\ \citenamefont {Li}}]{Farnell-et-al_2014}%
  \BibitemOpen
  \bibfield  {author} {\bibinfo {author} {\bibfnamefont {D.~J.~J.}\
  \bibnamefont {Farnell}}, \bibinfo {author} {\bibfnamefont {O.}~\bibnamefont
  {G\"otze}}, \bibinfo {author} {\bibfnamefont {J.}~\bibnamefont {Richter}},
  \bibinfo {author} {\bibfnamefont {R.~F.}\ \bibnamefont {Bishop}},\ and\
  \bibinfo {author} {\bibfnamefont {P.~H.~Y.}\ \bibnamefont {Li}},\ }\bibfield
  {title} {\bibinfo {title} {Quantum $s=\frac{1}{2}$ antiferromagnets on
  {A}rchimedean lattices: {T}he route from semiclassical magnetic order to
  nonmagnetic quantum states},\ }\href
  {https://doi.org/10.1103/PhysRevB.89.184407} {\bibfield  {journal} {\bibinfo
  {journal} {Phys.\ Rev.\ B}\ }\textbf {\bibinfo {volume} {89}},\ \bibinfo
  {pages} {184407} (\bibinfo {year} {2014})}\BibitemShut {NoStop}%
\bibitem [{\citenamefont {Li}\ and\ \citenamefont
  {Bishop}(2016)}]{Li-Bishop_2016}%
  \BibitemOpen
  \bibfield  {author} {\bibinfo {author} {\bibfnamefont {P.~H.~Y.}\
  \bibnamefont {Li}}\ and\ \bibinfo {author} {\bibfnamefont {R.~F.}\
  \bibnamefont {Bishop}},\ }\bibfield  {title} {\bibinfo {title} {Ground-state
  phases of the spin-$1$ ${J}_{1}\text{\ensuremath{-}}{J}_{2}$ {H}eisenberg
  antiferromagnet on the honeycomb lattice},\ }\href
  {https://doi.org/10.1103/PhysRevB.93.214438} {\bibfield  {journal} {\bibinfo
  {journal} {Phys.\ Rev.\ B}\ }\textbf {\bibinfo {volume} {93}},\ \bibinfo
  {pages} {214438} (\bibinfo {year} {2016})}\BibitemShut {NoStop}%
\bibitem [{\citenamefont {Bishop}\ \emph {et~al.}(2019)\citenamefont {Bishop},
  \citenamefont {Li}, \citenamefont {G\"otze},\ and\ \citenamefont
  {Richter}}]{Bishop-et-al_2019}%
  \BibitemOpen
  \bibfield  {author} {\bibinfo {author} {\bibfnamefont {R.~F.}\ \bibnamefont
  {Bishop}}, \bibinfo {author} {\bibfnamefont {P.~H.~Y.}\ \bibnamefont {Li}},
  \bibinfo {author} {\bibfnamefont {O.}~\bibnamefont {G\"otze}},\ and\ \bibinfo
  {author} {\bibfnamefont {J.}~\bibnamefont {Richter}},\ }\bibfield  {title}
  {\bibinfo {title} {Frustrated spin-$\frac{1}{2}$ {H}eisenberg magnet on a
  square-lattice bilayer: {H}igh-order study of the quantum critical behavior
  of the
  ${J}_{1}\text{\ensuremath{-}}{J}_{2}\text{\ensuremath{-}}{J}_{1}^{\ensuremath{\perp}}$
  model},\ }\href {https://doi.org/10.1103/PhysRevB.100.024401} {\bibfield
  {journal} {\bibinfo  {journal} {Phys.\ Rev.\ B}\ }\textbf {\bibinfo {volume}
  {100}},\ \bibinfo {pages} {024401} (\bibinfo {year} {2019})}\BibitemShut
  {NoStop}%
\bibitem [{\citenamefont {Farnell}\ \emph {et~al.}(2019)\citenamefont
  {Farnell}, \citenamefont {Bishop},\ and\ \citenamefont
  {Richter}}]{Farnell-et-al_2019}%
  \BibitemOpen
  \bibfield  {author} {\bibinfo {author} {\bibfnamefont {D.~J.~J.}\
  \bibnamefont {Farnell}}, \bibinfo {author} {\bibfnamefont {R.~F.}\
  \bibnamefont {Bishop}},\ and\ \bibinfo {author} {\bibfnamefont
  {J.}~\bibnamefont {Richter}},\ }\bibfield  {title} {\bibinfo {title}
  {Non-coplanar model states in quantum magnetism applications of the
  high-order coupled cluster method},\ }\href
  {https://doi.org/10.1007/s10955-019-02297-1} {\bibfield  {journal} {\bibinfo
  {journal} {J.\ Stat.\ Phys.}\ }\textbf {\bibinfo {volume} {176}},\ \bibinfo
  {pages} {180} (\bibinfo {year} {2019})}\BibitemShut {NoStop}%
\bibitem [{\citenamefont {Li}\ and\ \citenamefont
  {Bishop}(2022)}]{Li-Bishop_2022}%
  \BibitemOpen
  \bibfield  {author} {\bibinfo {author} {\bibfnamefont {P.~H.~Y.}\
  \bibnamefont {Li}}\ and\ \bibinfo {author} {\bibfnamefont {R.~F.}\
  \bibnamefont {Bishop}},\ }\bibfield  {title} {\bibinfo {title} {Frustrated
  spin-$\frac12$ {H}eisenberg magnet on an ${AA}$-stacked honeycomb bilayer:
  {H}igh-order study of the collinear magnetic phases of the
  ${J}_{1}\text{\ensuremath{-}}{J}_{2}\text{\ensuremath{-}}{J}_{1}^{\ensuremath{\perp}}$
  model},\ }\href {https://doi.org/10.1016/j.jmmm.2022.169307} {\bibfield
  {journal} {\bibinfo  {journal} {J.\ Magn.\ Magn.\ Mater.}\ }\textbf {\bibinfo
  {volume} {555}},\ \bibinfo {pages} {169307} (\bibinfo {year}
  {2022})}\BibitemShut {NoStop}%
\bibitem [{\citenamefont {Hellmann}(1935)}]{Hellmann_1935}%
  \BibitemOpen
  \bibfield  {author} {\bibinfo {author} {\bibfnamefont {H.}~\bibnamefont
  {Hellmann}},\ }\bibfield  {title} {\bibinfo {title} {Ein kombiniertes
  {N}äherungsverfahren zur {E}nergieberechnung im {V}ielelektronenproblem},\
  }\href@noop {} {\bibfield  {journal} {\bibinfo  {journal} {Acta Physicochim.\
  URSS}\ }\textbf {\bibinfo {volume} {1}},\ \bibinfo {pages} {913} (\bibinfo
  {year} {1934/1935})}\BibitemShut {NoStop}%
\bibitem [{\citenamefont {Feynman}(1939)}]{Feynman_1939}%
  \BibitemOpen
  \bibfield  {author} {\bibinfo {author} {\bibfnamefont {R.~P.}\ \bibnamefont
  {Feynman}},\ }\bibfield  {title} {\bibinfo {title} {Forces in molecules},\
  }\href {https://doi.org/10.1103/PhysRev.56.340} {\bibfield  {journal}
  {\bibinfo  {journal} {Phys.\ Rev.}\ }\textbf {\bibinfo {volume} {56}},\
  \bibinfo {pages} {340} (\bibinfo {year} {1939})}\BibitemShut {NoStop}%
\bibitem [{\citenamefont {Goldstone}(1957)}]{Goldstone_1957}%
  \BibitemOpen
  \bibfield  {author} {\bibinfo {author} {\bibfnamefont {J.}~\bibnamefont
  {Goldstone}},\ }\bibfield  {title} {\bibinfo {title} {Derivation of the
  {B}rueckner many-body theory},\ }\href
  {https://doi.org/10.1098/rspa.1957.0037} {\bibfield  {journal} {\bibinfo
  {journal} {Proc.\ R.\ Soc.\ London, Ser.\ A}\ }\textbf {\bibinfo {volume}
  {239}},\ \bibinfo {pages} {267} (\bibinfo {year} {1957})}\BibitemShut
  {NoStop}%
\bibitem [{Note1()}]{Note1}%
  \BibitemOpen
  \bibinfo {note} {We use the program package CCCM of D.~J.~J.~Farnell and J.
  Schulenburg, see \protect \url
  {https://www-e.ovgu.de/jschulen/ccm/}.}\BibitemShut {Stop}%
\bibitem [{\citenamefont {C\^onsoli}\ \emph {et~al.}(2020)\citenamefont
  {C\^onsoli}, \citenamefont {Janssen}, \citenamefont {Vojta},\ and\
  \citenamefont {Andrade}}]{Consoli2020}%
  \BibitemOpen
  \bibfield  {author} {\bibinfo {author} {\bibfnamefont {P.~M.}\ \bibnamefont
  {C\^onsoli}}, \bibinfo {author} {\bibfnamefont {L.}~\bibnamefont {Janssen}},
  \bibinfo {author} {\bibfnamefont {M.}~\bibnamefont {Vojta}},\ and\ \bibinfo
  {author} {\bibfnamefont {E.~C.}\ \bibnamefont {Andrade}},\ }\bibfield
  {title} {\bibinfo {title} {{Heisenberg-Kitaev model in a magnetic field:
  $1/S$ expansion}},\ }\href {https://doi.org/10.1103/PhysRevB.102.155134}
  {\bibfield  {journal} {\bibinfo  {journal} {Phys.\ Rev.\ B}\ }\textbf
  {\bibinfo {volume} {102}},\ \bibinfo {pages} {155134} (\bibinfo {year}
  {2020})}\BibitemShut {NoStop}%
\bibitem [{\citenamefont {Pohle}\ \emph {et~al.}(2023)\citenamefont {Pohle},
  \citenamefont {Shannon},\ and\ \citenamefont {Motome}}]{Pohle2023}%
  \BibitemOpen
  \bibfield  {author} {\bibinfo {author} {\bibfnamefont {R.}~\bibnamefont
  {Pohle}}, \bibinfo {author} {\bibfnamefont {N.}~\bibnamefont {Shannon}},\
  and\ \bibinfo {author} {\bibfnamefont {Y.}~\bibnamefont {Motome}},\
  }\bibfield  {title} {\bibinfo {title} {{Spin nematics meet spin liquids:
  Exotic quantum phases in the spin-1 bilinear-biquadratic model with Kitaev
  interactions}},\ }\href {https://doi.org/10.1103/PhysRevB.107.L140403}
  {\bibfield  {journal} {\bibinfo  {journal} {Phys.\ Rev.\ B}\ }\textbf
  {\bibinfo {volume} {107}},\ \bibinfo {pages} {L140403} (\bibinfo {year}
  {2023})}\BibitemShut {NoStop}%
\bibitem [{\citenamefont {Pohle}\ \emph {et~al.}(2024)\citenamefont {Pohle},
  \citenamefont {Shannon},\ and\ \citenamefont {Motome}}]{Pohle2024}%
  \BibitemOpen
  \bibfield  {author} {\bibinfo {author} {\bibfnamefont {R.}~\bibnamefont
  {Pohle}}, \bibinfo {author} {\bibfnamefont {N.}~\bibnamefont {Shannon}},\
  and\ \bibinfo {author} {\bibfnamefont {Y.}~\bibnamefont {Motome}},\
  }\href@noop {} {\bibinfo {title} {{Eight-color chiral spin liquid in the
  $S=1$ bilinear-biquadratic model with Kitaev interactions}}} (\bibinfo {year}
  {2024}),\ \Eprint {https://arxiv.org/abs/2404.11623} {arXiv:2404.11623
  [cond-mat.str-el]} \BibitemShut {NoStop}%
\end{thebibliography}

%apsrev4-2.bst 2019-01-14 (MD) hand-edited version of apsrev4-1.bst
%Control: key (0)
%Control: author (8) initials jnrlst
%Control: editor formatted (1) identically to author
%Control: production of article title (0) allowed
%Control: page (0) single
%Control: year (1) truncated
%Control: production of eprint (0) enabled
%

\end{document}